\documentclass[reprint,amsmath,amssymb,aps,prb,superscriptaddress,longbibliography]{revtex4-1}

\pdfoutput=1

\usepackage{hyperref,url}
\usepackage{amsmath}
\usepackage{amsfonts}
\usepackage{mathtools}
\usepackage[capitalise]{cleveref}
\usepackage[noline,boxruled,commentsnumbered,titlenumbered]{algorithm2e}
\usepackage{algpseudocode}
\usepackage{algorithmicx}

\usepackage{placeins}
\usepackage{mhchem}
\hypersetup{breaklinks,colorlinks=true,linkcolor=blue,citecolor=blue,urlcolor=blue,filecolor=blue}
\usepackage{graphicx}
\usepackage{dcolumn}
\usepackage{rotating}
\usepackage{etoolbox} 
\usepackage{subcaption}

\newcommand\mat\mathbf

\newcommand{\Columbia}{\affiliation{Department of Chemistry, Columbia University, New York, NY, USA}}
\newcommand{\Cal}{\affiliation{Department of Chemistry, University of California, Berkeley, CA, USA}}
\newcommand{\QChem}{\affiliation{Q-Chem Inc., Pleasanton, CA, USA}}
\begin{document}
\author{Joonho Lee}
\email{jl5653@columbia.edu}
\Columbia
\author{Adam Rettig}
\Cal
\author{Xintian Feng}
\QChem
\author{Evgeny Epifanovsky}
\QChem
\author{Martin Head-Gordon}
\Cal
\title{Faster Exact Exchange for Solids via occ-RI-K: Application to Combinatorially Optimized Range-Separated Hybrid Functionals for Simple Solids with Pseudopotentials Near the Basis Set Limit}

\begin{abstract}
In this work, we 
developed and showcased
the occ-RI-K algorithm
to compute the exact exchange contribution
in density functional calculations of solids near the basis set limit.
Within the gaussian planewave (GPW) density fitting, 
our algorithm achieves a 1-2 orders of magnitude speedup
compared to conventional GPW algorithms.
Since our algorithm is well-suited for simulations with large basis sets,
we applied it to 
12 hybrid density functionals with pseudopotentials and a large uncontracted basis set
to assess
their performance on band gaps of 25 simple solids near the basis set limit.
The largest calculation performed in this work involves
16 electrons and 350 basis functions in the unit cell utilizing a 6$\times$6$\times$6 k-mesh.
With 20-27\% exact exchange, global hybrid functionals (B3LYP, PBE0, revPBE0, B97-3, SCAN0) perform similarly with a root-mean-square-deviation (RMSD) of 0.61-0.77 eV while other global hybrid functionals such as M06-2X (2.02 eV) and MN15 (1.05 eV) show higher RMSD due to their increased fraction of exact exchange.
A short-range hybrid functional, HSE achieves a similar RMSD (0.76 eV) but shows a noticeable underestimation of band gaps due to the complete lack of long-range exchange.
We found that
two combinatorially optimized range-separated hybrid functionals, $\omega$B97X-rV (3.94 eV) and $\omega$B97M-rV (3.40 eV),
and the two other range separated hybrid functionals, CAM-B3LYP (2.41 eV) and CAM-QTP01 (4.16 eV),
significantly overestimate the band gap
because of their high fraction of long-range exact exchange.
Given the failure of $\omega$B97X-rV and $\omega$B97M-rV, 
we have yet to find a density functional that offers consistent performance for both molecules and solids.
Our algorithm development and density functional assessment will serve as a stepping stone towards
developing more accurate hybrid functionals and applying them to practical applications.
\end{abstract}
\maketitle
\newpage
\section{Introduction}
Accurate predictions of band gaps ($E_g$)
of semiconductors 
are often
at the center of
computational design of
new
functional materials
with applications to
transistors and photovoltaics.\cite{Jain2013Jul}
Due to its computational efficiency,
Kohn-Sham 
density functional theory (DFT)
has been the workhorse for
this task in modern electronic structure theory.\cite{Mardirossian2017Oct}
However, the accuracy of DFT
can be quite poor
for band gap problems.\cite{Perdew1985Mar} 
Local functionals (i.e., those functionals without exact exchange)
severely underestimate
band gaps\cite{Perdew1985Mar,Yakovkin2007Jun,Mori-Sanchez2008Apr}
whereas
hybrid functionals (i.e., those with exact exchange)
often
overestimate
band gaps.\cite{Heyd2005Nov,Mori-Sanchez2008Apr,Kim2009Jul,Jain2011Nov,Garza2016Oct}
Beyond DFT,
GW methods have been
extremely successful \cite{Hybertsen1986Oct,Reining2018May}
but their
computational
cost ultimately limits their applicability to relatively small solids.
Furthermore, it may suffer from ambiguity due to multiple solutions when attempting self-consistency.\cite{Tandetzky2015Sep}

For main group molecular applications, 
there has been great progress towards
finding statistically better density functionals.
A high-quality database with nearly 5000 reference relative energies was used to assess 200 density functionals.\cite{Mardirossian2017Oct}
Based on that, 
for each rung of density functional, we have identified
statistically best functionals.
Combinatorially optimized
density functionals
(B97M-rV,\cite{Mardirossian2015Feb,Mardirossian2017Jan} $\omega$B97X-rV,\cite{Mardirossian2014May,Mardirossian2017Jan} $\omega$B97M-rV\cite{Mardirossian2016Jun,Mardirossian2017Jan})
developed by the Head-Gordon group
clearly stood out in this benchmark study.
Each of these functionals is the best-performing density functional
among local density functionals, hybrid generalized gradient approximation (GGA) functionals, and
hybrid meta-GGA (mGGA) functionals, respectively. Other benchmark studies have reached similar conclusions,\cite{Goerigk:2017,Najibi:2018b} including for transition-metal containing systems.\cite{dohm2018comprehensive,Chan:2019}

In our recent paper,
we assessed 
the performance of B97M-rV
and 9 other local density functionals for computing the band gaps of 37 simple semiconductors,
using a large Gaussian basis set to reach the basis set limit.\cite{Lee2021Oct}
In that benchmark study,
B97M-rV\cite{Mardirossian2015Feb,Mardirossian2017Jan} was found to have
a root-mean-square-deviation (RMSD) of 1.18 eV and a mean-signed-error (MSD) of -0.85 eV,
significantly underestimating band gaps.
Nonetheless, B97M-rV and other modern mGGA functionals (SCAN,\cite{sun2015strongly} M06-L,\cite{Zhao2006Nov} MN15-L\cite{Yu2016Mar}) were found to be 
more accurate than local density approximation (LDA)\cite{slater1951simplification,perdew1981self} and PBE\cite{perdew1996generalized} functionals.
Motivated by this,
in this work,
we aim to assess the
performance of 
modern hybrid density functionals over the same benchmark set.

Assessments of density functionals 
should be performed at 
the basis set limit as most of them were trained at this limit.
The uncontracted basis set used in our previous study\cite{Lee2021Oct} is
fairly large
and this poses
computational challenges to 
the assessment of hybrid functionals near the basis set limit.
The computational bottleneck of hybrid functionals
is the evaluation of exact exchange
which scales (assuming spatial locality of Gaussian basis functions)
as
$\tilde{\mathcal O}(N_k^2 N^3)$ where $N_k$ is the number of k-points sampled (i.e. dependent on symmetry, but not on system size)
and 
$N$ denotes the size of the computational cell. In other words $N$ represents quantities such as the number of AOs in the computational cell, $n_{AO}$, or the number of real-space grid points, $N_g$, in the cell.
The cubic-scaling evaluation of exact exchange is far more expensive than
the linear-scaling evaluation of the Coulomb matrix, which scales as $\mathcal O (N_k N)$.\cite{Lee2021Oct}

To cope with the steep scaling of exact exchange, we extend the
occ-RI-K algorithm\cite{Manzer2015Jul} developed for molecules
to solids,
which achieves
a significant speedup compared to other algorithms
when a large basis set is used. We implement this new algorithm in Q-Chem.\cite{Epifanovsky2021Aug}
A similar technique known as the adaptively compressed exchange (ACE) algorithm has already been widely used
in planewaves codes.\cite{Lin2016May}
While it does not offer any scaling reduction, the occ-RI-K algorithm 
significantly reduces the prefactor of the exact exchange evaluation
and thereby may enable
extensive benchmark studies in the basis set limit such as those presented in this work.
While we focus on a particular density fitting scheme, the gaussian planewave (GPW) density fitting,\cite{lippert1997hybrid,vandevondele2005quickstep}
our occ-RI-K algorithm should be
applicable to
other periodic density fitting methods.\cite{Franchini2014May,Sun2017Oct}

We note that efficient evaluation of exact exchange for periodic systems has seen great progress
in many other available electronic structure packages.
Packages such as CRYSTAL,\cite{Dovesi2020May} CP2K,\cite{Kuhne2020May} TURBOMOLE,\cite{Balasubramani2020May} FHI-AIMS,\cite{Blum2022Aug} and PySCF\cite{sun2020recent} use Gaussian orbitals like Q-Chem.
These packages often support all-electron calculations that we do not consider in this work. Nonetheless, all-electron calculations can also greatly benefit from occ-RI-K for large basis set calculations as seen in the molecular case.\cite{Manzer2015Jul}
Other codes based on planewaves include QuantumESPRESSO,\cite{Giannozzi2020Apr} VASP,\cite{Hafner2008Oct} FLEUR,\cite{BibEntry2022Sep} and Wien2k.\cite{Blaha2020Feb}
In particular, the first two employ the ACE algorithm to speed up the exact exchange calculations greatly, similar in spirit to our attempt in this work.

This paper is organized as follows:
(1) we review the GPW density fitting scheme
and available exact exchange algorithms,
(2) we then present
the occ-RI-K algorithm for solids within the GPW scheme,
(3) we move to the timing benchmark of our occ-RI-K algorithm compared to other algorithms,
(4) we discuss the performance of hybrid functionals on the band gap benchmark set,
and (5) we then conclude.

\section{Theory}
\subsection{Review of the GPW implementation}\label{subsec:gpw}
We focus on an implementation of exact exchange within the
atomic Bloch orbital framework using
\begin{equation}
\psi_{\mu_\mathbf{k}}(\mathbf r)
=
\frac{1}{\sqrt{N_k}}
\sum_{\mathbf R} e^{i\mathbf k \cdot \mathbf R} \phi_\mu(\mathbf r - \mathbf R).
\end{equation}
where $\phi_\mu$ is the $\mu$-th atomic orbital, $\mathbf R$ is the direct lattice vector, $\mathbf k$ is the 
crystalline momentum, and
$\psi_{\mu_\mathbf{k}}$
is the $\mu$-th Bloch orbital at $\mathbf k$.

The exact exchange energy contribution to the total energy per unit cell in the atomic Bloch orbital basis is
given by
\begin{equation}
\frac{E_K}{N_k} = 
- \frac1{2N_k}
\sum_{\mathbf k_1, \mathbf k_2}
\sum_{\mu\nu\lambda\sigma}
P_{\mu\nu}^{\mathbf k_2}
(\nu_{\mathbf k_2}\lambda_{\mathbf k_1}|\sigma_{\mathbf k_1}\mu_{\mathbf k_2})
P_{\lambda\sigma}^{\mathbf{k}_1}
\end{equation}
where $N_k$ is the number of k-points,
$\mathbf P^\mathbf{k}$
is the density matrix at $\mathbf k$,
\begin{equation}
P^\mathbf{k}_{\mu\nu}
=
\sum_{i\in \text{occ}}
C_{\mu i}^{\mathbf k}
(C_{\nu i}^{\mathbf k})^*
\end{equation}
and 
$(\nu_{\mathbf k_2}\lambda_{\mathbf k_1}|\sigma_{\mathbf k_1}\mu_{\mathbf k_2})$ is defined as
\begin{align}\nonumber
&(\nu_{\mathbf k_2}\lambda_{\mathbf k_1}|\sigma_{\mathbf k_1}\mu_{\mathbf k_2})
=\\ \nonumber
&\int_{\Omega^\ell} \mathrm d\mathbf{r}_1
\int_{\Omega^\ell} \mathrm d\mathbf{r}_2
(\psi_{\nu_{\mathbf{k}_2}}(\mathbf r_1))^* 
\psi_{\lambda_{\mathbf{k}_1}}(\mathbf r_1) \\
&V_\text{coul}(|\mathbf{r_1-r_2}|)
(\psi_{\sigma_{\mathbf{k}_1}}(\mathbf r_2) )^*
\psi_{\mu_{\mathbf{k}_2}}(\mathbf r_2)
\end{align}
where 
$V_\text{coul}$ is the Coulomb operator kernel whose form depends on exchange-correlation functionals, 
$\Omega^\ell$ denotes the volume of the entire simulation cell (i.e., supercell), defined as
$\Omega^\ell = N_k \Omega$ with $\Omega$ being the volume of a unit cell.
The Fock matrix contribution from the exchange energy is
\begin{align}
K_{\nu\mu}^{\mathbf k_2}
&=
\frac
{\partial E_{K}}
{\partial P_{\mu\nu}^{\mathbf k_2}}=-
\sum_{\mathbf k_1}
\sum_{\sigma\lambda}
(\nu_{\mathbf k_2}\lambda_{\mathbf k_1}|\sigma_{\mathbf k_1}\mu_{\mathbf k_2})
P_{\lambda\sigma}^{\mathbf k_1}
\label{eq:exxao}
\end{align}

The algorithms developed and studied in this work are based on the GPW density fitting scheme popularized by 
Hutter and co-workers.\cite{lippert1997hybrid,vandevondele2005quickstep}
In essence, the GPW scheme expands the pair density of Bloch orbitals in terms of planewaves:
\begin{equation}
(\psi_{\sigma_{\mathbf k_1}}(\mathbf r))^* \psi_{\mu_{\mathbf k_2}} (\mathbf r)
=
\sum_{\mathbf G} Z_{\sigma_{\mathbf k_1}\mu_{\mathbf k_2}}^{\mathbf G}
e^{i(\mathbf G-\mathbf k_1 + \mathbf k_2)\cdot \mathbf r}
\label{eq:aodf}
\end{equation}
where we evaluate the density fitting coefficients via
a Fourier transform,
\begin{equation}
Z_{\sigma_{\mathbf k_1}\mu_{\mathbf k_2}}^{\mathbf G}
=
\frac{1}{\Omega}
\int_\Omega
\mathrm d \mathbf r 
(\psi_{\sigma_{\mathbf k_1}} (\mathbf r))^*
\psi_{\mu_{\mathbf k_2}} (\mathbf r)
e^{-i(\mathbf G - \mathbf k_1 + \mathbf k_2)\cdot \mathbf r}
\end{equation}
The density fitted result is then used to evaluate the Coulomb potential via an inverse Fourier transform
\begin{equation}
V_{\sigma_{\mathbf k_1}\mu_{\mathbf k_2}}(\mathbf r)
=
\sum_{\mathbf G}
V^{\mathbf G}_{\sigma_{\mathbf k_1}\mu_{\mathbf k_2}} e^{i(\mathbf G-\mathbf k_1 + \mathbf k_2)\cdot \mathbf r}
\label{eq:Vrao}
\end{equation}
where 
\begin{equation}
V^{\mathbf G}_{\sigma_{\mathbf k_1}\mu_{\mathbf k_2}}
=
\begin{cases}
f(|\mathbf G-\mathbf k_1 + \mathbf k_2|)
Z_{\sigma_{\mathbf k_1}\mu_{\mathbf k_2}}^{\mathbf G}
& \text{if}\: |\mathbf G-\mathbf k_1 + \mathbf k_2| > 0\\
\chi & \text{if}\: |\mathbf G-\mathbf k_1 + \mathbf k_2| = 0
\end{cases}
\label{eq:VGao}
\end{equation}
and the form of $f(x)$ and $\chi$ depend 
on the underlying Coulomb operator ($V_\text{coul}$).

We consider three forms of the Coulomb operator as necessary for global hybrid, short-range hybrid, and range separated hybrid functionals, respectively:
\begin{equation}
f(x) = 
\begin{cases}
\frac{4\pi}{x^2} & \text{if}\:V_\text{coul} = \frac{1}{|\mathbf r|}\\
\frac{4\pi}{x^2}(1-e^{-x^2/(4\omega^2)}) & \text{if}\:V_\text{coul} = \frac{\text{erfc}(\omega |\mathbf r|)}{|\mathbf r|}\\
\frac{4\pi}{x^2}e^{-x^2/(4\omega^2)} & \text{if}\:V_\text{coul} = \frac{\text{erf}(\omega |\mathbf r|)}{|\mathbf r|}
\end{cases}
\label{eq:recipcoul}
\end{equation}
$\chi$ is to correct the finite size effect and we use a simple Madelung constant correction\cite{Fraser1996Jan} in the case of
the unscreened and long-range Coulomb operators.
The short-range Coulomb operator does not diverge at $x=0$, so we use $f(x\rightarrow0)$:
\begin{equation}
\chi = \frac{\pi}{\omega^2}
\end{equation}
We also tested the truncated Coulomb operator employed by Spencer and Alavi which achieves this by combining 
the Coulomb and the long-range Coulomb operators: 
:\cite{Spencer2008May}
\begin{equation}
V_\text{coul}(\mathbf r) = \begin{cases}
\frac {1}{|\mathbf r|} \Theta(R_c - |\mathbf r|)\\
\frac {\text{erf}(\omega |\mathbf r|)}{|\mathbf r|} \Theta(R_c - |\mathbf r|)
\end{cases}
\end{equation}
where $\Theta$ is the Heaviside step function and the spherical cutoff $R_c$ is determined from $(4\pi/3)R_c^3 = \Omega^\ell$.
For range separated hybrids, we applied the truncated Coulomb operator strategy just to the long-range
contribution while treating the short-range contribution exactly.\cite{Kawashima2017Mar}
In reciprocal space, these transform into the following form (analogously to \cref{eq:recipcoul}):
\begin{equation}
f(x) =
\frac{4\pi}{x^2} (1-\cos(xR_c)),
\end{equation}
for the truncated Coulomb operator
and
\begin{align}\nonumber
f(x) &= 
-\frac{4\pi}{x^2} \mathrm{erf}(\omega R_c) \cos(x R_c) +\frac{2\pi}{x^2} e^{-\frac{x^2}{4\omega^2}}\\
&\times\left[
\text{erf}(\omega(R_c+\frac{ix}{2\omega^2}))
+
\text{erf}(\omega(R_c-\frac{ix}{2\omega^2}))
\right],
\label{eq:lrtrunc}
\end{align}
for the truncated long-range Coulomb operator.\cite{Kawashima2017Mar}
Both cases have well-defined $x\rightarrow 0$ limits and therefore one can use $\chi = f(x\rightarrow0)$.
The erf terms in \cref{eq:lrtrunc} diverge as $x\rightarrow \infty$ and the multiplicative exponential function decays to zero as $x\rightarrow \infty$.
These two terms cancel each other and produce a finite, well-behaved quantity in the end but some care is needed for
a numerically stable evaluation as described in \cref{asubsec:num}.

In GPW, the Fourier transforms are handled by the discrete Fourier transform as implemented in fast Fourier transform libraries.
The computational complexity of each Fourier transform call is $\mathcal O(N_g \log N_g)$ where $N_g$ is the number of grid points used in the unit cell.
Using the GPW density fitting, we consider a total of three algorithms in this work where all three yield exactly the same ground state energy and valence band (occupied orbital) energies.
While some of our algorithms are capable of avoiding the storage of $\psi_{\mu_\mathbf{k}}(\mathbf r)$ on the real-space grid,
for the descriptions below we assume that this tensor can be stored in memory.
We describe a strategy to avoid storing $\psi_{\mu_\mathbf{k}}(\mathbf r)$ within the occ-RI-K algorithm in \cref{sec:direct}.

\subsection{Atomic Orbital (AO)-RI-K algorithm}
In the AO-RI-K algorithm, 
our goal is to compute \cref{eq:exxao} as written and 
a pair of Bloch atomic orbitals as shown in \cref{eq:aodf} is density fitted.
A nice feature of this algorithm is that 
one can benefit from exploiting the sparse structure of 
Bloch atomic orbitals (\{$\psi_{\mu_{\mathbf k}}$\}) where
we can assume that only a small number of grid points
carry non-zero values for each Bloch atomic orbital.
Our scaling analysis will assume this as our implementation exploits this.

The AO-RI-K algorithm (shown in Algorithm \ref{algo:aogpw}) starts by forming the following intermediate:
\begin{equation}
\tilde{\psi}_{\sigma_{\mathbf k_1}}(\mathbf r)
=
\sum_\lambda
P_{\lambda\sigma}^{\mathbf k_1} \psi_{\lambda_{\mathbf k_1}}(\mathbf r)
\label{eq:aointerm}
\end{equation}
which costs $\mathcal O(N_k N_g)$ memory and $\mathcal O(N_k N_g n_\text{AO})$ compute where sparsity was used to remove the scaling with $n_\text{AO}$.
Looping over pairs of k-points ($\mathbf k_1$ and $\mathbf k_2$) and pairs of atomic orbital indices ($\mu_{\mathbf k_2}$, $\sigma_{\mathbf k_1}$), we evaluate  
\cref{eq:Vrao} with $\mathcal O(N_k^2 n_\text{AO}^2 N_g\log N_g)$ (i.e. cubic) effort,
where $n_\text{AO}$ is the number of atomic Bloch orbitals in the unit cell.
Within the inner loops over $\mathbf k_1$ and $\sigma_{\mathbf k_1}$, 
we accumulate the following intermediate (starting from zero),
\begin{equation}
\tilde{V}_{\mu_{\mathbf k_2}} (\mathbf r) = \sum_{\mathbf k_1}\sum_{\sigma_{\mathbf k_1}}
\tilde{\psi}_{\sigma_{\mathbf k_1}}(\mathbf r)V_{\sigma_{\mathbf k_1}\mu_{\mathbf k_2}}(\mathbf r)
\label{eq:aointerm2}
\end{equation}
which scales as $\mathcal O(N_k^2 n_\text{AO}^2N_g)$ and
we accumulate the final exchange matrix contribution, 
\begin{equation}
K_{\nu\mu}^{\mathbf{k}_2}  = -\sum_{\mathbf r}\tilde{V}_{\mu_{\mathbf k_2}} (\mathbf r) (\psi_{\nu_{\mathbf k_2}}(\mathbf r))^*
\label{eq:aok}
\end{equation}
which scales as $\mathcal O(N_k N_g)$ after sparsity was used to remove the dependence of the scaling on $n_\text{AO}$.
The bottleneck of this algorithm is executing the FFT, which scales as $\mathcal O(N_k^2 n_\text{AO}^2 N_g\log N_g)$.

\begin{algorithm}[H]
\label{algo:aogpw}
\DontPrintSemicolon
Perform \cref{eq:aointerm}.\tcp*{$\mathcal O(N_k N_g)$}
\For(\tcp*[f]{Parallel loop.}){$\mathbf k_2=1$ \KwTo $N_k$}
{
\For(\tcp*[f]{Parallel loop.}){$\mu_{\mathbf k_2}=1$ \KwTo $n_\text{AO}$}
{
\For{$\mathbf k_1=1$ \KwTo $N_k$}
{
\For{$\sigma_{\mathbf k_1}=1$ \KwTo $n_\text{AO}$}
{
Form \cref{eq:Vrao}. \tcp*{$\mathcal O(N_k^2 n_\text{AO}^2 N_g \log N_g)$}
Execute \cref{eq:aointerm2} to obtain $\tilde{V}$. \tcp*[f]{$\mathcal O(N_k^2 n_\text{AO}^2N_g)$}
}}
Execute \cref{eq:aok}. \tcp*{$\mathcal O(N_kN_g)$}
}}
\caption{AO-RI-K algorithm.}
\end{algorithm}

\subsection{Molecular Orbital (MO)-RI algorithm}
In the MO-RI algorithm,\cite{Neese2002Aug,Weigend2002Sep,Ye2022Mar}
we compute the exchange matrix via 
\begin{align}
K_{\nu\mu}^{\mathbf k_2}
&=
-\sum_{\mathbf k_1}
\sum_{i\in\text{occ}}
(\nu_{\mathbf k_2}i_{\mathbf k_1}|i_{\mathbf k_1}\mu_{\mathbf k_2})
\label{eq:exxmo}
\end{align}
where an occupied orbital is defined as
\begin{equation}
\psi_{i_{\mathbf k}}(\mathbf r)
=
\sum_{\mu}
C_{\mu i}^{\mathbf k}
\psi_{\mu_{\mathbf k}}(\mathbf r)
\label{eq:occmo}
\end{equation}
We form this intermediate at the cost of $\mathcal O(N_k N_g n_\text{occ})$ (with sparsity) operations, where $n_\text{occ}$ is the number of occupied orbitals and store this in memory. This $N_k N_g n_\text{occ}$ memory requirement scales quadratically with cell size.

In the MO-RI algorithm,
we density fit the $(\psi_{i_{\mathbf k_1}}(\mathbf r))^*\psi_{\mu_{\mathbf k_2}}(\mathbf r)$ products.
Looping over pairs of k-points, occupied orbital indices, and atomic orbital indices, 
the overall cubic cost of density fitting will scale as $\mathcal O(N_k^2 n_\text{occ}n_\text{AO}N_g\log N_g)$.
This suggests an immediate cost reduction from AO-RI to MO-RI is obtained by a factor of $n_\text{AO}/n_\text{occ}$, which can be a significant speedup when one considers a relatively large (such as triple-zeta or larger) basis set.
This speedup can be roughly a factor of 5 for triple-zeta quality bases and becomes larger as the basis set size is increased (keeping the system size fixed).

\begin{algorithm}[h!]
\SetAlgoNoLine\DontPrintSemicolon
Perform \cref{eq:occmo}.\tcp*{$\mathcal O(N_k N_gn_\text{occ})$}
\For(\tcp*[f]{Parallel loop.}){$\mathbf k_2=1$ \KwTo $N_k$}
{
\For(\tcp*[f]{Parallel loop.}){$\mu_{\mathbf k_2}=1$ \KwTo $n_\text{AO}$}
{
\For{$\mathbf k_1=1$ \KwTo $N_k$}
{
\For{$i_{\mathbf k_1}=1$ \KwTo $n_\text{occ}$}
{
Form $V_{i_{\mathbf k_1}\mu_{\mathbf k_2}}(\mathbf r)$. \tcp*{$\mathcal O(N_k^2 n_\text{AO}n_\text{occ} N_g \log N_g)$}
Execute \cref{eq:mointerm2} to obtain $W$. 
\tcp*[f]{$\mathcal O(N_k^2 n_\text{AO}n_\text{occ}N_g)$}
}}
Execute \cref{eq:aok} with $W$. \tcp*{$\mathcal O(N_kN_g)$}
}}
\caption{MO-RI-K algorithm.}
\label{algo:mogpw}
\end{algorithm}

The overall MO-RI algorithm, summarized in Algorithm \ref{algo:mogpw}, is similar to the AO-RI algorithm.
One loops over a pair of k-points ($\mathbf k_1, \mathbf k_2$), occupied orbital indices $i_{\mathbf k_1}$, and atomic orbital indices $\mu_{\mathbf k_2}$ and forms
the Coulomb potential $V_{i_{\mathbf k_1}\mu_{\mathbf k_2}}(\mathbf r)$ that arises from the density, $(\psi_{i_{\mathbf k_1}}(\mathbf r))^*\psi_{\mu_{\mathbf k_2}}(\mathbf r)$.
One then accumulates the following intermediate in the inner loop (i.e. the loops over $\mathbf k_1$ and $i_{\mathbf k_1}$):
\begin{equation}
W_{\mu_{\mathbf k_2}} (\mathbf r) = 
\sum_{\mathbf k_1}
\sum_{i_{\mathbf k_1}}
{\psi}_{i_{\mathbf k_1}}(\mathbf r)V_{i_{\mathbf k_1}\mu_{\mathbf k_2}}(\mathbf r)
\label{eq:mointerm2}
\end{equation}
with compute cost scaling as 
$\mathcal O(N_k^2 n_\text{AO}n_\text{occ}N_g)$.
The K-matrix accumulation is done the same way as \cref{eq:aok} with the intermediate in \cref{eq:mointerm2} in the outer loop with the same cost of $\mathcal O(N_k n_\text{AO}^2)$.
Similar to the AO-RI algorithm, the FFT calls were found to be the bottleneck, with cubic scaling compute cost of $\mathcal O(N_k^2 n_\text{AO}n_\text{occ}N_g\log N_g)$.

\subsection{Occupied orbital (occ)-RI-K algorithm}\label{sec:occrik}
The occ-RI-K algorithm\cite{Manzer2015Jul} speeds up evaluation of the exact exchange operator by \textit{ignoring} its component in the virtual space. In other words, denoting the occupied orbital space projector as $\hat{P}$ and the unoccupied orbital space projector as  $\hat{Q}$,
one can approximate
\begin{equation}
\hat{K}
\simeq
\hat{P}\hat{K}\hat{P}
+
\hat{P}\hat{K}\hat{Q}
+
\hat{Q}\hat{K}\hat{P}
\label{eq:occrikop}
\end{equation}
ignoring $\hat{Q}\hat{K}\hat{Q}$. 
This approximation is {\it exact} when considering quantities that depend only on occupied orbitals such as the self-consistent field (SCF) energy, the valence band energies and orbitals, and of course the density matrix.

Using the same idea, we will compute only part of the exchange matrix,
\begin{align}
K_{\nu j}^{\mathbf k_2}
&=
-\sum_{\mathbf k_1}
\sum_{i\in\text{occ}}
(\nu_{\mathbf k_2}i_{\mathbf k_1}|i_{\mathbf k_1}j_{\mathbf k_2})
\label{eq:occrik}
\end{align}
to obtain the AO-occupied block of $\mathbf K$.
This amounts to the computation of $(\hat{P}+\hat{Q})\hat{K}\hat{P}$ which can be used to obtain
\cref{eq:occrikop} with simple matrix multiplications for each k-point.\cite{Manzer2015Jul}
Since the computational bottleneck of AO-RI-K and MO-RI-K is the FFT step, 
our goal is to reduce the prefactor for this step using the same intuition as occ-RI-K.

In the occ-RI-K algorithm, shown in Algorithm \ref{algo:occrikgpw},
one first forms the intermediates in \cref{eq:occmo} and loops over a pair of k-points ($\mathbf k_1, \mathbf k_2$) and a pair of occupied orbitals ($i_{\mathbf k_1}, j_{\mathbf k_2}$).
The density, $(\psi_{i_{\mathbf k_1}}(\mathbf r))^*\psi_{j_{\mathbf k_2}}(\mathbf r)$, will be fitted by planewaves and
the corresponding Coulomb potential, 
$V_{i_{\mathbf k_1} j_{\mathbf k_2}}(\mathbf r)$
is formed at  
$\mathcal O(N_k^2n_\text{occ}^2 N_g\log N_g)$ cost.
Similarly to the other GPW algorithms, in the inner loops ($\mathbf k_1, i_{\mathbf k_1}$)
one accumulates the following intermediate:
\begin{equation}
\tilde{W}_{j_{\mathbf k_2}}(\mathbf r)
=
\sum_{\mathbf k_1}
\sum_{i_{\mathbf k_1}}
\psi_{i_{\mathbf k_1}}(\mathbf r)
V_{i_{\mathbf k_1} j_{\mathbf k_2}}(\mathbf r)
\label{eq:occrikinterm}
\end{equation}
with $\mathcal O(N_k^2 n_\text{occ}^2 N_g)$ compute cost.
We assume that we have enough memory to hold   $\tilde{W}(\mathbf r)$, imposing an $\mathcal O(N_k n_\text{occ} N_g)$ quadratic-scaling storage requirement (significantly smaller than required to hold $\psi_{\nu_{\mathbf k_2}}(\mathbf r)$).
After obtaining $\tilde{W}$, we compute
\begin{equation}
K_{\nu j}^{\mathbf{k}_2}  = -\sum_{\mathbf r}\tilde{W}_{j_{\mathbf k_2}} (\mathbf r) (\psi_{\nu_{\mathbf k_2}}(\mathbf r))^*
\label{eq:occrik}
\end{equation}
at $\mathcal O(N_k n_\text{occ} N_g)$ cost, assuming sparsity of $\psi_{\nu_{\mathbf k_2}}(\mathbf r)$.
Compared to the MO-RI-K algorithm, we achieve a clear $n_\text{AO}/n_\text{occ}$ speed-up in all steps in the loop.
Most importantly, the number of FFT calls is reduced from $N_k^2n_\text{AO}n_\text{occ}$ to $N_k^2n_\text{occ}^2$.

\begin{algorithm}[H]
\label{algo:occrikgpw}
\SetAlgoNoLine\DontPrintSemicolon
Perform \cref{eq:occmo}.\tcp*{$\mathcal O(N_k N_g n_\text{occ})$}
\For(\tcp*[f]{Parallel loop.}){$\mathbf k_2=1$ \KwTo $N_k$}
{
\For(\tcp*[f]{Parallel loop.}){$j_{\mathbf k_2}=1$ \KwTo $n_\text{occ}$}
{
\For(\tcp*[f]{Parallel loop.}){$\mathbf k_1=1$ \KwTo $N_k$}
{
\For(\tcp*[f]{Parallel loop.}){$i_{\mathbf k_1}=1$ \KwTo $n_\text{occ}$}
{
Form $V_{i_{\mathbf k_1}j_{\mathbf k_2}}(\mathbf r)$. \tcp*{$\mathcal O(N_k^2 n_\text{occ}^2 N_g \log N_g)$}
Execute \cref{eq:occrikinterm} to obtain $\tilde{W}$. \tcp*[f]{$\mathcal O(N_k^2 n_\text{occ}^2N_g)$}
}}
}
}
Execute \cref{eq:occrik}. \tcp*{$\mathcal O(N_kN_gn_\text{occ})$}
\caption{occ-RI-K algorithm.}
\end{algorithm}

In some applications, one may want to compute the first few conduction bands (unoccupied orbitals).
This is particularly important when one tries to compute the band gap.
In that case, one can simply extend 
the occ-RI-K algorithm to compute the first few conduction bands {\it exactly}.
We write $\hat{Q} = \hat{R} + \hat{S}$ where $\hat{R}$ is the projector onto the space spanned by
conduction bands of our interest and $\hat{S}$ is the projector onto the rest of the conduction bands.
Then, we can approximate $\hat{K}$ by
\begin{align}\nonumber
\hat{K}
&\simeq
\hat{P}\hat{K}\hat{P}
+
\hat{P}\hat{K}\hat{Q}
+
\hat{Q}\hat{K}\hat{P}\\
&+\hat{R}\hat{K}\hat{R}
+\hat{R}\hat{K}\hat{Q}
+\hat{Q}\hat{K}\hat{R}
\end{align}
This only needs the evaluation of
$K_{\nu p}^{\mathbf k_2}$ where $p$ includes valence bands (occupied orbitals) and desired conduction bands (unoccupied orbitals) at $\mathbf k_2$.
However, when
$\{\mathbf k_1\}$ (i.e., those used for the ground state calculations) and $\{\mathbf k_2\}$ (those used for the band calculations) in \cref{eq:occrik}
are different,
the occ-RI-K algorithm described above
is no longer applicable because 
one does not have orbitals available for $\{\mathbf k_2\}$.
As a workaround, one may append $\{\mathbf k_1\}$ with $\{\mathbf k_2\}$ for the ground state calculations
or
employ Wannier interpolation\cite{Mostofi2008May} to obtain orbitals at $\{\mathbf k_2\}$ from orbitals at $\{\mathbf k_1\}$.

\subsection{Integral-direct strategies}\label{sec:direct}

The memory requirement for storing the basis function on grid points ($\psi_{\mu_{\mathbf k}}(\mathbf r)$) scales as $\mathcal O(N_gN_k)$
assuming the sparsity of the underlying basis functions.
In practice, the sparsity may not be effective with a relatively tight threshold until we reach a very large computational cell. 
In such cases, the required memory can scale as $\mathcal O(N_gN_kn_{AO})$ which can be quite sizable.
If this memory consumption is unaffordable, one needs to resort to an ``integral-direct'' strategy where
one does not store $\psi_{\mu_{\mathbf k}}(\mathbf r)$ in memory, but instead computes them on-the-fly.

This leads to a small modification of Algorithm \ref{algo:occrikgpw} as shown in Algorithm \ref{algo:direct}.
The only difference is that one repeatedly computes 
$\psi_{i_{\mathbf k}}(\mathbf r)$
adding an extra 
computational cost of $\mathcal O(N_k^2 N_g n_\text{occ}^2)$.
This step is not more expensive than other parts of the algorithm.
In our implementation, depending on available memory, the integral-direct algorithm is triggered.
A similar strategy has been explored in Gaussian density fitting recently.\cite{Bintrim2022Sep}

\begin{algorithm}[H]
\label{algo:direct}
\SetAlgoNoLine\DontPrintSemicolon
\For{$\mathbf k_2=1$ \KwTo $N_k$}
{
\For(\tcp*[f]{Parallel loop.}){$j_{\mathbf k_2}=1$ \KwTo $n_\text{occ}$}
{
\For(\tcp*[f]{Parallel loop.}){$\mathbf k_1=1$ \KwTo $N_k$}
{
\For(\tcp*[f]{Parallel loop.}){$i_{\mathbf k_1}=1$ \KwTo $n_\text{occ}$}
{
Perform \cref{eq:occmo} for $i_{\mathbf k_1}$,$j_{\mathbf k_2}$.\tcp*{$\mathcal O(N^2_k N_g n_\text{occ}^2)$}
Form $V_{i_{\mathbf k_1}j_{\mathbf k_2}}(\mathbf r)$. \tcp*{$\mathcal O(N_k^2 n_\text{occ}^2 N_g \log N_g)$}
Execute \cref{eq:occrikinterm} to obtain $\tilde{W}$. \tcp*[f]{$\mathcal O(N_k^2 n_\text{occ}^2N_g)$}
}}
Execute \cref{eq:occrik} for $j_{\mathbf{k}_2}$. \tcp*{$\mathcal O(N_kN_gn_\text{occ})$}
}
}
\caption{Integral-direct occ-RI-K algorithm.}
\end{algorithm}

\section{Computational details}
\begin{table}[h!]
\begin{tabular}{|c|c|c|c|cc|}
\hline
Functional      & Year & Hybrid type & Ingredients & \multicolumn{1}{c|}{$c_{x,sr}$} & $c_{x,lr}$ \\ \hline
B3LYP\cite{Becke1993Apr}           & 1993 & GH          & GGA         & \multicolumn{2}{c|}{0.20}                    \\ \hline
PBE0 \cite{Perdew1996Dec}           & 1996 & GH          & GGA         & \multicolumn{2}{c|}{0.25}                    \\ \hline
revPBE0\cite{Zhang1998Jan}         & 1998 & GH          & GGA         & \multicolumn{2}{c|}{0.25}                    \\ \hline
B97-3\cite{Keal2005Sep}           & 2005 & GH          & GGA         & \multicolumn{2}{c|}{0.269288}                \\ \hline
M06-2X\cite{Zhao2008May}          & 2008 & GH          & mGGA        & \multicolumn{2}{c|}{0.54}                    \\ \hline
MN15\cite{Yu2016Jul}            & 2016 & GH          & mGGA        & \multicolumn{2}{c|}{0.44}                    \\ \hline
SCAN0\cite{Hui2016Jan}            & 2016 & GH          & mGGA        & \multicolumn{2}{c|}{0.25}                    \\ \hline
HSE\cite{Heyd2003May,Krukau2006Dec,Heyd2006Jun,Henderson2008May}             & 2008 & RSH         & GGA         & \multicolumn{1}{c|}{0.25}       & 0.00       \\ \hline
CAM-B3LYP \cite{Yanai2004Jul}      & 2004 & RSH         & GGA         & \multicolumn{1}{c|}{0.19}       & 0.65       \\ \hline
$\omega$B97X-rV\cite{Mardirossian2014May} & 2014 & RSH         & GGA         & \multicolumn{1}{c|}{0.167}      & 1.00       \\ \hline
$\omega$B97M-rV\cite{Mardirossian2016Jun} & 2016 & RSH         & mGGA        & \multicolumn{1}{c|}{0.15}       & 1.00       \\ \hline
CAM-QTP01\cite{Jin2016Jul} & 2016 & RSH & GGA &  \multicolumn{1}{c|}{0.23}       & 1.00 \\ \hline
\end{tabular}
\caption{Summary of 12 density functionals investigated in this work.
$c_{x,sr}$ is the coefficient for the short-range exact exchange
and
$c_{x,lr}$ is the coefficient for the long-range exact exchange.
}
\label{tab:func}
\end{table}

We consider a total of 12 density functionals in this work.
There are seven global hybrid (GH) functionals
and four range separated hybrid (RSH) functionals,
with a range of different amount of exact exchange and year of development.
For GGA GH functionals, 
B3LYP,\cite{Becke1993Apr} PBE0,\cite{Perdew1996Dec} revPBE0,\cite{Zhang1998Jan} and B97-3\cite{Keal2005Sep} were considered.
For mGGA GH functionals,
M06-2X,\cite{Zhao2008May} MN15,\cite{Yu2016Jul} and SCAN0\cite{Hui2016Jan} were considered.
For RSH functionals, we consider
a short-range functional (HSE\cite{Heyd2003May,Krukau2006Dec,Heyd2006Jun,Henderson2008May})
and
four
long-range corrected density functionals (CAM-B3LYP,\cite{Yanai2004Jul} $\omega$B97X-rV,\cite{Mardirossian2014May} $\omega$B97M-rV,\cite{Mardirossian2016Jun}CAM-QTP01\cite{Jin2016Jul}).
We do not consider dispersion corrections such as D2, D3, and D3(BJ)\cite{Grimme2006Nov,Grimme2010Apr,Grimme2011May} in this work
because they do not affect the band gaps at all.
In practical applications besides the band gap, all of the aforementioned functionals, except $\omega$B97X-rV and $\omega$B97M-rV,
should be supplemented by dispersion corrections.
All our calculations were performed with a development version of Q-Chem.\cite{Epifanovsky2021Aug}
For relatively well studied functionals, PBE0 and HSE, we compare our band gaps against literature values in Refs. \citenum{Garza2016Oct,Borlido2019Sep} and found an excellent agreement (see \cref{fig:tran}).

We summarize these functionals in \cref{tab:func} along with their fraction of short-range ($c_{x,\text{sr}}$) and long-range ($c_{x,\text{lr}}$) exact exchange.
One key feature of $\omega$B97X-rV, $\omega$B97M-rV, and CAM-QTP01
is that
they include 
the long-range exact exchange contribution up to
100$\%$. 
Intuitively, this can be worrisome for band gap applications
because
in the long-range there is no Coulomb screening present in the method (like in Hartree-Fock theory).
Another interesting remark about CAM-QTP01 is that this is an RSH functional fitted to experimental ionization potentials, which may be a useful property for improving the band gaps.\cite{Jin2016Jul}
We will see how these manifest in the band gap benchmark later.

We used a large uncontracted basis set developed in our previous paper (unc-def2-QZVP-GTH)\cite{Lee2021Oct}
to ensure that we obtain band gaps near the basis set limit.
We used the GTH-PBE pseudopotential for all functionals considered in this work due to the lack of
functional-specific GTH pseudopotentials for these functionals.\cite{Goedecker1996Jul,Hartwigsen1998Aug}
We took the geometry and experimental band gaps of 25 solids from ref. \citenum{Lee2021Oct} (also see references therein).

As it was tested for local functionals,\cite{Lee2021Oct} the band gap change due to the pseudopotential is expected to be
much smaller than the band gap error energy scale that we will discuss here.
We used  $6\times6\times6$ Monkhorst-Pack $\mathbf k$-mesh which is sufficient to reach
the thermodynamic limit 
for systems discussed in this work.
For our
GPW calculations, we followed the same $E_\text{cut}$ value as our previous study.\cite{Lee2021Oct}
Namely, we used $E_\text{cut}$ of 1500 eV for every solid considered in this work. 
To measure the remaining basis set incompleteness error,
we compared 
the $\Gamma$-point band gap of B3LYP computed by
our code against those from
QuantumESPRESSO.\cite{Giannozzi2020Apr} 
We used a kinetic energy cutoff of 400 Ry for every system.
The basis set error of our band gap calculations with GTOs is smaller than 10 meV, which
is consistent with what we found for local density functionals.\cite{Lee2021Oct}
This comparison is available in \cref{tab:gap3}.
We also compare the total energy in \cref{tab:total} and confirm that the basis set error in total energy with a $6\times6\times6$ $\mathbf k$-mesh is less than 0.5 m$E_h$ per cell.

We used finite size correction strategies described in \cref{subsec:gpw} for handling the divergence of exact exchange term. For the ground state SCF calculations, we used the simple Madelung constant correction.\cite{Fraser1996Jan} While this correction scheme smoothly converged the ground state SCF energies to the thermodynamic limit up to the k-mesh of 6$\times$6$\times$6, the subsequent band structure calculations showed erratic discontinuities in the resulting bands. We confirmed that this is due to the residual size effect so we switched to the truncated Coulomb operator technique\cite{Spencer2008May} when computing bands.
Nonetheless, the band gaps using two different correction schemes are in a qualitative agreement as can be seen in \cref{tab:gap1} and \cref{tab:gap2}.

The largest calculation that we performed in this work involves up to 350 basis functions and 16 electrons in the central unit cell (i.e., AlN)
with 6$\times$6$\times$6 k-mesh.

\section{Results and Discussion}
\subsection{Timing benchmark}
\begin{figure}
    \centering
    \begin{subfigure}[t]{\linewidth}
        \includegraphics[scale=0.35]{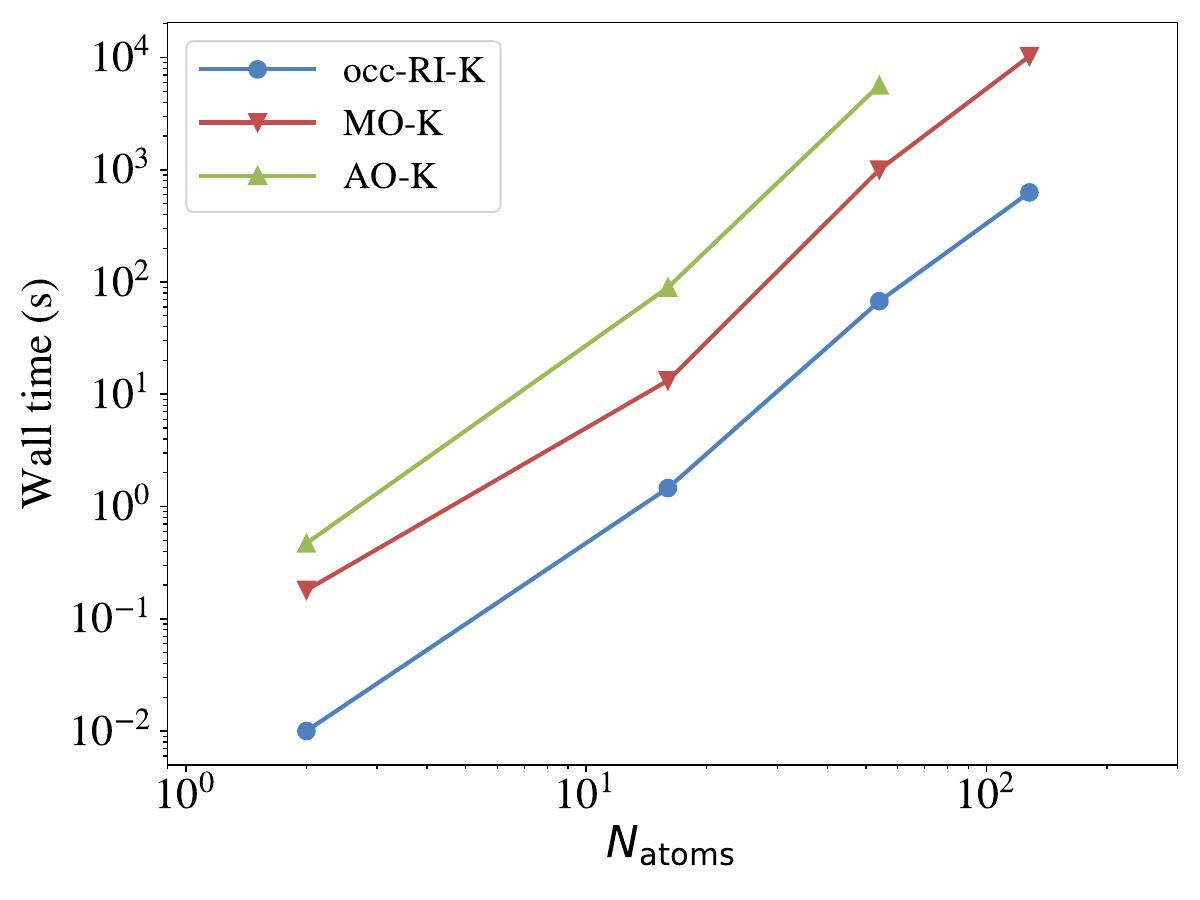}
        \caption{}
        \label{fig:gpt_timing}
    \end{subfigure}
    \\
    \begin{subfigure}[t]{\linewidth}
        \includegraphics[scale=0.35]{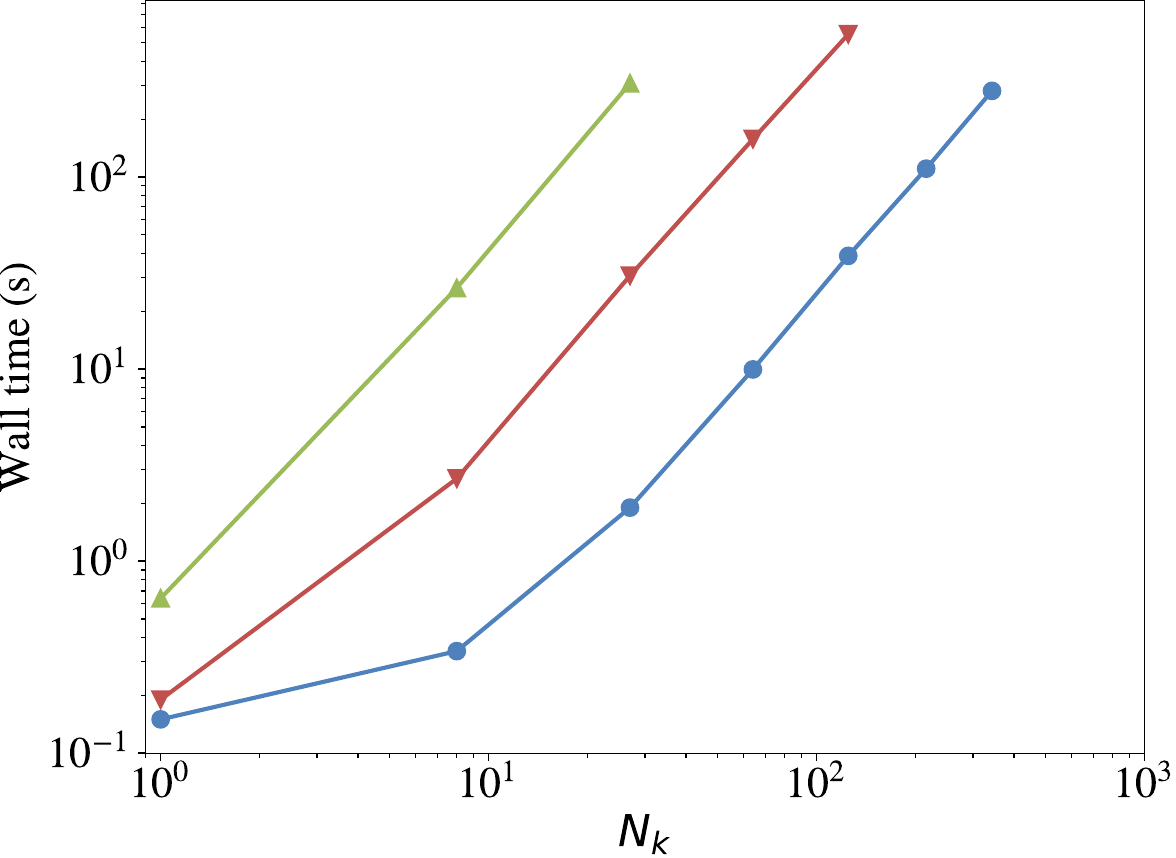}
        \caption{}
        \label{fig:kpts_timing}
    \end{subfigure}
    \caption{Wall time (seconds) of a single exchange-matrix build for the three exchange algorithms presented for diamond (a) $\Gamma$-point calculations as a function of the number of atoms in the super cell and (b) $\mathbf k$-point calculations as a function of the number of k-points. 
Given the same total number of C atoms, all methods in (a) and (b) yield the same total energy per C atom.
    }
    \label{fig:timing}
\end{figure}

We benchmarked the compute time of each exact exchange algorithm on a single test case, diamond with QZV2P-GTH basis set.\cite{VandeVondele2007Sep} 
Diamond is chosen because it is a representative semiconductor, and QZV2P-GTH is employed so that the benefit of occ-RI-K can be highlighted.
We tested the scaling with respect to system size as well as number of k-points. All calculations were done on 32 cores using two AMD Opteron 6376 processors.

In \cref{tab:Egamma} and \cref{tab:Ekpoint},
we present the Hartree-Fock total energies per atom
of diamond with varying supercell size (\cref{tab:Egamma}) and $\mathbf k$-mesh (\cref{tab:Ekpoint}).
Comparing the two tables, we illustrate the equivalence of the supercell and $\mathbf k$-point implementations
for the same number of atoms.
Furthermore, we show that our occ-RI-K implementation makes no additional approximations beyond AO-K and MO-K
as emphasized in \cref{sec:occrik}. AO-K, MO-K, and occ-RI-K energies agree with each other for the same number of atoms.

We analyzed the scaling with respect to system size via supercell $\Gamma$-point calculations. \cref{fig:gpt_timing} shows the wall time of each exchange algorithm as a function of the number of atoms included in the supercell. We see that AO-K quickly becomes intractable for large systems; a 3$\times$3$\times$3 supercell, corresponding to 54 atoms per unit cell in the calculation, is about the limit of this algorithm for the diamond system. The MO-K algorithm is over five times faster for all supercells considered. This allows calculations with two to three times the number of electrons as the AO-K algorithm. Finally, the occ-RI-K algorithm provides an additional speedup over MO-K of almost 15 for most supercells considered, allowing even larger calculations.
Overall, occ-RI-K achieves nearly two orders of magnitude speedup compared to the AO-K algorithm.
Furthermore, the slope of occ-RI-K in the log-log plot in \cref{fig:gpt_timing} suggests that
the algorithm scales as $\mathcal O(N^{2.9})$ which confirms the cubic-scaling with respect to system size as noted in \cref{sec:occrik}.

We additionally analyzed the performance of each algorithm where we fix the size of the unit cell (two carbon atoms per cell) and vary  the number of k-points. For these we find that MO-K offers roughly a factor of 10 speedup over AO-K and occ-RI-K further speeds this up by an additional factor of roughly 15.
The speedup provided by occ-RI-K is more than two orders of magnitude speedup compared to the AO-K algorithm.
The slope of occ-RI-K algorithm in \cref{fig:kpts_timing} confirms $\mathcal O(N_k^{2.0})$ scaling consistent with our scaling analysis presented in \cref{sec:occrik}.
We note that calculations with small $\mathbf k$-meshes as well as small supercells can be unphysical in that the finite size error can be substantial.
We, nonetheless, performed these calculations to analyze computational scaling.

While more practical application of exact exchange will likely be much more difficult than our prototypical example, diamond, we see that occ-RI-K offers substantial speedups over alternatives, allowing calculations with significantly more electrons and $\mathbf k$-points.
\begin{table}[]
\begin{tabular}{|r|lll|}
\hline
\multicolumn{1}{|c|}{}       & \multicolumn{3}{c|}{Supercell}                                                                  \\ \hline
\multicolumn{1}{|c|}{N$_\text{atoms}$} & \multicolumn{1}{c|}{AO-K}        & \multicolumn{1}{c|}{MO-K}        & \multicolumn{1}{c|}{occ-RI-K} \\ \hline
2                            & \multicolumn{1}{l|}{-5.1913973} & \multicolumn{1}{l|}{-5.1913973} & -5.1913973                  \\ \hline
16                           & \multicolumn{1}{l|}{-5.5159556} & \multicolumn{1}{l|}{-5.5159556} & -5.5159556                  \\ \hline
54                           & \multicolumn{1}{c|}{N/A}           & \multicolumn{1}{l|}{-5.5436244} & -5.5436244                  \\ \hline
\end{tabular}
\caption{Hartree-Fock total energies per atom ($E_h$) using the $\Gamma$-point implementation for various supercell sizes.
N/A means not available.
}
\label{tab:Egamma}
\end{table}

\begin{table}[]
\begin{tabular}{|r|lll|}
\hline
                             & \multicolumn{3}{c|}{$\mathbf k$-point}                                                                    \\ \hline
\multicolumn{1}{|c|}{N$_\text{atoms}$} & \multicolumn{1}{c|}{AO-K}        & \multicolumn{1}{c|}{MO-K}        & \multicolumn{1}{c|}{occ-RI-K} \\ \hline
2                            & \multicolumn{1}{l|}{-5.1913973} & \multicolumn{1}{l|}{-5.1913973} & -5.1913973                  \\ \hline
16                           & \multicolumn{1}{l|}{-5.5159556} & \multicolumn{1}{l|}{-5.5159556} & -5.5159556                  \\ \hline
54                           & \multicolumn{1}{l|}{-5.5436244} & \multicolumn{1}{l|}{-5.5436244} & -5.5436244                  \\ \hline
128                          & \multicolumn{1}{c|}{N/A}           & \multicolumn{1}{l|}{-5.5460132} & -5.5460132                  \\ \hline
250                          & \multicolumn{1}{c|}{N/A}           & \multicolumn{1}{l|}{-5.5456075} & -5.5456075                  \\ \hline
432                          & \multicolumn{1}{c|}{N/A}           & \multicolumn{1}{c|}{N/A}           & -5.5450982                  \\ \hline
686                          & \multicolumn{1}{c|}{N/A}           & \multicolumn{1}{c|}{N/A}           & -5.5447456                  \\ \hline
\end{tabular}
\caption{Hartree-Fock total energies per atom ($E_h$) using the $\mathbf k$-point implementation for various $\mathbf k$-mesh sizes (i.e., the number of atoms).
N/A means not available.}
\label{tab:Ekpoint}
\end{table}

\begin{figure}[h!]
    \centering
    \scalebox{0.41}{\includegraphics{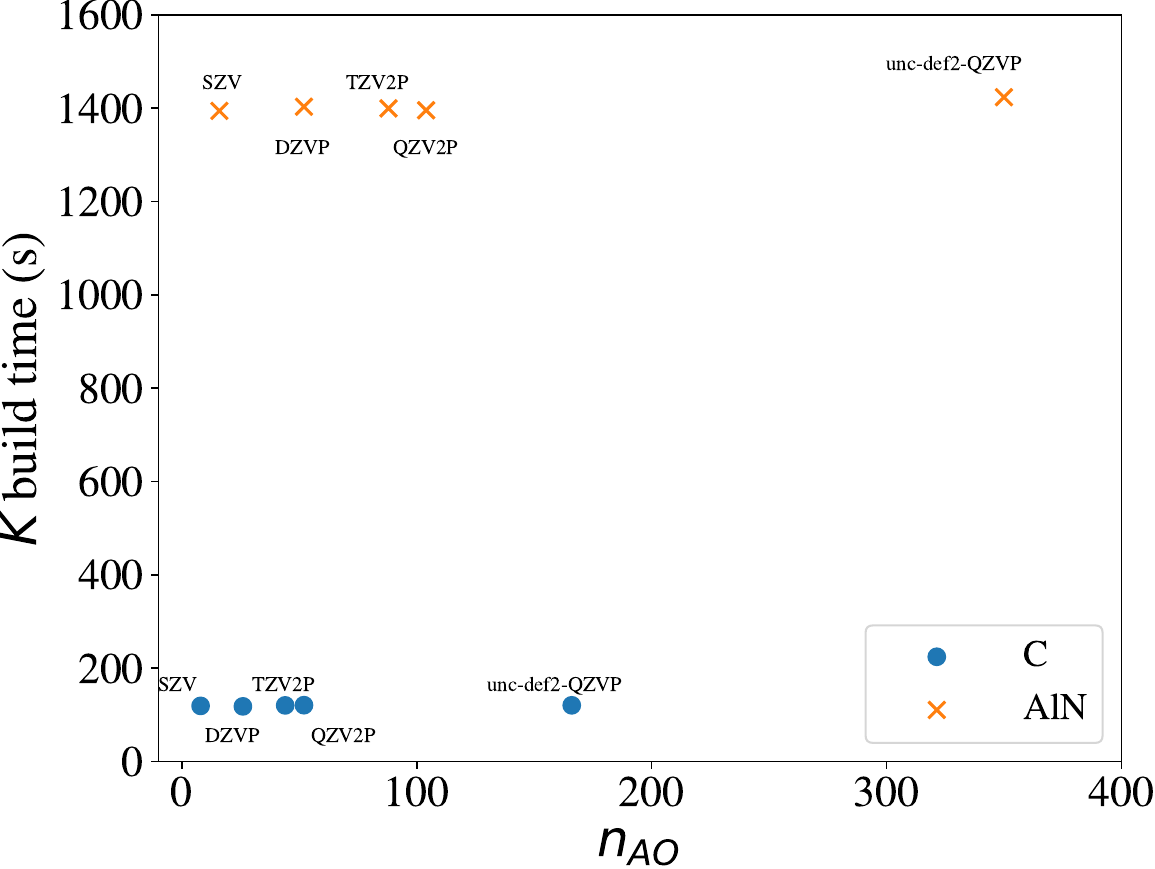}}
    \caption{
    Wall time (seconds) of a single K-matrix build with varying basis set sizes. AlN and C have 16 and 8 electrons in the unit cell, respectively.
    }
    \label{fig:basis_timing}
\end{figure}

We also performed a set of timing benchmark calculations of diamond and AlN with a 6$\times$6$\times$6 $\mathbf k$-mesh as a function of the basis set size (SZV-GTH, DZVP-GTH, TZV2P-GTH, QZV2P-GTH, unc-def2-QZVP-GTH.) In \cref{fig:basis_timing}, the corresponding timing results are presented.
We observe nearly no basis set size dependence in our K-matrix build time due to the fact that the number of expensive FFT calls is independent of the number of basis functions. This highlights the utility of occ-RI-K even further.

\subsection{Band gap assessment}\label{sec:bandgap}
\begin{figure}[h!]
    \centering
    \scalebox{0.48}{\includegraphics{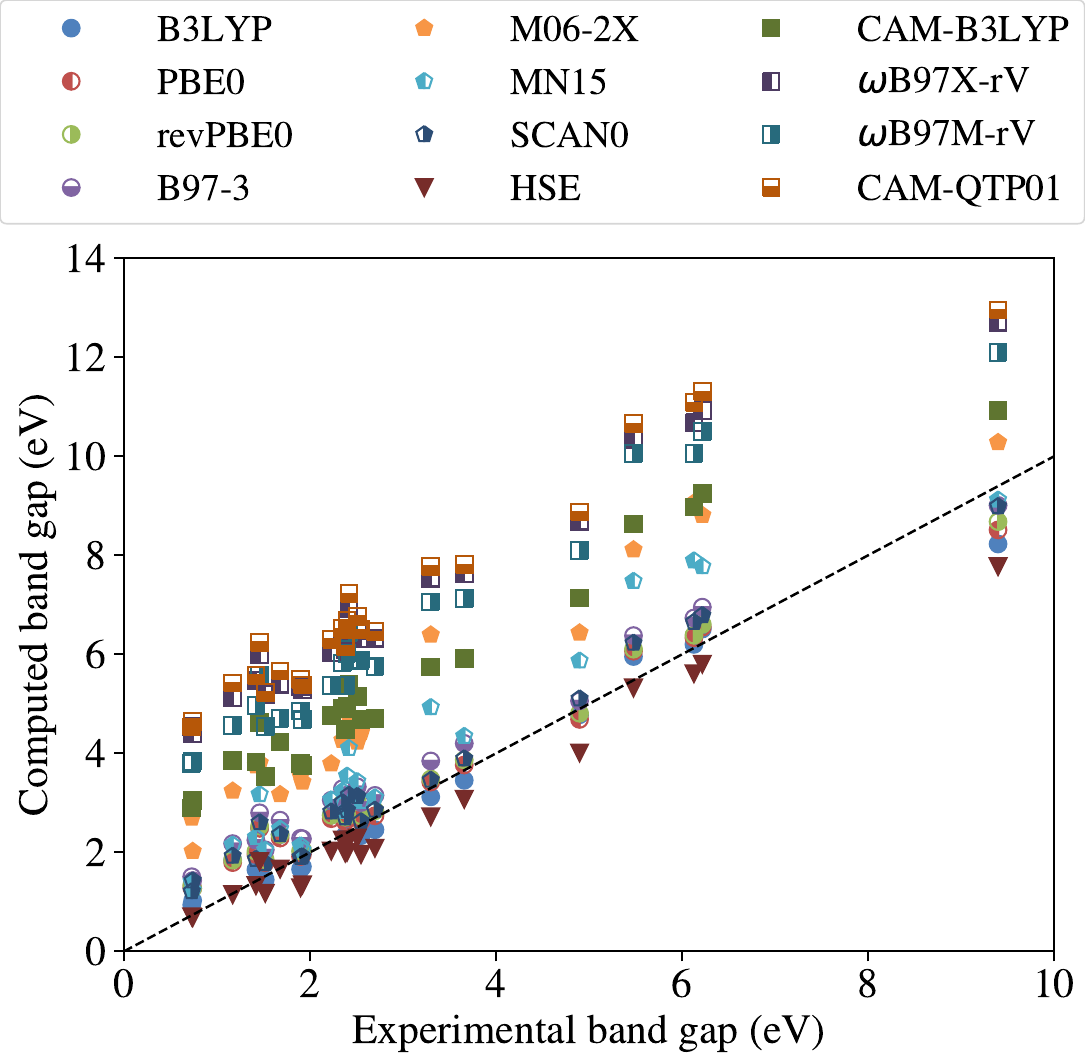}}
    \caption{
Scatter plot of computed band gap (eV) versus experimental band gap (eV).
Round markers are used for GGA global hybrids,
pentagons are used for mGGA global hybrids,
triangles are used for short-range hybrid functional (HSE),
and
squares are used for long-range corrected range separated functionals.
The black dotted line is guide for the eye.
    }
    \label{fig:scatter}
\end{figure}

We first discuss the overall band gap distribution of each functional
as shown in \cref{fig:scatter}.
Along the dotted line of $y=x$, 
we observe that round and triangle data points are relatively well aligned.
These are GGA GH functionals and HSE, respectively. 
It is widely accepted that
HSE
performs well for
band gap problems, but
the good performance of GGA GH functionals
is not so well-known.\cite{Garza2016Oct,Borlido2019Sep}
However, some deterioration of the good performance of both these classes of functionals is noticable in \cref{fig:scatter} 
for larger band gap materials (above 6 eV).

Given they are more recently developed functionals, mGGA GH functionals (pentagons) are quite disappointing.
M06-2X and MN15 have a high fraction of exact exchange ($\sim$50\%).
This higher fraction of exact exchange compared to other GGA GHs (all about $\sim$25\%)
seems to be the cause for an overall overestimation of the band gaps.
With 25\% of exact exchange, SCAN0 performs better than M06-2X and MN15, but
it still seems slightly worse than GGA GHs.

Lastly, the performance of long-range corrected functionals (squares)
is catastrophic
with the tendency of overestimating band gaps for all materials considered here.
The short-range exact exchange is only 15\%--20\% in these functionals,
which is even less than HSE (25\%).
This gross overestimation of band gaps is likely due to the large fraction of 
long-range exact exchange.
CAM-B3LYP has 65\% of long-range exact exchange while both of the combinatorially optimized functionals and CAM-QTP01 have 100\% long-range exact exchange.

\begin{figure*}[!ht]
    \centering
    \scalebox{0.55}{\includegraphics{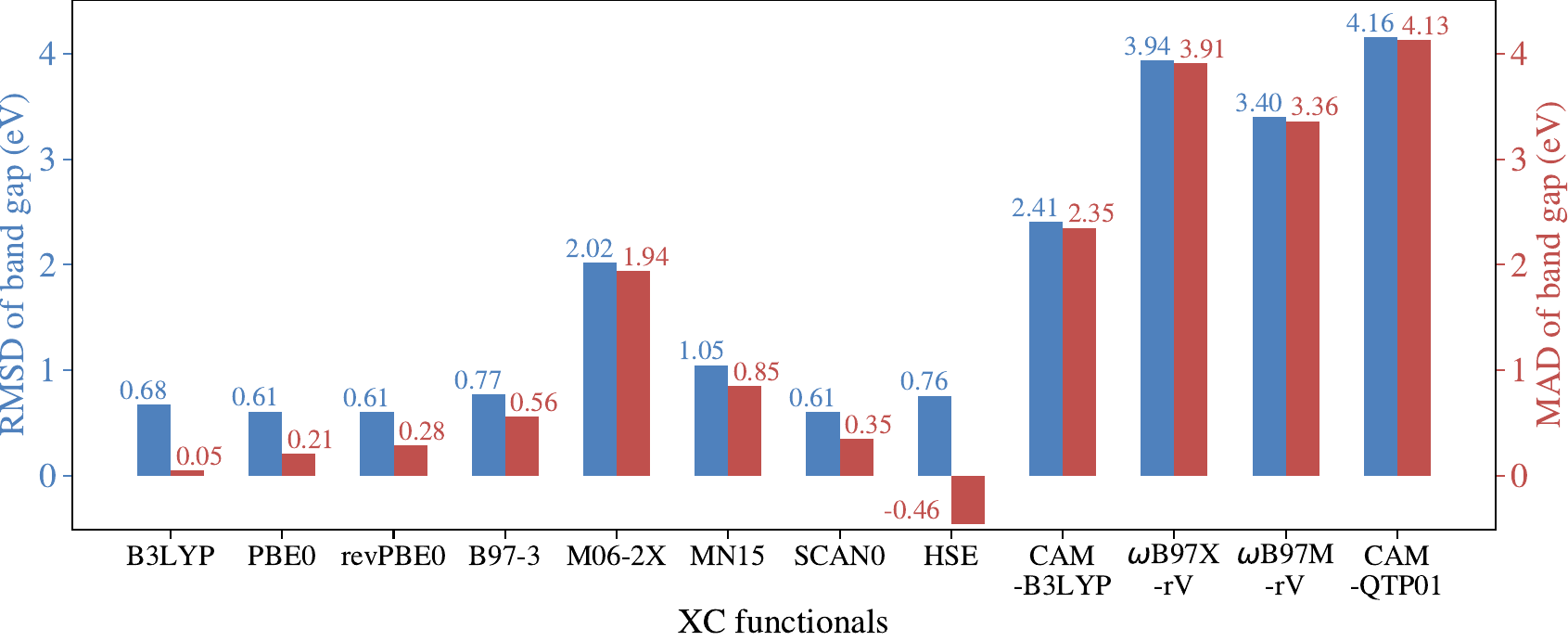}}
    \caption{
    Band gap (eV) comparison over 25 solids between DFT (12 different functionals) and experiments: Blue: root-mean-square-deviation (RMSD) of DFT band gaps (eV) with respect to those of experiments and Red: mean-average-deviation (MAD) of DFT band gaps (eV) with respect to those of experiments.
    }
    \label{fig:bar}
\end{figure*}

We obtain a more global perspective by inspecting the statistical data presented in \cref{fig:bar}.
In terms of root-mean-square-deviation (RMSD),
B3LYP, PBE0, revPBE0, B97-3, SCAN0, and HSE are all quite comparable (0.61-0.77 eV).
Other functionals including MN15, M06-2X, CAM-B3LYP, $\omega$B97X-rV, $\omega$B97M-rV, and CAM-QTP01
are significantly worse than these functionals.
The worst performing functional is CAM-QTP01 (4.16 eV) and the second worst performing functional is $\omega$B97X-rV (3.94 eV).
In terms of mean-average-deviation (MAD), another interesting trend arises. 
The HSE functional has a noticeable, negative MAD, which is likely due to
the lack of long-range exact exchange.
Other functionals with a higher fraction of exact exchange
show positive MAD values.
Given these data, following the combinatorial design strategy, it may be beneficial to develop a variant of $\omega$B97X-rV or $\omega$B97M-rV where
the long-range exact exchange is limited to less than $25\%$.
Examining the difference between maximum deviation and minimum deviation, 
we found more modern functionals such as M06-2X (2.54 eV) may benefit more from error cancellation in practice 
than B3LYP (3.78 eV).
The raw data for plots presented here are available in \cref{tab:gap2}.

\subsection {Outlook for future functional developments}\label{subsec:outlook}
\begin{figure}[ht!]
    \centering
    \scalebox{0.44}{\includegraphics{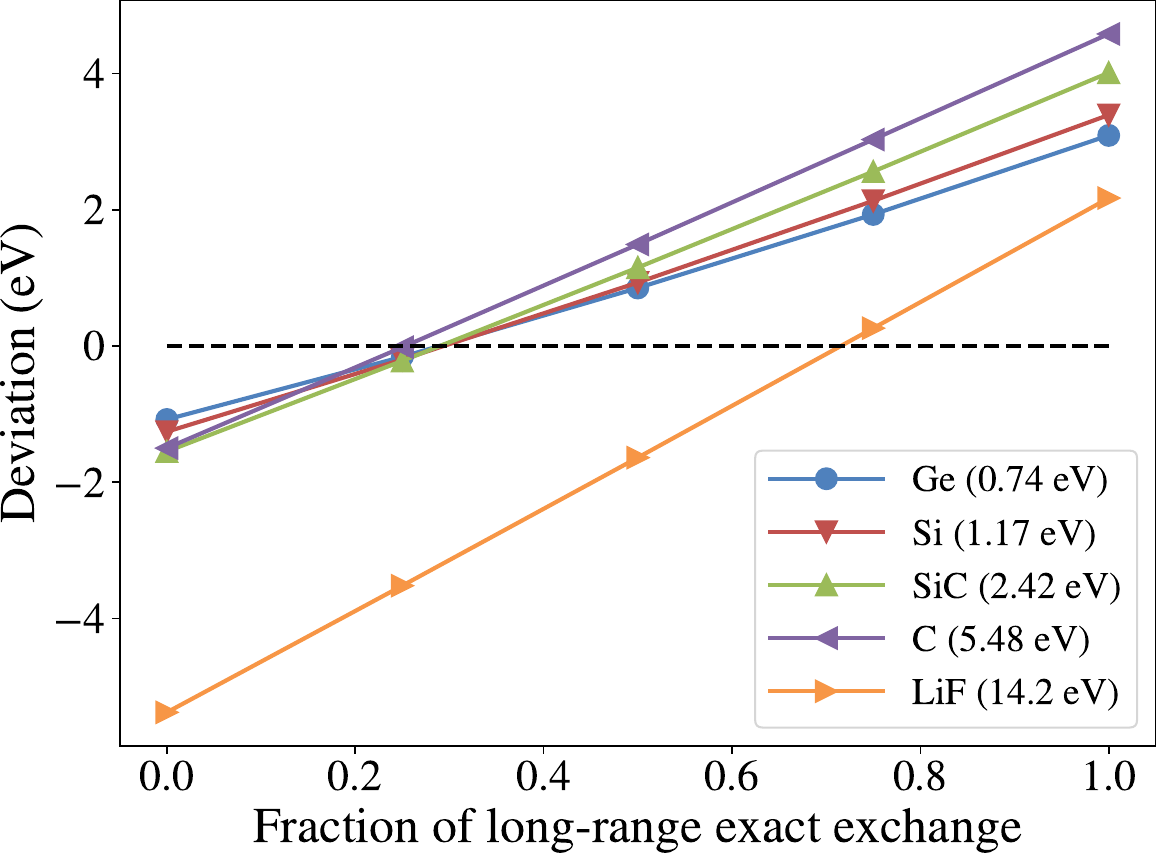}}
    \caption{
    Deviation (eV) of the computed band gaps from $\omega$B97M-rV
    with respect to experimental band gaps (given in parentheses)
    as a function of the 
    fraction of long-range exact exchange ($c_{x,lr}$).
    }
    \label{fig:scan}
\end{figure}
\begin{figure}[ht!]
    \centering
    \scalebox{0.48}{\includegraphics{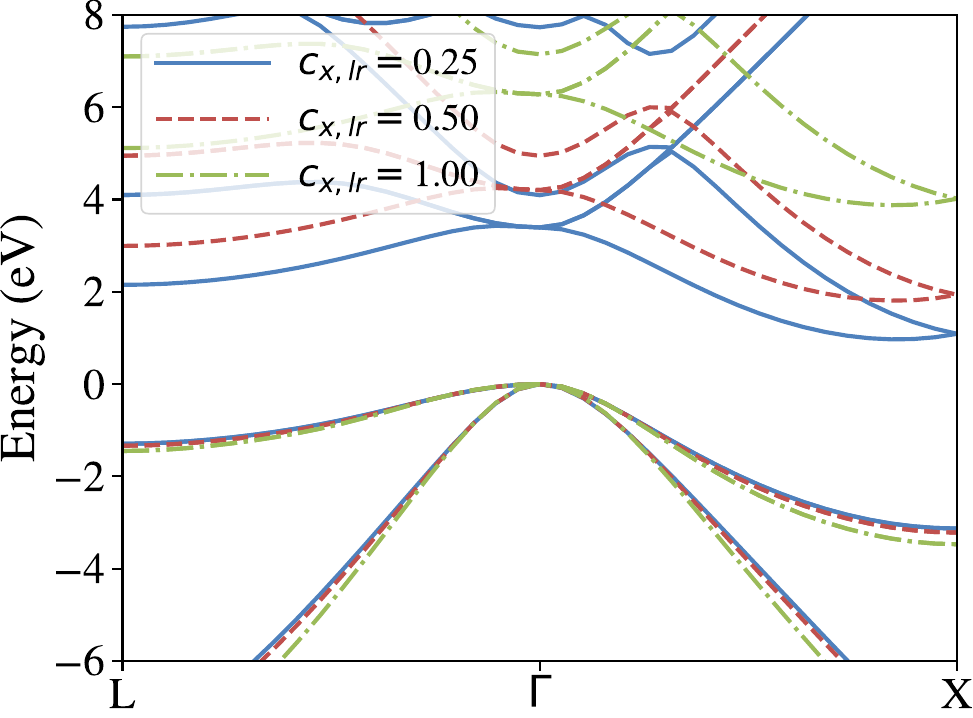}}
    \caption{
    Bands of Si computed from $\omega$B97M-rV
    as a function of the 
    fraction of long-range exact exchange ($c_{x,lr}$).
    The $\Gamma$-point valence band maximum is shifted to zero for comparisons.
    }
    \label{fig:si_bands}
\end{figure}
To gain more insights into functional developments,
we examine the effect of the fraction of long-range exact exchange (i.e., $c_{x,lr}$) in $\omega$B97M-rV on five solids (Ge, Si, SiC, C, LiF), whose
experimental band gaps range from 0.74 eV to 14.2 eV.
These results are presented in \cref{fig:scan}.
Despite the fact that we relaxed orbitals for each of $c_{x,lr}$ values,
the change in band gaps shows a completely linear behavior with respect to $c_{x,lr}$.
This is observed in nearly all bands (\cref{fig:si_bands}), not just in the frontier bands.
The band structure shows a nearly constant shift for different $c_{x,lr}$ values.

The most striking aspect of this plot is that
the optimal $c_{x,lr}$ for four solids (Ge, Si, SiC, and C) is around 0.25--0.3
whereas
the optimal value for LiF is near 0.75. 
$c_{x,lr}$ of 0.25--0.3 is close to the fraction of exact exchange in
the GH functionals that perform well as discussed in \cref{sec:bandgap}.
Qualitatively,
large gap materials 
do not benefit as much from
screening
and rather
more long-range exact exchange
is desirable.
This is qualitatively similar to what we see from molecular systems that typically exhibit large gaps.
This has been previously pointed out several others in literature.\cite{Civalleri2012Oct,Koller2013Oct,Skone2014May,Garza2016Oct}

Given these observations,
there are two potential ways for future functional developments
that can perform well for both solids and molecules.
The first is that one may
combinatorially optimize a density functional
with short-range, middle-range, and long-range exact exchange.\cite{Henderson2007Dec,Henderson2008Aug,Janesko2009Jan}
The idea is that one should not have a too high fraction of long-range exact exchange
for small-to-medium-gap materials, but
one would  need a large fraction of middle-range exact exchange for good performance on
large-gap materials and molecules.
The second idea is to develop a system-specific density functional that would vary
the fraction of exact exchange depending on the system.
This is closely related to
the dielectric-dependent hybrid functionals developed by Galli and co-workers,\cite{Skone2014May} 
but its performance on molecular systems has not been extensively assessed yet.\cite{Brawand2016Oct}
The last is to minimize the quasiparticle energy correction from G0W0 by tuning the fraction of exact exchange following the work by Atalla {\it et al}.\cite{Atalla2013Oct,Pinheiro2015Nov,Marom2017Jan}
This approach has shown promising accuracy for molecules and solids.

\section{Conclusions}\label{sec:conclusions}
In this work,
the occ-RI-K algorithm\cite{Manzer2015Jul}, which was originally developed for and has been successfully applied to molecules, has been extended to evaluate exact exchange in solid-state applications.
Within the GPW density fitting scheme,\cite{lippert1997hybrid,vandevondele2005quickstep} we showed that
the occ-RI-K algorithm 
achieves a nearly 1-2 orders of magnitude speedup
compared to other
conventional ways of computing the exact exchange contribution.
With the efficient occ-RI-K algorithm,
we were able to assess
the performance of a total of 12 hybrid density functionals 
for computing the band gap of 25 simple solids.

From the benchmark, we found that 
better performing density functionals were global hybrid functionals (B3LYP, PBE0, revPBE0, B97-3, SCAN0)
where the fraction of exact exchange is between 0.20 and 0.27.
A short-range hybrid functional, HSE, was found to underestimate the band gaps quite significantly compared to other hybrid functionals,
consistent with a previous study.\cite{Garza2016Oct} 
Minnesota functionals, M06-2X and MN15, are known for
their good performance on main group chemistry benchmarks, but
their band gaps were found to be severely overestimated due to their relatively high fraction of exact exchange.
Long-range corrected density functionals (CAM-B3LYP, $\omega$B97X-rV, $\omega$B97M-rV, CAM-QTP01)
all grossly overestimate the band gaps due to their high fraction of long-range exact exchange.
We also found that the optimal fraction of long-range exact exchange in $\omega$B97M-rV
needs to vary significantly depending on materials. 

Our work leaves a lot of room for future algorithmic developments, functional assessments, and functional developments.
For algorithms,
even with the occ-RI-K algorithm, the formal scaling of $\tilde{\mathcal O}(N_k^2 N^3)$ can be too expensive for more realistic solids.
By combining with tensor hypercontraction,\cite{Hohenstein2012,lee2019systematically} one can reduce this cost to $\tilde{\mathcal O}(N_k N^3)$.\cite{Wu2022Jan}
This algorithm will enable routine application of hybrid functionals to materials that require a large k-mesh. 
We are currently developing and investigating this algorithm.
Furthermore, an all-electron implementation of occ-RI-K will eliminate pseudopotential errors.
Such an implementation will make a relative efficiency comparison possible against 
other all-electron exchange algorithms\cite{Euwema1973Jan,Baraille1998Dec,Gillan2008Oct,Paier2009Nov,Civalleri2010Mar}
as well as
all-electron band gap data.\cite{Koller2013Oct,Tran2017May,Tran2018Feb}
For functional assessments,
we did not cover examples where small molecules are
interacting with the surface of solids, which
is commonly found in heterogeneous catalysis.
We expect our combinatorially optimized density functionals to perform well for barrier heights and adsorption energies at surfaces, but
there are only limited benchmark data points available.\cite{MallikarjunSharada2017Sep} 
For functional developments,
we noted that
mid-range exact exchange functionals,\cite{Henderson2007Dec,Henderson2008Aug,Janesko2009Jan}
functionals with system-dependent fraction of exact exchange\cite{Skone2014May} 
and
functionals that minimize the quasiparticle correction of G0W0 \cite{Atalla2013Oct,Pinheiro2015Nov,Marom2017Jan}
could be worth exploring further in the future.
Local hybrid functionals \cite{Maier2019Jan} and optimally tuned range-separated hybrids \cite{Dahvyd2021Aug} are also good alternatives to investigate further.
With the combined effort of algorithmic improvements and density functional developments and assessments, 
we hope to increase the predictive power and scalability of modern density functionals for simulations of molecules and materials.

\section{Note}
Towards the completion of this manuscript, a related work on using occ-RI-K for $\Gamma$-point calculations appeared on arXiv.\cite{Sharma2022Jul}

\section{Data Availability}
We provide inputs for QuantumESPRESSO in the Supplementary Materials, which contains geometry information about solids studied here.

\section{Acknowledgment}
We thank Leo Cunha for his initial assistance with QuantumESPRESSO calculations.
This work was supported by the National Institutes of Health SBIR program through Grant No. 2R44GM128480-02A1.
JL thanks David Reichman for support.
\section{Conflict of Interest}
E.E. and M.H.-G. are part-owners of Q-Chem, Inc.

\section{Appendix}
\appendix

\setcounter{equation}{0}
\renewcommand{\theequation}{A\arabic{equation}}
\setcounter{table}{0}
\renewcommand{\thetable}{A\Roman{table}}
\setcounter{figure}{0}
\renewcommand{\thefigure}{A\Roman{figure}}
\setcounter{subsection}{0}
\renewcommand{\thesubsection}{A\arabic{subsection}}

\subsection{Numerically stable evaluation of truncated long-range Coulomb operator} \label{asubsec:num}
In the second term in \cref{eq:lrtrunc}, we observe
\begin{align}
    \lim_{x\rightarrow \infty} e^{-x^2/4\omega^2} &\rightarrow 0\\
    \lim_{x\rightarrow \infty} \text{erf}(\omega(R_c+\frac{ix}{2\omega^2})) &\rightarrow \infty
\end{align}
These terms cancel out giving a finite result but the individual terms quickly exceed double precision even for moderate grid sizes. The error function can be expanded about $\infty$, giving:
\begin{equation}
    \text{erf}(x)  = 1 - \frac{e^{-x^2}}{\sqrt{\pi} x}\sum_{n=0}^\infty (-1)^n\frac{(2n - 1)!!}{(2x^2)^n} \approx 1 - \frac{e^{-x^2}}{\sqrt{\pi} x}
\end{equation}
Substituting this expression into \cref{eq:lrtrunc} cancels out the problematic terms leading to a more numerically stable form:
\begin{align}\nonumber
    f(x) &\approx \frac{4\pi}{x^2}
    \bigg[\frac{e^{-\omega^2 R_c^2}}{\sqrt{\pi}(\omega^2 R_c^2 + \mathbf{G}^2/4\omega^2)} \bigg(\omega R_c \cos (x R_c)\\ 
    &- \frac{x}{2 \omega} \sin (x R_c)  \bigg) - \cos (x R_c)\text{erf} (\omega R_c) \bigg]
\end{align}

\begin{table*}
\begin{tabular}{crrrrrrrrrrrrr}
\hline\hline
Name & \multicolumn{1}{c}{B3LYP} & \multicolumn{1}{c}{PBE0} & \multicolumn{1}{c}{revPBE0} & \multicolumn{1}{c}{B97-3} & \multicolumn{1}{c}{M06-2X} & \multicolumn{1}{c}{MN15} & \multicolumn{1}{c}{SCAN0} & \multicolumn{1}{c}{HSE} & \multicolumn{1}{c}{\begin{tabular}[c]{@{}c@{}}CAM\\ -B3LYP\end{tabular}} & \multicolumn{1}{c}{\begin{tabular}[c]{@{}c@{}}$\omega$B97X\\ -rV\end{tabular}} & \multicolumn{1}{c}{\begin{tabular}[c]{@{}c@{}}$\omega$B97M\\ -rV\end{tabular}} 
& \multicolumn{1}{c}{\begin{tabular}[c]{@{}c@{}} CAM\\ -QTP01\end{tabular}}
& \multicolumn{1}{c}{Exp.} \\
\hline\hline
    C &   5.98 &  6.09 &     6.13 &   6.39 &    7.85 &  7.29 &   6.26 &  5.32 &       8.22 &             9.54 &             9.25 &       9.86 &  5.48 \\
   Si &   1.83 &  1.75 &     1.79 &   2.11 &    2.97 &  1.96 &   1.88 &  1.15 &       3.52 &             4.53 &             3.97 &       4.82 &  1.17 \\
   Ge &   1.05 &  1.27 &     1.27 &   1.41 &    1.84 &  1.11 &   1.42 &  0.69 &       2.78 &             3.88 &             3.32 &       4.12 &  0.74 \\
  SiC &   2.91 &  2.84 &     2.87 &   3.02 &    4.37 &  3.79 &   3.03 &  2.26 &       4.91 &             6.08 &             5.58 &       6.38 &  2.42 \\
   BN &   6.46 &  6.45 &     6.51 &   6.83 &    8.41 &  7.46 &   6.70 &  5.80 &       8.71 &            10.02 &             9.59 &      10.40 &  6.22 \\
   BP &   2.72 &  2.68 &     2.72 &   2.98 &    4.07 &  3.35 &   2.84 &  2.00 &       4.59 &             5.73 &             5.34 &       6.01 &  2.40 \\
  BAs &   2.51 &  2.49 &     2.53 &   2.78 &    3.58 &  3.00 &   2.61 &  1.82 &       4.28 &             5.38 &             4.95 &       5.63 &  1.46 \\
  AlP &   2.92 &  2.85 &     2.91 &   3.20 &    3.91 &  3.18 &   3.03 &  2.29 &       4.74 &             5.78 &             5.16 &       6.05 &  2.51 \\
 AlAs &   2.69 &  2.58 &     2.65 &   2.94 &    3.50 &  2.83 &   2.73 &  2.03 &       4.40 &             5.37 &             4.72 &       5.64 &  2.23 \\
 AlSb &   2.31 &  2.22 &     2.29 &   2.56 &    2.92 &  2.28 &   2.29 &  1.67 &       3.90 &             4.83 &             4.15 &       5.09 &  1.68 \\
 $\beta$-GaN &   2.95 &  3.19 &     3.25 &   3.59 &    5.83 &  4.48 &   3.24 &  2.72 &       5.06 &             6.46 &             5.98 &       6.69 &  3.30 \\
  GaP &   2.77 &  2.84 &     2.92 &   3.20 &    3.98 &  3.03 &   2.94 &  2.25 &       4.54 &             5.69 &             5.21 &       5.90 &  2.35 \\
 GaAs &   1.43 &  1.74 &     1.80 &   1.99 &    3.25 &  1.88 &   1.74 &  1.18 &       3.22 &             4.65 &             4.00 &       4.69 &  1.52 \\
 GaSb &   0.94 &  1.26 &     1.31 &   1.47 &    2.50 &  1.27 &   1.20 &  0.70 &       2.63 &             4.05 &             3.31 &       4.05 &  0.73 \\
  InP &   1.61 &  1.89 &     1.95 &   2.17 &    3.49 &  2.12 &   1.80 &  1.33 &       3.50 &             4.88 &             4.38 &       4.99 &  1.42 \\
  ZnS &   3.38 &  3.66 &     3.76 &   4.08 &    5.59 &  4.10 &   3.79 &  3.08 &       5.50 &             6.88 &             6.39 &       7.07 &  3.66 \\
 ZnSe &   2.40 &  2.65 &     2.75 &   3.04 &    4.38 &  2.89 &   2.77 &  2.09 &       4.33 &             5.65 &             5.07 &       5.78 &  2.70 \\
 ZnTe &   2.30 &  2.55 &     2.64 &   2.88 &    4.06 &  2.64 &   2.63 &  1.99 &       4.15 &             5.49 &             4.79 &       5.58 &  2.38 \\
  CdS &   2.27 &  2.53 &     2.64 &   2.89 &    4.11 &  2.81 &   2.53 &  1.96 &       4.28 &             5.61 &             5.18 &       5.79 &  2.55 \\
 CdSe &   1.60 &  1.83 &     1.94 &   2.18 &    3.27 &  1.94 &   1.82 &  1.29 &       3.44 &             4.70 &             4.21 &       4.84 &  1.90 \\
 CdTe &   1.66 &  1.88 &     1.98 &   2.19 &    3.17 &  1.87 &   1.85 &  1.35 &       3.45 &             4.71 &             4.11 &       4.81 &  1.92 \\
  LiH &   4.81 &  4.72 &     4.85 &   5.10 &    6.50 &  5.93 &   5.14 &  4.01 &       7.21 &             8.56 &             8.01 &       8.74 &  4.90 \\
  LiF &  11.47 & 11.89 &    12.09 &  12.32 &   14.19 & 13.25 &  12.35 & 11.44 &      13.98 &            15.48 &            14.98 &      15.88 & 14.20 \\
 LiCl &   8.02 &  8.26 &     8.42 &   8.73 &    9.72 &  8.66 &   8.73 &  7.78 &      10.25 &            11.66 &            11.05 &      11.91 &  9.40 \\
  AlN &   6.09 &  6.17 &     6.22 &   6.55 &    8.58 &  7.51 &   6.49 &  5.61 &       8.37 &             9.70 &             9.07 &      10.10 &  6.13 \\
\hline
\hline
RMSD & 0.73 & 0.64 & 0.63 & 0.75 & 1.73 & 0.88 & 0.62 & 0.76 & 2.04 & 3.24 & 2.72 & 3.46 & N/A \\
MAD & -0.01 & 0.12 & 0.19 & 0.45 & 1.63 & 0.61 & 0.26 & -0.46 & 1.94 & 3.20 & 2.66 & 3.42 & N/A \\
MAX & 1.05 & 1.03 & 1.07 & 1.32 & 2.53 & 1.81 & 1.15 & 0.36 & 2.82 & 4.06 & 3.77 & 4.38 & N/A \\
MIN & -2.73 & -2.31 & -2.11 & -1.88 & -0.01 & -0.95 & -1.85 & -2.76 & -0.22 & 1.28 & 0.78 & 1.69 & N/A \\
\hline\hline
\end{tabular}
\caption{Experimental and theoretical band gaps (eV) from various functionals over 25 solids.
The singularity treatment for exact exchange was performed via the Madelung correction.
N/A means ``not available''.
RMSD, MAD, MAX and MIN denote, respectively,
root-mean-square-deviation,
mean-average-deviation,
maximum signed deviation,
and
minimum signed deviation
in reference to experimental values.
All calculations were based on SCF calculations with 6$\times$6$\times$6 $\mathbf k$-mesh.
\label{tab:gap1}}
\end{table*}

\begin{table*}
\begin{tabular}{crrrrrrrrrrrrr}
\hline\hline
Name & \multicolumn{1}{c}{B3LYP} & \multicolumn{1}{c}{PBE0} & \multicolumn{1}{c}{revPBE0} & \multicolumn{1}{c}{B97-3} & \multicolumn{1}{c}{M06-2X} & \multicolumn{1}{c}{MN15} & \multicolumn{1}{c}{SCAN0} & \multicolumn{1}{c}{HSE} & \multicolumn{1}{c}{\begin{tabular}[c]{@{}c@{}}CAM\\ -B3LYP\end{tabular}} & \multicolumn{1}{c}{\begin{tabular}[c]{@{}c@{}}$\omega$B97X\\ -rV\end{tabular}} 
& \multicolumn{1}{c}{\begin{tabular}[c]{@{}c@{}}$\omega$B97M\\ -rV\end{tabular}}
& \multicolumn{1}{c}{\begin{tabular}[c]{@{}c@{}}CAM\\ -QTP01\end{tabular}} 
& \multicolumn{1}{c}{Exp.} \\
\hline\hline
    C &   5.96 &  6.07 &     6.11 &   6.38 &    8.12 &  7.48 &   6.23 &  5.32 &       8.63 &            10.35 &            10.06 &      10.66 &  5.48 \\
   Si &   1.85 &  1.79 &     1.84 &   2.18 &    3.24 &  2.16 &   1.93 &  1.15 &       3.87 &             5.13 &             4.57 &       5.42 &  1.17 \\
   Ge &   1.03 &  1.28 &     1.27 &   1.43 &    2.02 &  1.24 &   1.43 &  0.69 &       3.05 &             4.41 &             3.83 &       4.65 &  0.74 \\
  SiC &   3.00 &  2.97 &     3.01 &   3.17 &    4.76 &  4.11 &   3.16 &  2.26 &       5.40 &             6.93 &             6.43 &       7.23 &  2.42 \\
   BN &   6.51 &  6.55 &     6.61 &   6.96 &    8.81 &  7.78 &   6.79 &  5.80 &       9.25 &            10.93 &            10.50 &      11.31 &  6.22 \\
   BP &   2.70 &  2.69 &     2.73 &   3.02 &    4.33 &  3.55 &   2.85 &  2.00 &       4.96 &             6.40 &             6.01 &       6.68 &  2.40 \\
  BAs &   2.50 &  2.49 &     2.53 &   2.80 &    3.81 &  3.18 &   2.61 &  1.82 &       4.62 &             6.00 &             5.57 &       6.25 &  1.46 \\
  AlP &   3.00 &  2.96 &     3.02 &   3.33 &    4.23 &  3.44 &   3.14 &  2.29 &       5.14 &             6.49 &             5.87 &       6.77 &  2.51 \\
 AlAs &   2.75 &  2.68 &     2.75 &   3.06 &    3.79 &  3.07 &   2.82 &  2.03 &       4.77 &             6.03 &             5.38 &       6.31 &  2.23 \\
 AlSb &   2.36 &  2.29 &     2.36 &   2.65 &    3.17 &  2.48 &   2.37 &  1.67 &       4.23 &             5.40 &             4.71 &       5.67 &  1.68 \\
 $\beta$-GaN &   3.12 &  3.42 &     3.48 &   3.85 &    6.40 &  4.93 &   3.46 &  2.72 &       5.75 &             7.54 &             7.06 &       7.78 &  3.30 \\
  GaP &   2.81 &  2.92 &     2.99 &   3.29 &    4.27 &  3.24 &   3.01 &  2.25 &       4.91 &             6.31 &             5.83 &       6.52 &  2.35 \\
 GaAs &   1.45 &  1.79 &     1.85 &   2.05 &    3.50 &  2.05 &   1.78 &  1.18 &       3.53 &             5.20 &             4.54 &       5.24 &  1.52 \\
 GaSb &   0.94 &  1.29 &     1.34 &   1.51 &    2.70 &  1.41 &   1.21 &  0.70 &       2.90 &             4.54 &             3.79 &       4.54 &  0.73 \\
  InP &   1.65 &  1.96 &     2.02 &   2.25 &    3.76 &  2.31 &   1.87 &  1.33 &       3.83 &             5.47 &             4.96 &       5.58 &  1.42 \\
  ZnS &   3.46 &  3.77 &     3.87 &   4.20 &    5.92 &  4.35 &   3.90 &  3.08 &       5.92 &             7.63 &             7.13 &       7.82 &  3.66 \\
 ZnSe &   2.46 &  2.74 &     2.84 &   3.15 &    4.68 &  3.11 &   2.86 &  2.09 &       4.70 &             6.33 &             5.75 &       6.46 &  2.70 \\
 ZnTe &   2.35 &  2.62 &     2.71 &   2.97 &    4.33 &  2.84 &   2.71 &  1.99 &       4.48 &             6.09 &             5.38 &       6.17 &  2.38 \\
  CdS &   2.34 &  2.63 &     2.74 &   3.01 &    4.42 &  3.04 &   2.63 &  1.96 &       4.68 &             6.32 &             5.88 &       6.50 &  2.55 \\
 CdSe &   1.66 &  1.92 &     2.03 &   2.29 &    3.55 &  2.14 &   1.91 &  1.29 &       3.80 &             5.36 &             4.86 &       5.49 &  1.90 \\
 CdTe &   1.71 &  1.96 &     2.06 &   2.28 &    3.42 &  2.05 &   1.92 &  1.35 &       3.76 &             5.29 &             4.68 &       5.38 &  1.92 \\
  LiH &   4.78 &  4.69 &     4.81 &   5.06 &    6.44 &  5.87 &   5.11 &  4.01 &       7.14 &             8.69 &             8.10 &       8.88 &  4.90 \\
  LiF &  11.73 & 12.22 &    12.42 &  12.68 &   14.91 & 13.83 &  12.68 & 11.44 &      14.85 &            16.82 &            16.32 &      17.23 & 14.20 \\
 LiCl &   8.23 &  8.52 &     8.68 &   9.01 &   10.29 &  9.12 &   8.99 &  7.78 &      10.93 &            12.71 &            12.10 &      12.96 &  9.40 \\
  AlN &   6.20 &  6.34 &     6.39 &   6.74 &    9.06 &  7.90 &   6.66 &  5.61 &       8.98 &            10.69 &            10.06 &      11.09 &  6.13 \\
\hline
\hline
RMSD & 0.68 & 0.61 & 0.61 & 0.77 & 2.02 & 1.05 & 0.61 & 0.76 & 2.41 & 3.94 & 3.40 & 4.16 & N/A \\
MAD & 0.05 & 0.21 & 0.28 & 0.56 & 1.94 & 0.85 & 0.35 & -0.46 & 2.35 & 3.91 & 3.36 & 4.13 & N/A \\
MAX & 1.04 & 1.03 & 1.07 & 1.34 & 3.10 & 2.00 & 1.15 & 0.36 & 3.16 & 4.87 & 4.58 & 5.18 & N/A \\
MIN & -2.46 & -1.98 & -1.78 & -1.52 & 0.71 & -0.37 & -1.52 & -2.76 & 0.65 & 2.62 & 2.12 & 3.03 & N/A \\
\hline\hline
\end{tabular}
\caption{Same as \cref{tab:gap1} except that this is using the truncated Coulomb scheme for the singularity correction.
\label{tab:gap2}}
\end{table*}

\begin{table*}
\begin{tabular}{|c|r|r|r|r|r|r|r|r|}
\hline
Name & \multicolumn{1}{c|}{$N_k^{1/3}$=1} & \multicolumn{1}{c|}{$N_k^{1/3}$=2} & \multicolumn{1}{c|}{$N_k^{1/3}$=3} 
& \multicolumn{1}{c|}{$N_k^{1/3}$=4} & \multicolumn{1}{c|}{$N_k^{1/3}$=5} & \multicolumn{1}{c|}{$N_k^{1/3}$=6} 
& \multicolumn{1}{c|}{$N_k^{1/3}$=7} 
& \multicolumn{1}{c|}{Q-Chem, $N_k^{1/3}$=6} \\ \hline
C    & -10.315592 & -11.22904  & -11.316087 & -11.329812 & -11.332389 & -11.332836 & -11.332859 & -11.332728 \\ \hline
Si   & -7.263846  & -7.7424678 & -7.8042559 & -7.8168333 & -7.8200165 & -7.8208858 & -7.8211188 & -7.820662  \\ \hline
Ge   & -7.1307138 & -7.7096594 & -7.7819087 & -7.7978528 & -7.8023892 & -7.8038689 & -7.8043957 & -7.803826  \\ \hline
SiC  & -9.0057586 & -9.5266777 & -9.5911035 & -9.6020953 & -9.6043135 & -9.6047434 & -9.6047916 & -9.604381  \\ \hline
BN   & -11.984863 & -12.759566 & -12.825463 & -12.833995 & -12.835224 & -12.835311 & -12.835242 & -12.835082 \\ \hline
BP   & -8.6770283 & -9.31314   & -9.3866072 & -9.4002713 & -9.4034178 & -9.4041736 & -9.4043326 & -9.403920  \\ \hline
BAs  & -8.3285878 & -8.9973475 & -9.072129  & -9.0863517 & -9.0897338 & -9.0905892 & -9.0907905 & -9.090522  \\ \hline
AlP  & -8.2351585 & -8.6134039 & -8.6587502 & -8.6664663 & -8.6680014 & -8.6682884 & -8.6683125 & -8.668096  \\ \hline
AlAs & -7.9029732 & -8.3130158 & -8.3623186 & -8.3712477 & -8.373195  & -8.3736278 & -8.3737044 & -8.373506  \\ \hline
AlSb & -7.0993312 & -7.507275  & -7.5556889 & -7.5650053 & -7.5672091 & -7.5677636 & -7.5678933 & -7.567673  \\ \hline
$\beta$-GaN & -83.619295 & -84.116855 & -84.160975 & -84.16707  & -84.167925 & -84.167964 & -84.167894 & -84.168307 \\ \hline
GaP  & -80.225323 & -80.728226 & -80.784072 & -80.794673 & -80.797133 & -80.797732 & -80.797862 & -80.798163 \\ \hline
GaAs & -79.902226 & -80.432168 & -80.493022 & -80.505338 & -80.50848  & -80.509371 & -80.509633 & -80.509833 \\ \hline
GaSb &   N/C         & -79.635938 & -79.697619 & -79.71142  & -79.714836 & -79.715891 & -79.716242 & -79.716115 \\ \hline
InP  & -62.105167 & -62.553162 & -62.598143 & -62.606302 & -62.608092 & -62.608495 & -62.608565 & -62.608354 \\ \hline
ZnS  & -70.012286 & -70.391394 & -70.422686 & -70.426918 & -70.427461 & -70.427451 & -70.427381 & -70.427359 \\ \hline
ZnSe & -69.183122 & -69.588633 & -69.625613 & -69.631328 & -69.63231  & -69.632438 & -69.632414 & -69.632411 \\ \hline
ZnTe & -67.877895 & -68.289422 & -68.330126 & -68.337139 & -68.338569 & -68.338851 & -68.33888  & -68.338818 \\ \hline
CdS  & -55.692652 & -56.017378 & -56.039946 & -56.042664 & -56.042894 & -56.042816 & -56.042732 & -56.042425 \\ \hline
CdSe &    N/C        & -55.22138  & -55.249459 & -55.253457 & -55.254045 & -55.254076 & -55.254027 & -55.253732 \\ \hline
CdTe &      N/C      & -53.930771 & -53.962107 & -53.967055 & -53.96794  & -53.968066 & -53.968049 & -53.967760 \\ \hline
LiH  & -8.3559441 & -8.0875882 & -8.1087375 & -8.1063673 & -8.1064386 & -8.1062722 & -8.1062233 & -8.106026  \\ \hline
LiF  & -31.659096 & -31.869633 & -31.876959 & -31.87666  & -31.876383 & -31.876244 & -31.876171 & -31.875991 \\ \hline
LiCl & -22.358903 & -22.570483 & -22.584203 & -22.584804 & -22.584638 & -22.584509 & -22.584435 & -22.584162 \\ \hline
AlN  & -23.819271 & -24.174478 & -24.206023 & -24.209691 & -24.209945 & -24.209861 & -24.209777 & -24.209097 \\ \hline
\end{tabular}
\caption{B3LYP total energies per cell with varying $N_k$ using QuantumESPRESSO with kinetic energy cutoff of 400 Ry and $N_k^{1/3}=6$ total energies obtained from Q-Chem.
N/C means not converged.
\label{tab:total}}
\end{table*}

\begin{table*}
\begin{tabular}{|c|r|r|r|r|r|r|r|r|}
\hline
Name & \multicolumn{1}{c|}{$N_k^{1/3}$=1} & \multicolumn{1}{c|}{$N_k^{1/3}$=2} & \multicolumn{1}{c|}{$N_k^{1/3}$=3} 
& \multicolumn{1}{c|}{$N_k^{1/3}$=4} & \multicolumn{1}{c|}{$N_k^{1/3}$=5} & \multicolumn{1}{c|}{$N_k^{1/3}$=6} 
& \multicolumn{1}{c|}{$N_k^{1/3}$=7} 
& \multicolumn{1}{c|}{Q-Chem, $N_k^{1/3}$=6} \\ \hline
C    & 8.84  & 8.02  & 7.71  & 7.58  & 7.52  & 7.49  & 7.47  & 7.49  \\ \hline
Si   & 4.88  & 4.28  & 4.08  & 4.00  & 3.96  & 3.93  & 3.92  & 3.93  \\ \hline
Ge   & 2.63  & 1.74  & 1.39  & 1.21  & 1.11  & 1.05  & 1.01  & 1.05  \\ \hline
SiC  & 8.67  & 8.43  & 8.23  & 8.14  & 8.11  & 8.09  & 8.08  & 8.09  \\ \hline
BN   & 10.25 & 11.00 & 11.03 & 11.00 & 10.97 & 10.96 & 10.95 & 10.96 \\ \hline
BP   & 5.96  & 5.33  & 5.10  & 5.00  & 4.95  & 4.93  & 4.92  & 4.93  \\ \hline
BAs  & 5.65  & 5.05  & 4.83  & 4.73  & 4.68  & 4.66  & 4.65  & 4.66  \\ \hline
AlP  & 5.55  & 5.11  & 4.93  & 4.85  & 4.81  & 4.79  & 4.78  & 4.79  \\ \hline
AlAs & 4.28  & 3.77  & 3.56  & 3.46  & 3.41  & 3.38  & 3.37  & 3.38  \\ \hline
AlSb & 3.86  & 3.25  & 3.02  & 2.91  & 2.85  & 2.83  & 2.81  & 2.82  \\ \hline
$\beta$-GaN & 3.04  & 3.35  & 3.28  & 3.21  & 3.17  & 3.14  & 3.12  & 3.14  \\ \hline
GaP  & 3.81  & 3.36  & 3.15  & 3.04  & 2.98  & 2.95  & 2.93  & 2.95  \\ \hline
GaAs & 2.45  & 1.96  & 1.72  & 1.59  & 1.52  & 1.47  & 1.44  & 1.47  \\ \hline
GaSb & N/C  & 0.66  & 0.21  & 1.09  & 1.01  & 0.96  & 0.93  & 0.96  \\ \hline
InP  & 2.25  & 2.02  & 1.85  & 1.76  & 1.70  & 1.67  & 1.65  & 1.67  \\ \hline
ZnS  & 3.51  & 3.66  & 3.58  & 3.52  & 3.49  & 3.47  & 3.46  & 3.47  \\ \hline
ZnSe & 2.65  & 2.72  & 2.61  & 2.54  & 2.50  & 2.47  & 2.46  & 2.47  \\ \hline
ZnTe & 2.78  & 2.65  & 2.51  & 2.43  & 2.39  & 2.36  & 2.34  & 2.36  \\ \hline
CdS  & 2.11  & 2.50  & 2.45  & 2.40  & 2.37  & 2.35  & 2.34  & 2.35  \\ \hline
CdSe & N/C   & 1.88  & 1.81  & 1.74  & 1.70  & 1.68  & 1.66  & 1.68  \\ \hline
CdTe & N/C   & 1.97  & 1.86  & 1.79  & 1.75  & 1.73  & 1.71  & 1.73  \\ \hline
LiH  & 23.58 & 23.54 & 23.47 & 23.47 & 23.47 & 23.47 & 23.47 & 23.47 \\ \hline
LiF  & 10.29 & 11.74 & 11.78 & 11.76 & 11.75 & 11.74 & 11.74 & 11.74 \\ \hline
LiCl & 7.92  & 8.30  & 8.29  & 8.26  & 8.25  & 8.24  & 8.24  & 8.24  \\ \hline
AlN  & 6.01  & 6.19  & 6.22  & 6.21  & 6.21  & 6.20  & 6.20  & 6.20 \\ \hline
\end{tabular}
\caption{B3LYP band gaps at $\Gamma$ based on SCF calculations with varying $N_k$ using QuantumESPRESSO with kinetic energy cutoff of 400 Ry and $N_k^{1/3}=6$ band gaps obtained from Q-Chem.
N/C means not converged.
\label{tab:gap3}}
\end{table*}

\begin{figure}[ht!]
    \centering
    \scalebox{0.4}{\includegraphics{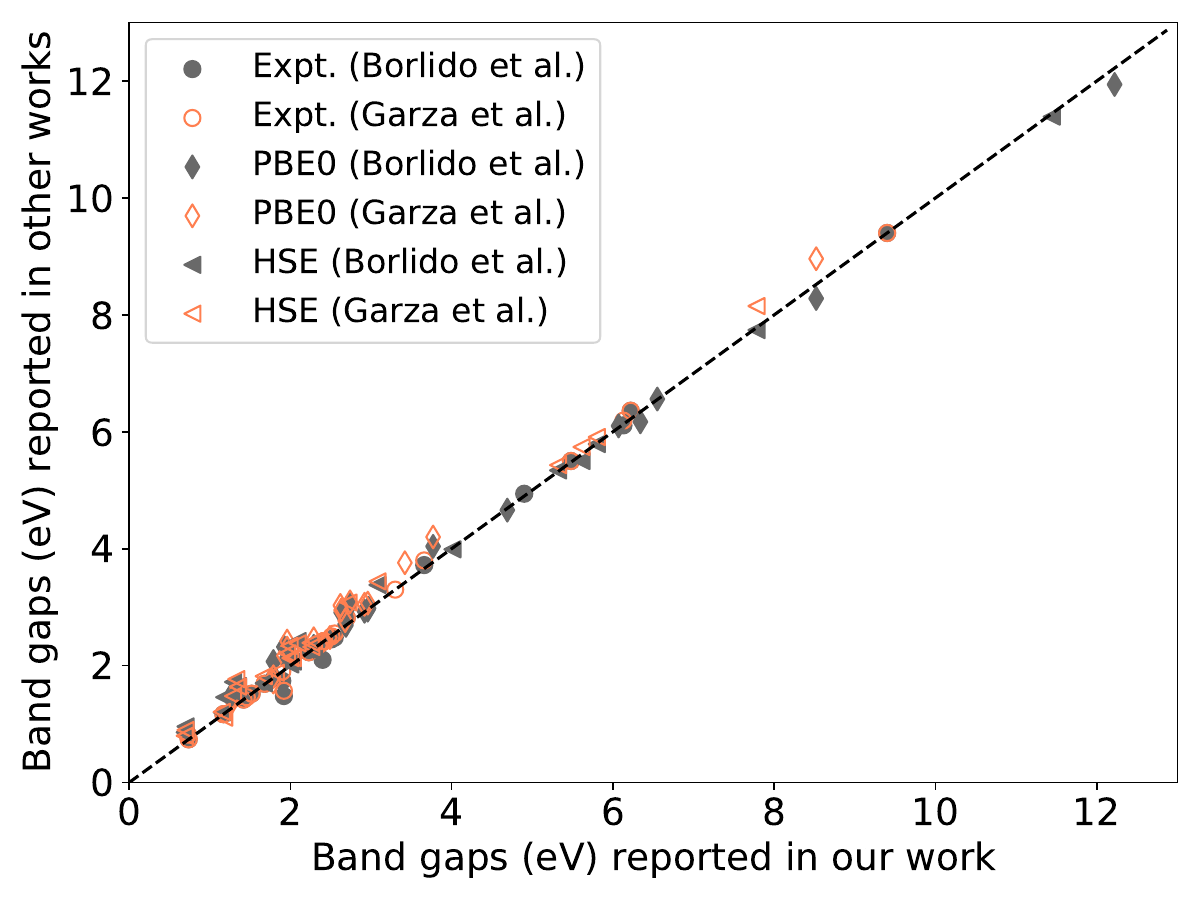}}
    \caption{
Comparison
among band gaps reported here,
those reported in the work of Garza {\it et al.} (Ref. \citenum{Garza2016Oct}),
and those reported in the work of Borlido {\it et al.} (Ref. \citenum{Borlido2019Sep}.)
    }
    \label{fig:tran}
\end{figure}

\bibliography{refs}

\begin{thebibliography}{95}%
\makeatletter
\providecommand \@ifxundefined [1]{%
 \@ifx{#1\undefined}
}%
\providecommand \@ifnum [1]{%
 \ifnum #1\expandafter \@firstoftwo
 \else \expandafter \@secondoftwo
 \fi
}%
\providecommand \@ifx [1]{%
 \ifx #1\expandafter \@firstoftwo
 \else \expandafter \@secondoftwo
 \fi
}%
\providecommand \natexlab [1]{#1}%
\providecommand \enquote  [1]{``#1''}%
\providecommand \bibnamefont  [1]{#1}%
\providecommand \bibfnamefont [1]{#1}%
\providecommand \citenamefont [1]{#1}%
\providecommand \href@noop [0]{\@secondoftwo}%
\providecommand \href [0]{\begingroup \@sanitize@url \@href}%
\providecommand \@href[1]{\@@startlink{#1}\@@href}%
\providecommand \@@href[1]{\endgroup#1\@@endlink}%
\providecommand \@sanitize@url [0]{\catcode `\\12\catcode `\$12\catcode
  `\&12\catcode `\#12\catcode `\^12\catcode `\_12\catcode `\%12\relax}%
\providecommand \@@startlink[1]{}%
\providecommand \@@endlink[0]{}%
\providecommand \url  [0]{\begingroup\@sanitize@url \@url }%
\providecommand \@url [1]{\endgroup\@href {#1}{\urlprefix }}%
\providecommand \urlprefix  [0]{URL }%
\providecommand \Eprint [0]{\href }%
\providecommand \doibase [0]{http://dx.doi.org/}%
\providecommand \selectlanguage [0]{\@gobble}%
\providecommand \bibinfo  [0]{\@secondoftwo}%
\providecommand \bibfield  [0]{\@secondoftwo}%
\providecommand \translation [1]{[#1]}%
\providecommand \BibitemOpen [0]{}%
\providecommand \bibitemStop [0]{}%
\providecommand \bibitemNoStop [0]{.\EOS\space}%
\providecommand \EOS [0]{\spacefactor3000\relax}%
\providecommand \BibitemShut  [1]{\csname bibitem#1\endcsname}%
\let\auto@bib@innerbib\@empty
\bibitem [{\citenamefont {Jain}\ \emph {et~al.}(2013)\citenamefont {Jain},
  \citenamefont {Ong}, \citenamefont {Hautier}, \citenamefont {Chen},
  \citenamefont {Richards}, \citenamefont {Dacek}, \citenamefont {Cholia},
  \citenamefont {Gunter}, \citenamefont {Skinner}, \citenamefont {Ceder},\ and\
  \citenamefont {Persson}}]{Jain2013Jul}%
  \BibitemOpen
  \bibfield  {author} {\bibinfo {author} {\bibfnamefont {Anubhav}\ \bibnamefont
  {Jain}}, \bibinfo {author} {\bibfnamefont {Shyue~Ping}\ \bibnamefont {Ong}},
  \bibinfo {author} {\bibfnamefont {Geoffroy}\ \bibnamefont {Hautier}},
  \bibinfo {author} {\bibfnamefont {Wei}\ \bibnamefont {Chen}}, \bibinfo
  {author} {\bibfnamefont {William~Davidson}\ \bibnamefont {Richards}},
  \bibinfo {author} {\bibfnamefont {Stephen}\ \bibnamefont {Dacek}}, \bibinfo
  {author} {\bibfnamefont {Shreyas}\ \bibnamefont {Cholia}}, \bibinfo {author}
  {\bibfnamefont {Dan}\ \bibnamefont {Gunter}}, \bibinfo {author}
  {\bibfnamefont {David}\ \bibnamefont {Skinner}}, \bibinfo {author}
  {\bibfnamefont {Gerbrand}\ \bibnamefont {Ceder}}, \ and\ \bibinfo {author}
  {\bibfnamefont {Kristin~A.}\ \bibnamefont {Persson}},\ }\bibfield  {title}
  {\enquote {\bibinfo {title} {{Commentary: The Materials Project: A materials
  genome approach to accelerating materials innovation}},}\ }\href {\doibase
  10.1063/1.4812323} {\bibfield  {journal} {\bibinfo  {journal} {APL Mater.}\
  }\textbf {\bibinfo {volume} {1}},\ \bibinfo {pages} {011002} (\bibinfo {year}
  {2013})}\BibitemShut {NoStop}%
\bibitem [{\citenamefont {Mardirossian}\ and\ \citenamefont
  {Head-Gordon}(2017)}]{Mardirossian2017Oct}%
  \BibitemOpen
  \bibfield  {author} {\bibinfo {author} {\bibfnamefont {Narbe}\ \bibnamefont
  {Mardirossian}}\ and\ \bibinfo {author} {\bibfnamefont {Martin}\ \bibnamefont
  {Head-Gordon}},\ }\bibfield  {title} {\enquote {\bibinfo {title} {{Thirty
  years of density functional theory in computational chemistry: an overview
  and extensive assessment of 200 density functionals}},}\ }\href {\doibase
  10.1080/00268976.2017.1333644} {\bibfield  {journal} {\bibinfo  {journal}
  {Mol. Phys.}\ }\textbf {\bibinfo {volume} {115}},\ \bibinfo {pages}
  {2315--2372} (\bibinfo {year} {2017})}\BibitemShut {NoStop}%
\bibitem [{\citenamefont {Perdew}(1985)}]{Perdew1985Mar}%
  \BibitemOpen
  \bibfield  {author} {\bibinfo {author} {\bibfnamefont {John~P.}\ \bibnamefont
  {Perdew}},\ }\bibfield  {title} {\enquote {\bibinfo {title} {{Density
  functional theory and the band gap problem}},}\ }\href {\doibase
  10.1002/qua.560280846} {\bibfield  {journal} {\bibinfo  {journal} {Int. J.
  Quantum Chem.}\ }\textbf {\bibinfo {volume} {28}},\ \bibinfo {pages}
  {497--523} (\bibinfo {year} {1985})}\BibitemShut {NoStop}%
\bibitem [{\citenamefont {Yakovkin}\ and\ \citenamefont
  {Dowben}(2007)}]{Yakovkin2007Jun}%
  \BibitemOpen
  \bibfield  {author} {\bibinfo {author} {\bibfnamefont {I.~N.}\ \bibnamefont
  {Yakovkin}}\ and\ \bibinfo {author} {\bibfnamefont {P.~A.}\ \bibnamefont
  {Dowben}},\ }\bibfield  {title} {\enquote {\bibinfo {title} {{THE PROBLEM OF
  THE BAND GAP IN LDA CALCULATIONS}},}\ }\href {\doibase
  10.1142/S0218625X07009499} {\bibfield  {journal} {\bibinfo  {journal} {Surf.
  Rev. Lett.}\ }\textbf {\bibinfo {volume} {14}},\ \bibinfo {pages} {481--487}
  (\bibinfo {year} {2007})}\BibitemShut {NoStop}%
\bibitem [{\citenamefont {Mori-S{\ifmmode\acute{a}\else\'{a}\fi}nchez}\ \emph
  {et~al.}(2008)\citenamefont {Mori-S{\ifmmode\acute{a}\else\'{a}\fi}nchez},
  \citenamefont {Cohen},\ and\ \citenamefont {Yang}}]{Mori-Sanchez2008Apr}%
  \BibitemOpen
  \bibfield  {author} {\bibinfo {author} {\bibfnamefont {Paula}\ \bibnamefont
  {Mori-S{\ifmmode\acute{a}\else\'{a}\fi}nchez}}, \bibinfo {author}
  {\bibfnamefont {Aron~J.}\ \bibnamefont {Cohen}}, \ and\ \bibinfo {author}
  {\bibfnamefont {Weitao}\ \bibnamefont {Yang}},\ }\bibfield  {title} {\enquote
  {\bibinfo {title} {{Localization and Delocalization Errors in Density
  Functional Theory and Implications for Band-Gap Prediction}},}\ }\href
  {\doibase 10.1103/PhysRevLett.100.146401} {\bibfield  {journal} {\bibinfo
  {journal} {Phys. Rev. Lett.}\ }\textbf {\bibinfo {volume} {100}},\ \bibinfo
  {pages} {146401} (\bibinfo {year} {2008})}\BibitemShut {NoStop}%
\bibitem [{\citenamefont {Heyd}\ \emph {et~al.}(2005)\citenamefont {Heyd},
  \citenamefont {Peralta}, \citenamefont {Scuseria},\ and\ \citenamefont
  {Martin}}]{Heyd2005Nov}%
  \BibitemOpen
  \bibfield  {author} {\bibinfo {author} {\bibfnamefont {Jochen}\ \bibnamefont
  {Heyd}}, \bibinfo {author} {\bibfnamefont {Juan~E.}\ \bibnamefont {Peralta}},
  \bibinfo {author} {\bibfnamefont {Gustavo~E.}\ \bibnamefont {Scuseria}}, \
  and\ \bibinfo {author} {\bibfnamefont {Richard~L.}\ \bibnamefont {Martin}},\
  }\bibfield  {title} {\enquote {\bibinfo {title} {{Energy band gaps and
  lattice parameters evaluated with the Heyd-Scuseria-Ernzerhof screened hybrid
  functional}},}\ }\href {\doibase 10.1063/1.2085170} {\bibfield  {journal}
  {\bibinfo  {journal} {J. Chem. Phys.}\ }\textbf {\bibinfo {volume} {123}},\
  \bibinfo {pages} {174101} (\bibinfo {year} {2005})}\BibitemShut {NoStop}%
\bibitem [{\citenamefont {Kim}\ \emph {et~al.}(2009)\citenamefont {Kim},
  \citenamefont {Hummer},\ and\ \citenamefont {Kresse}}]{Kim2009Jul}%
  \BibitemOpen
  \bibfield  {author} {\bibinfo {author} {\bibfnamefont {Yoon-Suk}\
  \bibnamefont {Kim}}, \bibinfo {author} {\bibfnamefont {Kerstin}\ \bibnamefont
  {Hummer}}, \ and\ \bibinfo {author} {\bibfnamefont {Georg}\ \bibnamefont
  {Kresse}},\ }\bibfield  {title} {\enquote {\bibinfo {title} {{Accurate band
  structures and effective masses for InP, InAs, and InSb using hybrid
  functionals}},}\ }\href {\doibase 10.1103/PhysRevB.80.035203} {\bibfield
  {journal} {\bibinfo  {journal} {Phys. Rev. B}\ }\textbf {\bibinfo {volume}
  {80}},\ \bibinfo {pages} {035203} (\bibinfo {year} {2009})}\BibitemShut
  {NoStop}%
\bibitem [{\citenamefont {Jain}\ \emph {et~al.}(2011)\citenamefont {Jain},
  \citenamefont {Chelikowsky},\ and\ \citenamefont {Louie}}]{Jain2011Nov}%
  \BibitemOpen
  \bibfield  {author} {\bibinfo {author} {\bibfnamefont {Manish}\ \bibnamefont
  {Jain}}, \bibinfo {author} {\bibfnamefont {James~R.}\ \bibnamefont
  {Chelikowsky}}, \ and\ \bibinfo {author} {\bibfnamefont {Steven~G.}\
  \bibnamefont {Louie}},\ }\bibfield  {title} {\enquote {\bibinfo {title}
  {{Reliability of Hybrid Functionals in Predicting Band Gaps}},}\ }\href
  {\doibase 10.1103/PhysRevLett.107.216806} {\bibfield  {journal} {\bibinfo
  {journal} {Phys. Rev. Lett.}\ }\textbf {\bibinfo {volume} {107}},\ \bibinfo
  {pages} {216806} (\bibinfo {year} {2011})}\BibitemShut {NoStop}%
\bibitem [{\citenamefont {Garza}\ and\ \citenamefont
  {Scuseria}(2016)}]{Garza2016Oct}%
  \BibitemOpen
  \bibfield  {author} {\bibinfo {author} {\bibfnamefont {Alejandro~J.}\
  \bibnamefont {Garza}}\ and\ \bibinfo {author} {\bibfnamefont {Gustavo~E.}\
  \bibnamefont {Scuseria}},\ }\bibfield  {title} {\enquote {\bibinfo {title}
  {{Predicting Band Gaps with Hybrid Density Functionals}},}\ }\href {\doibase
  10.1021/acs.jpclett.6b01807} {\bibfield  {journal} {\bibinfo  {journal} {J.
  Phys. Chem. Lett.}\ }\textbf {\bibinfo {volume} {7}},\ \bibinfo {pages}
  {4165--4170} (\bibinfo {year} {2016})}\BibitemShut {NoStop}%
\bibitem [{\citenamefont {Hybertsen}\ and\ \citenamefont
  {Louie}(1986)}]{Hybertsen1986Oct}%
  \BibitemOpen
  \bibfield  {author} {\bibinfo {author} {\bibfnamefont {Mark~S.}\ \bibnamefont
  {Hybertsen}}\ and\ \bibinfo {author} {\bibfnamefont {Steven~G.}\ \bibnamefont
  {Louie}},\ }\bibfield  {title} {\enquote {\bibinfo {title} {{Electron
  correlation in semiconductors and insulators: Band gaps and quasiparticle
  energies}},}\ }\href {\doibase 10.1103/PhysRevB.34.5390} {\bibfield
  {journal} {\bibinfo  {journal} {Phys. Rev. B}\ }\textbf {\bibinfo {volume}
  {34}},\ \bibinfo {pages} {5390--5413} (\bibinfo {year} {1986})}\BibitemShut
  {NoStop}%
\bibitem [{\citenamefont {Reining}(2018)}]{Reining2018May}%
  \BibitemOpen
  \bibfield  {author} {\bibinfo {author} {\bibfnamefont {Lucia}\ \bibnamefont
  {Reining}},\ }\bibfield  {title} {\enquote {\bibinfo {title} {{The GW
  approximation: content, successes and limitations}},}\ }\href {\doibase
  10.1002/wcms.1344} {\bibfield  {journal} {\bibinfo  {journal} {WIREs Comput.
  Mol. Sci.}\ }\textbf {\bibinfo {volume} {8}},\ \bibinfo {pages} {e1344}
  (\bibinfo {year} {2018})}\BibitemShut {NoStop}%
\bibitem [{\citenamefont {Tandetzky}\ \emph {et~al.}(2015)\citenamefont
  {Tandetzky}, \citenamefont {Dewhurst}, \citenamefont {Sharma},\ and\
  \citenamefont {Gross}}]{Tandetzky2015Sep}%
  \BibitemOpen
  \bibfield  {author} {\bibinfo {author} {\bibfnamefont {F.}~\bibnamefont
  {Tandetzky}}, \bibinfo {author} {\bibfnamefont {J.~K.}\ \bibnamefont
  {Dewhurst}}, \bibinfo {author} {\bibfnamefont {S.}~\bibnamefont {Sharma}}, \
  and\ \bibinfo {author} {\bibfnamefont {E.~K.~U.}\ \bibnamefont {Gross}},\
  }\bibfield  {title} {\enquote {\bibinfo {title} {{Multiplicity of solutions
  to $GW$-type approximations}},}\ }\href {\doibase 10.1103/PhysRevB.92.115125}
  {\bibfield  {journal} {\bibinfo  {journal} {Phys. Rev. B}\ }\textbf {\bibinfo
  {volume} {92}},\ \bibinfo {pages} {115125} (\bibinfo {year}
  {2015})}\BibitemShut {NoStop}%
\bibitem [{\citenamefont {Mardirossian}\ and\ \citenamefont
  {Head-Gordon}(2015)}]{Mardirossian2015Feb}%
  \BibitemOpen
  \bibfield  {author} {\bibinfo {author} {\bibfnamefont {Narbe}\ \bibnamefont
  {Mardirossian}}\ and\ \bibinfo {author} {\bibfnamefont {Martin}\ \bibnamefont
  {Head-Gordon}},\ }\bibfield  {title} {\enquote {\bibinfo {title} {{Mapping
  the genome of meta-generalized gradient approximation density functionals:
  The search for B97M-V}},}\ }\href {\doibase 10.1063/1.4907719} {\bibfield
  {journal} {\bibinfo  {journal} {J. Chem. Phys.}\ }\textbf {\bibinfo {volume}
  {142}},\ \bibinfo {pages} {074111} (\bibinfo {year} {2015})}\BibitemShut
  {NoStop}%
\bibitem [{\citenamefont {Mardirossian}\ \emph {et~al.}(2017)\citenamefont
  {Mardirossian}, \citenamefont {Ruiz~Pestana}, \citenamefont {Womack},
  \citenamefont {Skylaris}, \citenamefont {Head-Gordon},\ and\ \citenamefont
  {Head-Gordon}}]{Mardirossian2017Jan}%
  \BibitemOpen
  \bibfield  {author} {\bibinfo {author} {\bibfnamefont {Narbe}\ \bibnamefont
  {Mardirossian}}, \bibinfo {author} {\bibfnamefont {Luis}\ \bibnamefont
  {Ruiz~Pestana}}, \bibinfo {author} {\bibfnamefont {James~C.}\ \bibnamefont
  {Womack}}, \bibinfo {author} {\bibfnamefont {Chris-Kriton}\ \bibnamefont
  {Skylaris}}, \bibinfo {author} {\bibfnamefont {Teresa}\ \bibnamefont
  {Head-Gordon}}, \ and\ \bibinfo {author} {\bibfnamefont {Martin}\
  \bibnamefont {Head-Gordon}},\ }\bibfield  {title} {\enquote {\bibinfo {title}
  {{Use of the rVV10 Nonlocal Correlation Functional in the B97M-V Density
  Functional: Defining B97M-rV and Related Functionals}},}\ }\href {\doibase
  10.1021/acs.jpclett.6b02527} {\bibfield  {journal} {\bibinfo  {journal} {J.
  Phys. Chem. Lett.}\ }\textbf {\bibinfo {volume} {8}},\ \bibinfo {pages}
  {35--40} (\bibinfo {year} {2017})}\BibitemShut {NoStop}%
\bibitem [{\citenamefont {Mardirossian}\ and\ \citenamefont
  {Head-Gordon}(2014)}]{Mardirossian2014May}%
  \BibitemOpen
  \bibfield  {author} {\bibinfo {author} {\bibfnamefont {Narbe}\ \bibnamefont
  {Mardirossian}}\ and\ \bibinfo {author} {\bibfnamefont {Martin}\ \bibnamefont
  {Head-Gordon}},\ }\bibfield  {title} {\enquote {\bibinfo {title}
  {{{$\omega$}B97X-V: A 10-parameter, range-separated hybrid, generalized
  gradient approximation density functional with nonlocal correlation, designed
  by a survival-of-the-fittest strategy}},}\ }\href {\doibase
  10.1039/C3CP54374A} {\bibfield  {journal} {\bibinfo  {journal} {Phys. Chem.
  Chem. Phys.}\ }\textbf {\bibinfo {volume} {16}},\ \bibinfo {pages}
  {9904--9924} (\bibinfo {year} {2014})}\BibitemShut {NoStop}%
\bibitem [{\citenamefont {Mardirossian}\ and\ \citenamefont
  {Head-Gordon}(2016)}]{Mardirossian2016Jun}%
  \BibitemOpen
  \bibfield  {author} {\bibinfo {author} {\bibfnamefont {Narbe}\ \bibnamefont
  {Mardirossian}}\ and\ \bibinfo {author} {\bibfnamefont {Martin}\ \bibnamefont
  {Head-Gordon}},\ }\bibfield  {title} {\enquote {\bibinfo {title}
  {{{$\omega$}B97M-V: A combinatorially optimized, range-separated hybrid,
  meta-GGA density functional with VV10 nonlocal correlation}},}\ }\href
  {\doibase 10.1063/1.4952647} {\bibfield  {journal} {\bibinfo  {journal} {J.
  Chem. Phys.}\ }\textbf {\bibinfo {volume} {144}},\ \bibinfo {pages} {214110}
  (\bibinfo {year} {2016})}\BibitemShut {NoStop}%
\bibitem [{\citenamefont {Goerigk}\ \emph {et~al.}(2017)\citenamefont
  {Goerigk}, \citenamefont {Hansen}, \citenamefont {Bauer}, \citenamefont
  {Ehrlich}, \citenamefont {Najibi},\ and\ \citenamefont
  {Grimme}}]{Goerigk:2017}%
  \BibitemOpen
  \bibfield  {author} {\bibinfo {author} {\bibfnamefont {L.}~\bibnamefont
  {Goerigk}}, \bibinfo {author} {\bibfnamefont {A.}~\bibnamefont {Hansen}},
  \bibinfo {author} {\bibfnamefont {C.}~\bibnamefont {Bauer}}, \bibinfo
  {author} {\bibfnamefont {S.}~\bibnamefont {Ehrlich}}, \bibinfo {author}
  {\bibfnamefont {A.}~\bibnamefont {Najibi}}, \ and\ \bibinfo {author}
  {\bibfnamefont {S.}~\bibnamefont {Grimme}},\ }\bibfield  {title} {\enquote
  {\bibinfo {title} {A look at the density functional theory zoo with the
  advanced {GMTKN55} database for general main group thermochemistry, kinetics
  and noncovalent interactions},}\ }\href {\doibase 10.1039/C7CP04913G}
  {\bibfield  {journal} {\bibinfo  {journal} {Phys. Chem. Chem. Phys.}\
  }\textbf {\bibinfo {volume} {19}},\ \bibinfo {pages} {32184--32215} (\bibinfo
  {year} {2017})}\BibitemShut {NoStop}%
\bibitem [{\citenamefont {Najibi}\ and\ \citenamefont
  {Goerigk}(2018)}]{Najibi:2018b}%
  \BibitemOpen
  \bibfield  {author} {\bibinfo {author} {\bibfnamefont {A.}~\bibnamefont
  {Najibi}}\ and\ \bibinfo {author} {\bibfnamefont {L.}~\bibnamefont
  {Goerigk}},\ }\bibfield  {title} {\enquote {\bibinfo {title} {The nonlocal
  kernel in van der {Waals} density functionals as an additive correction: {An}
  extensive analysis with special emphasis on the {B97M}-{V} and
  $\omega${B97M}-{V} approaches},}\ }\href {\doibase 10.1021/acs.jctc.8b00842}
  {\bibfield  {journal} {\bibinfo  {journal} {J. Chem. Theory Comput.}\
  }\textbf {\bibinfo {volume} {14}},\ \bibinfo {pages} {5725--5738} (\bibinfo
  {year} {2018})}\BibitemShut {NoStop}%
\bibitem [{\citenamefont {Dohm}\ \emph {et~al.}(2018)\citenamefont {Dohm},
  \citenamefont {Hansen}, \citenamefont {Steinmetz}, \citenamefont {Grimme},\
  and\ \citenamefont {Checinski}}]{dohm2018comprehensive}%
  \BibitemOpen
  \bibfield  {author} {\bibinfo {author} {\bibfnamefont {Sebastian}\
  \bibnamefont {Dohm}}, \bibinfo {author} {\bibfnamefont {Andreas}\
  \bibnamefont {Hansen}}, \bibinfo {author} {\bibfnamefont {Marc}\ \bibnamefont
  {Steinmetz}}, \bibinfo {author} {\bibfnamefont {Stefan}\ \bibnamefont
  {Grimme}}, \ and\ \bibinfo {author} {\bibfnamefont {Marek~P}\ \bibnamefont
  {Checinski}},\ }\bibfield  {title} {\enquote {\bibinfo {title} {Comprehensive
  thermochemical benchmark set of realistic closed-shell metal organic
  reactions},}\ }\href@noop {} {\bibfield  {journal} {\bibinfo  {journal} {J.
  Chem. Theory Comput.}\ }\textbf {\bibinfo {volume} {14}},\ \bibinfo {pages}
  {2596--2608} (\bibinfo {year} {2018})}\BibitemShut {NoStop}%
\bibitem [{\citenamefont {Chan}\ \emph {et~al.}(2019)\citenamefont {Chan},
  \citenamefont {Gill},\ and\ \citenamefont {Kimura}}]{Chan:2019}%
  \BibitemOpen
  \bibfield  {author} {\bibinfo {author} {\bibfnamefont {B.}~\bibnamefont
  {Chan}}, \bibinfo {author} {\bibfnamefont {P.~M.~W.}\ \bibnamefont {Gill}}, \
  and\ \bibinfo {author} {\bibfnamefont {M.}~\bibnamefont {Kimura}},\
  }\bibfield  {title} {\enquote {\bibinfo {title} {Assessment of {DFT} methods
  for transition metals with the {TMC151} compilation of data sets and
  comparison with accuracies for main-group chemistry},}\ }\href {\doibase
  10.1021/acs.jctc.9b00239} {\bibfield  {journal} {\bibinfo  {journal} {J.
  Chem. Theory Comput.}\ }\textbf {\bibinfo {volume} {15}},\ \bibinfo {pages}
  {3610--3622} (\bibinfo {year} {2019})}\BibitemShut {NoStop}%
\bibitem [{\citenamefont {Lee}\ \emph {et~al.}(2021)\citenamefont {Lee},
  \citenamefont {Feng}, \citenamefont {Cunha}, \citenamefont {Gonthier},
  \citenamefont {Epifanovsky},\ and\ \citenamefont {Head-Gordon}}]{Lee2021Oct}%
  \BibitemOpen
  \bibfield  {author} {\bibinfo {author} {\bibfnamefont {Joonho}\ \bibnamefont
  {Lee}}, \bibinfo {author} {\bibfnamefont {Xintian}\ \bibnamefont {Feng}},
  \bibinfo {author} {\bibfnamefont {Leonardo~A.}\ \bibnamefont {Cunha}},
  \bibinfo {author} {\bibfnamefont
  {J{\ifmmode\acute{e}\else\'{e}\fi}r{\ifmmode\hat{o}\else\^{o}\fi}me~F.}\
  \bibnamefont {Gonthier}}, \bibinfo {author} {\bibfnamefont {Evgeny}\
  \bibnamefont {Epifanovsky}}, \ and\ \bibinfo {author} {\bibfnamefont
  {Martin}\ \bibnamefont {Head-Gordon}},\ }\bibfield  {title} {\enquote
  {\bibinfo {title} {{Approaching the basis set limit in Gaussian-orbital-based
  periodic calculations with transferability: Performance of pure density
  functionals for simple semiconductors}},}\ }\href {\doibase
  10.1063/5.0069177} {\bibfield  {journal} {\bibinfo  {journal} {J. Chem.
  Phys.}\ }\textbf {\bibinfo {volume} {155}},\ \bibinfo {pages} {164102}
  (\bibinfo {year} {2021})}\BibitemShut {NoStop}%
\bibitem [{\citenamefont {Sun}\ \emph {et~al.}(2015)\citenamefont {Sun},
  \citenamefont {Ruzsinszky},\ and\ \citenamefont {Perdew}}]{sun2015strongly}%
  \BibitemOpen
  \bibfield  {author} {\bibinfo {author} {\bibfnamefont {Jianwei}\ \bibnamefont
  {Sun}}, \bibinfo {author} {\bibfnamefont {Adrienn}\ \bibnamefont
  {Ruzsinszky}}, \ and\ \bibinfo {author} {\bibfnamefont {John~P}\ \bibnamefont
  {Perdew}},\ }\bibfield  {title} {\enquote {\bibinfo {title} {Strongly
  constrained and appropriately normed semilocal density functional},}\
  }\href@noop {} {\bibfield  {journal} {\bibinfo  {journal} {Phys. Rev. Lett.}\
  }\textbf {\bibinfo {volume} {115}},\ \bibinfo {pages} {036402} (\bibinfo
  {year} {2015})}\BibitemShut {NoStop}%
\bibitem [{\citenamefont {Zhao}\ and\ \citenamefont
  {Truhlar}(2006)}]{Zhao2006Nov}%
  \BibitemOpen
  \bibfield  {author} {\bibinfo {author} {\bibfnamefont {Yan}\ \bibnamefont
  {Zhao}}\ and\ \bibinfo {author} {\bibfnamefont {Donald~G.}\ \bibnamefont
  {Truhlar}},\ }\bibfield  {title} {\enquote {\bibinfo {title} {{A new local
  density functional for main-group thermochemistry, transition metal bonding,
  thermochemical kinetics, and noncovalent interactions}},}\ }\href {\doibase
  10.1063/1.2370993} {\bibfield  {journal} {\bibinfo  {journal} {J. Chem.
  Phys.}\ }\textbf {\bibinfo {volume} {125}},\ \bibinfo {pages} {194101}
  (\bibinfo {year} {2006})}\BibitemShut {NoStop}%
\bibitem [{\citenamefont {Yu}\ \emph {et~al.}(2016{\natexlab{a}})\citenamefont
  {Yu}, \citenamefont {He},\ and\ \citenamefont {Truhlar}}]{Yu2016Mar}%
  \BibitemOpen
  \bibfield  {author} {\bibinfo {author} {\bibfnamefont {Haoyu~S.}\
  \bibnamefont {Yu}}, \bibinfo {author} {\bibfnamefont {Xiao}\ \bibnamefont
  {He}}, \ and\ \bibinfo {author} {\bibfnamefont {Donald~G.}\ \bibnamefont
  {Truhlar}},\ }\bibfield  {title} {\enquote {\bibinfo {title} {{MN15-L: A New
  Local Exchange-Correlation Functional for Kohn{\textendash}Sham Density
  Functional Theory with Broad Accuracy for Atoms, Molecules, and Solids}},}\
  }\href {\doibase 10.1021/acs.jctc.5b01082} {\bibfield  {journal} {\bibinfo
  {journal} {J. Chem. Theory Comput.}\ }\textbf {\bibinfo {volume} {12}},\
  \bibinfo {pages} {1280--1293} (\bibinfo {year}
  {2016}{\natexlab{a}})}\BibitemShut {NoStop}%
\bibitem [{\citenamefont {Slater}(1951)}]{slater1951simplification}%
  \BibitemOpen
  \bibfield  {author} {\bibinfo {author} {\bibfnamefont {John~C}\ \bibnamefont
  {Slater}},\ }\bibfield  {title} {\enquote {\bibinfo {title} {A simplification
  of the hartree-fock method},}\ }\href@noop {} {\bibfield  {journal} {\bibinfo
   {journal} {Phys. Rev.}\ }\textbf {\bibinfo {volume} {81}},\ \bibinfo {pages}
  {385} (\bibinfo {year} {1951})}\BibitemShut {NoStop}%
\bibitem [{\citenamefont {Perdew}\ and\ \citenamefont
  {Zunger}(1981)}]{perdew1981self}%
  \BibitemOpen
  \bibfield  {author} {\bibinfo {author} {\bibfnamefont {John~P}\ \bibnamefont
  {Perdew}}\ and\ \bibinfo {author} {\bibfnamefont {Alex}\ \bibnamefont
  {Zunger}},\ }\bibfield  {title} {\enquote {\bibinfo {title} {Self-interaction
  correction to density-functional approximations for many-electron systems},}\
  }\href@noop {} {\bibfield  {journal} {\bibinfo  {journal} {Phys. Rev. B}\
  }\textbf {\bibinfo {volume} {23}},\ \bibinfo {pages} {5048} (\bibinfo {year}
  {1981})}\BibitemShut {NoStop}%
\bibitem [{\citenamefont {Perdew}\ \emph
  {et~al.}(1996{\natexlab{a}})\citenamefont {Perdew}, \citenamefont {Burke},\
  and\ \citenamefont {Ernzerhof}}]{perdew1996generalized}%
  \BibitemOpen
  \bibfield  {author} {\bibinfo {author} {\bibfnamefont {John~P}\ \bibnamefont
  {Perdew}}, \bibinfo {author} {\bibfnamefont {Kieron}\ \bibnamefont {Burke}},
  \ and\ \bibinfo {author} {\bibfnamefont {Matthias}\ \bibnamefont
  {Ernzerhof}},\ }\bibfield  {title} {\enquote {\bibinfo {title} {Generalized
  gradient approximation made simple},}\ }\href@noop {} {\bibfield  {journal}
  {\bibinfo  {journal} {Phys. Rev. Lett.}\ }\textbf {\bibinfo {volume} {77}},\
  \bibinfo {pages} {3865} (\bibinfo {year} {1996}{\natexlab{a}})}\BibitemShut
  {NoStop}%
\bibitem [{\citenamefont {Manzer}\ \emph {et~al.}(2015)\citenamefont {Manzer},
  \citenamefont {Horn}, \citenamefont {Mardirossian},\ and\ \citenamefont
  {Head-Gordon}}]{Manzer2015Jul}%
  \BibitemOpen
  \bibfield  {author} {\bibinfo {author} {\bibfnamefont {Samuel}\ \bibnamefont
  {Manzer}}, \bibinfo {author} {\bibfnamefont {Paul~R.}\ \bibnamefont {Horn}},
  \bibinfo {author} {\bibfnamefont {Narbe}\ \bibnamefont {Mardirossian}}, \
  and\ \bibinfo {author} {\bibfnamefont {Martin}\ \bibnamefont {Head-Gordon}},\
  }\bibfield  {title} {\enquote {\bibinfo {title} {{Fast, accurate evaluation
  of exact exchange: The occ-RI-K algorithm}},}\ }\href {\doibase
  10.1063/1.4923369} {\bibfield  {journal} {\bibinfo  {journal} {J. Chem.
  Phys.}\ }\textbf {\bibinfo {volume} {143}},\ \bibinfo {pages} {024113}
  (\bibinfo {year} {2015})}\BibitemShut {NoStop}%
\bibitem [{\citenamefont {Epifanovsky}\ \emph {et~al.}(2021)\citenamefont
  {Epifanovsky}, \citenamefont {Gilbert}, \citenamefont {Feng}, \citenamefont
  {Lee}, \citenamefont {Mao}, \citenamefont {Mardirossian}, \citenamefont
  {Pokhilko}, \citenamefont {White}, \citenamefont {Coons}, \citenamefont
  {Dempwolff}, \citenamefont {Gan}, \citenamefont {Hait}, \citenamefont {Horn},
  \citenamefont {Jacobson}, \citenamefont {Kaliman}, \citenamefont {Kussmann},
  \citenamefont {Lange}, \citenamefont {Lao}, \citenamefont {Levine},
  \citenamefont {Liu}, \citenamefont {McKenzie}, \citenamefont {Morrison},
  \citenamefont {Nanda}, \citenamefont {Plasser}, \citenamefont {Rehn},
  \citenamefont {Vidal}, \citenamefont {You}, \citenamefont {Zhu},
  \citenamefont {Alam}, \citenamefont {Albrecht}, \citenamefont {Aldossary},
  \citenamefont {Alguire}, \citenamefont {Andersen}, \citenamefont {Athavale},
  \citenamefont {Barton}, \citenamefont {Begam}, \citenamefont {Behn},
  \citenamefont {Bellonzi}, \citenamefont {Bernard}, \citenamefont {Berquist},
  \citenamefont {Burton}, \citenamefont {Carreras}, \citenamefont
  {Carter-Fenk}, \citenamefont {Chakraborty}, \citenamefont {Chien},
  \citenamefont {Closser}, \citenamefont {Cofer-Shabica}, \citenamefont
  {Dasgupta}, \citenamefont {de~Wergifosse}, \citenamefont {Deng},
  \citenamefont {Diedenhofen}, \citenamefont {Do}, \citenamefont {Ehlert},
  \citenamefont {Fang}, \citenamefont {Fatehi}, \citenamefont {Feng},
  \citenamefont {Friedhoff}, \citenamefont {Gayvert}, \citenamefont {Ge},
  \citenamefont {Gidofalvi}, \citenamefont {Goldey}, \citenamefont {Gomes},
  \citenamefont {Gonz{\ifmmode\acute{a}\else\'{a}\fi}lez-Espinoza},
  \citenamefont {Gulania}, \citenamefont {Gunina}, \citenamefont
  {Hanson-Heine}, \citenamefont {Harbach}, \citenamefont {Hauser},
  \citenamefont {Herbst}, \citenamefont
  {Hern{\ifmmode\acute{a}\else\'{a}\fi}ndez~Vera}, \citenamefont {Hodecker},
  \citenamefont {Holden}, \citenamefont {Houck}, \citenamefont {Huang},
  \citenamefont {Hui}, \citenamefont {Huynh}, \citenamefont {Ivanov},
  \citenamefont {J{\ifmmode\acute{a}\else\'{a}\fi}sz}, \citenamefont {Ji},
  \citenamefont {Jiang}, \citenamefont {Kaduk}, \citenamefont
  {K{\ifmmode\ddot{a}\else\"{a}\fi}hler}, \citenamefont {Khistyaev},
  \citenamefont {Kim}, \citenamefont {Kis}, \citenamefont {Klunzinger},
  \citenamefont {Koczor-Benda}, \citenamefont {Koh}, \citenamefont {Kosenkov},
  \citenamefont {Koulias}, \citenamefont {Kowalczyk}, \citenamefont {Krauter},
  \citenamefont {Kue}, \citenamefont {Kunitsa}, \citenamefont {Kus},
  \citenamefont {Ladj{\ifmmode\acute{a}\else\'{a}\fi}nszki}, \citenamefont
  {Landau}, \citenamefont {Lawler}, \citenamefont {Lefrancois}, \citenamefont
  {Lehtola}, \citenamefont {Li}, \citenamefont {Li}, \citenamefont {Liang},
  \citenamefont {Liebenthal}, \citenamefont {Lin}, \citenamefont {Lin},
  \citenamefont {Liu}, \citenamefont {Liu}, \citenamefont {Loipersberger},
  \citenamefont {Luenser}, \citenamefont {Manjanath}, \citenamefont {Manohar},
  \citenamefont {Mansoor}, \citenamefont {Manzer}, \citenamefont {Mao},
  \citenamefont {Marenich}, \citenamefont {Markovich}, \citenamefont {Mason},
  \citenamefont {Maurer}, \citenamefont {McLaughlin}, \citenamefont {Menger},
  \citenamefont {Mewes}, \citenamefont {Mewes}, \citenamefont {Morgante},
  \citenamefont {Mullinax}, \citenamefont {Oosterbaan}, \citenamefont {Paran},
  \citenamefont {Paul}, \citenamefont {Paul}, \citenamefont
  {Pavo{\ifmmode\check{s}\else\v{s}\fi}evi{\ifmmode\acute{c}\else\'{c}\fi}},
  \citenamefont {Pei}, \citenamefont {Prager}, \citenamefont {Proynov},
  \citenamefont {R{\ifmmode\acute{a}\else\'{a}\fi}k}, \citenamefont
  {Ramos-Cordoba}, \citenamefont {Rana}, \citenamefont {Rask}, \citenamefont
  {Rettig}, \citenamefont {Richard}, \citenamefont {Rob}, \citenamefont
  {Rossomme}, \citenamefont {Scheele}, \citenamefont {Scheurer}, \citenamefont
  {Schneider}, \citenamefont {Sergueev}, \citenamefont {Sharada}, \citenamefont
  {Skomorowski}, \citenamefont {Small}, \citenamefont {Stein}, \citenamefont
  {Su}, \citenamefont {Sundstrom}, \citenamefont {Tao}, \citenamefont
  {Thirman}, \citenamefont {Tornai}, \citenamefont {Tsuchimochi}, \citenamefont
  {Tubman}, \citenamefont {Veccham}, \citenamefont {Vydrov}, \citenamefont
  {Wenzel}, \citenamefont {Witte}, \citenamefont {Yamada}, \citenamefont {Yao},
  \citenamefont {Yeganeh}, \citenamefont {Yost}, \citenamefont {Zech},
  \citenamefont {Zhang}, \citenamefont {Zhang}, \citenamefont {Zhang},
  \citenamefont {Zuev}, \citenamefont {Aspuru-Guzik}, \citenamefont {Bell},
  \citenamefont {Besley}, \citenamefont {Bravaya}, \citenamefont {Brooks},
  \citenamefont {Casanova}, \citenamefont {Chai}, \citenamefont {Coriani},
  \citenamefont {Cramer}, \citenamefont {Cserey}, \citenamefont {DePrince},
  \citenamefont {DiStasio}, \citenamefont {Dreuw}, \citenamefont {Dunietz},
  \citenamefont {Furlani}, \citenamefont {Goddard}, \citenamefont
  {Hammes-Schiffer}, \citenamefont {Head-Gordon}, \citenamefont {Hehre},
  \citenamefont {Hsu}, \citenamefont {Jagau}, \citenamefont {Jung},
  \citenamefont {Klamt}, \citenamefont {Kong}, \citenamefont {Lambrecht},
  \citenamefont {Liang}, \citenamefont {Mayhall}, \citenamefont {McCurdy},
  \citenamefont {Neaton}, \citenamefont {Ochsenfeld}, \citenamefont {Parkhill},
  \citenamefont {Peverati}, \citenamefont {Rassolov}, \citenamefont {Shao},
  \citenamefont {Slipchenko}, \citenamefont {Stauch}, \citenamefont {Steele},
  \citenamefont {Subotnik}, \citenamefont {Thom}, \citenamefont {Tkatchenko},
  \citenamefont {Truhlar}, \citenamefont {Van~Voorhis}, \citenamefont
  {Wesolowski}, \citenamefont {Whaley}, \citenamefont {Woodcock}, \citenamefont
  {Zimmerman}, \citenamefont {Faraji}, \citenamefont {Gill}, \citenamefont
  {Head-Gordon}, \citenamefont {Herbert},\ and\ \citenamefont
  {Krylov}}]{Epifanovsky2021Aug}%
  \BibitemOpen
  \bibfield  {author} {\bibinfo {author} {\bibfnamefont {Evgeny}\ \bibnamefont
  {Epifanovsky}}, \bibinfo {author} {\bibfnamefont {Andrew T.~B.}\ \bibnamefont
  {Gilbert}}, \bibinfo {author} {\bibfnamefont {Xintian}\ \bibnamefont {Feng}},
  \bibinfo {author} {\bibfnamefont {Joonho}\ \bibnamefont {Lee}}, \bibinfo
  {author} {\bibfnamefont {Yuezhi}\ \bibnamefont {Mao}}, \bibinfo {author}
  {\bibfnamefont {Narbe}\ \bibnamefont {Mardirossian}}, \bibinfo {author}
  {\bibfnamefont {Pavel}\ \bibnamefont {Pokhilko}}, \bibinfo {author}
  {\bibfnamefont {Alec~F.}\ \bibnamefont {White}}, \bibinfo {author}
  {\bibfnamefont {Marc~P.}\ \bibnamefont {Coons}}, \bibinfo {author}
  {\bibfnamefont {Adrian~L.}\ \bibnamefont {Dempwolff}}, \bibinfo {author}
  {\bibfnamefont {Zhengting}\ \bibnamefont {Gan}}, \bibinfo {author}
  {\bibfnamefont {Diptarka}\ \bibnamefont {Hait}}, \bibinfo {author}
  {\bibfnamefont {Paul~R.}\ \bibnamefont {Horn}}, \bibinfo {author}
  {\bibfnamefont {Leif~D.}\ \bibnamefont {Jacobson}}, \bibinfo {author}
  {\bibfnamefont {Ilya}\ \bibnamefont {Kaliman}}, \bibinfo {author}
  {\bibfnamefont {J{\ifmmode\ddot{o}\else\"{o}\fi}rg}\ \bibnamefont
  {Kussmann}}, \bibinfo {author} {\bibfnamefont {Adrian~W.}\ \bibnamefont
  {Lange}}, \bibinfo {author} {\bibfnamefont {Ka~Un}\ \bibnamefont {Lao}},
  \bibinfo {author} {\bibfnamefont {Daniel~S.}\ \bibnamefont {Levine}},
  \bibinfo {author} {\bibfnamefont {Jie}\ \bibnamefont {Liu}}, \bibinfo
  {author} {\bibfnamefont {Simon~C.}\ \bibnamefont {McKenzie}}, \bibinfo
  {author} {\bibfnamefont {Adrian~F.}\ \bibnamefont {Morrison}}, \bibinfo
  {author} {\bibfnamefont {Kaushik~D.}\ \bibnamefont {Nanda}}, \bibinfo
  {author} {\bibfnamefont {Felix}\ \bibnamefont {Plasser}}, \bibinfo {author}
  {\bibfnamefont {Dirk~R.}\ \bibnamefont {Rehn}}, \bibinfo {author}
  {\bibfnamefont {Marta~L.}\ \bibnamefont {Vidal}}, \bibinfo {author}
  {\bibfnamefont {Zhi-Qiang}\ \bibnamefont {You}}, \bibinfo {author}
  {\bibfnamefont {Ying}\ \bibnamefont {Zhu}}, \bibinfo {author} {\bibfnamefont
  {Bushra}\ \bibnamefont {Alam}}, \bibinfo {author} {\bibfnamefont
  {Benjamin~J.}\ \bibnamefont {Albrecht}}, \bibinfo {author} {\bibfnamefont
  {Abdulrahman}\ \bibnamefont {Aldossary}}, \bibinfo {author} {\bibfnamefont
  {Ethan}\ \bibnamefont {Alguire}}, \bibinfo {author} {\bibfnamefont
  {Josefine~H.}\ \bibnamefont {Andersen}}, \bibinfo {author} {\bibfnamefont
  {Vishikh}\ \bibnamefont {Athavale}}, \bibinfo {author} {\bibfnamefont
  {Dennis}\ \bibnamefont {Barton}}, \bibinfo {author} {\bibfnamefont {Khadiza}\
  \bibnamefont {Begam}}, \bibinfo {author} {\bibfnamefont {Andrew}\
  \bibnamefont {Behn}}, \bibinfo {author} {\bibfnamefont {Nicole}\ \bibnamefont
  {Bellonzi}}, \bibinfo {author} {\bibfnamefont {Yves~A.}\ \bibnamefont
  {Bernard}}, \bibinfo {author} {\bibfnamefont {Eric~J.}\ \bibnamefont
  {Berquist}}, \bibinfo {author} {\bibfnamefont {Hugh G.~A.}\ \bibnamefont
  {Burton}}, \bibinfo {author} {\bibfnamefont {Abel}\ \bibnamefont {Carreras}},
  \bibinfo {author} {\bibfnamefont {Kevin}\ \bibnamefont {Carter-Fenk}},
  \bibinfo {author} {\bibfnamefont {Romit}\ \bibnamefont {Chakraborty}},
  \bibinfo {author} {\bibfnamefont {Alan~D.}\ \bibnamefont {Chien}}, \bibinfo
  {author} {\bibfnamefont {Kristina~D.}\ \bibnamefont {Closser}}, \bibinfo
  {author} {\bibfnamefont {Vale}\ \bibnamefont {Cofer-Shabica}}, \bibinfo
  {author} {\bibfnamefont {Saswata}\ \bibnamefont {Dasgupta}}, \bibinfo
  {author} {\bibfnamefont {Marc}\ \bibnamefont {de~Wergifosse}}, \bibinfo
  {author} {\bibfnamefont {Jia}\ \bibnamefont {Deng}}, \bibinfo {author}
  {\bibfnamefont {Michael}\ \bibnamefont {Diedenhofen}}, \bibinfo {author}
  {\bibfnamefont {Hainam}\ \bibnamefont {Do}}, \bibinfo {author} {\bibfnamefont
  {Sebastian}\ \bibnamefont {Ehlert}}, \bibinfo {author} {\bibfnamefont
  {Po-Tung}\ \bibnamefont {Fang}}, \bibinfo {author} {\bibfnamefont {Shervin}\
  \bibnamefont {Fatehi}}, \bibinfo {author} {\bibfnamefont {Qingguo}\
  \bibnamefont {Feng}}, \bibinfo {author} {\bibfnamefont {Triet}\ \bibnamefont
  {Friedhoff}}, \bibinfo {author} {\bibfnamefont {James}\ \bibnamefont
  {Gayvert}}, \bibinfo {author} {\bibfnamefont {Qinghui}\ \bibnamefont {Ge}},
  \bibinfo {author} {\bibfnamefont {Gergely}\ \bibnamefont {Gidofalvi}},
  \bibinfo {author} {\bibfnamefont {Matthew}\ \bibnamefont {Goldey}}, \bibinfo
  {author} {\bibfnamefont {Joe}\ \bibnamefont {Gomes}}, \bibinfo {author}
  {\bibfnamefont {Cristina~E.}\ \bibnamefont
  {Gonz{\ifmmode\acute{a}\else\'{a}\fi}lez-Espinoza}}, \bibinfo {author}
  {\bibfnamefont {Sahil}\ \bibnamefont {Gulania}}, \bibinfo {author}
  {\bibfnamefont {Anastasia~O.}\ \bibnamefont {Gunina}}, \bibinfo {author}
  {\bibfnamefont {Magnus W.~D.}\ \bibnamefont {Hanson-Heine}}, \bibinfo
  {author} {\bibfnamefont {Phillip H.~P.}\ \bibnamefont {Harbach}}, \bibinfo
  {author} {\bibfnamefont {Andreas}\ \bibnamefont {Hauser}}, \bibinfo {author}
  {\bibfnamefont {Michael~F.}\ \bibnamefont {Herbst}}, \bibinfo {author}
  {\bibfnamefont {Mario}\ \bibnamefont
  {Hern{\ifmmode\acute{a}\else\'{a}\fi}ndez~Vera}}, \bibinfo {author}
  {\bibfnamefont {Manuel}\ \bibnamefont {Hodecker}}, \bibinfo {author}
  {\bibfnamefont {Zachary~C.}\ \bibnamefont {Holden}}, \bibinfo {author}
  {\bibfnamefont {Shannon}\ \bibnamefont {Houck}}, \bibinfo {author}
  {\bibfnamefont {Xunkun}\ \bibnamefont {Huang}}, \bibinfo {author}
  {\bibfnamefont {Kerwin}\ \bibnamefont {Hui}}, \bibinfo {author}
  {\bibfnamefont {Bang~C.}\ \bibnamefont {Huynh}}, \bibinfo {author}
  {\bibfnamefont {Maxim}\ \bibnamefont {Ivanov}}, \bibinfo {author}
  {\bibfnamefont
  {{\ifmmode\acute{A}\else\'{A}\fi}d{\ifmmode\acute{a}\else\'{a}\fi}m}\
  \bibnamefont {J{\ifmmode\acute{a}\else\'{a}\fi}sz}}, \bibinfo {author}
  {\bibfnamefont {Hyunjun}\ \bibnamefont {Ji}}, \bibinfo {author}
  {\bibfnamefont {Hanjie}\ \bibnamefont {Jiang}}, \bibinfo {author}
  {\bibfnamefont {Benjamin}\ \bibnamefont {Kaduk}}, \bibinfo {author}
  {\bibfnamefont {Sven}\ \bibnamefont {K{\ifmmode\ddot{a}\else\"{a}\fi}hler}},
  \bibinfo {author} {\bibfnamefont {Kirill}\ \bibnamefont {Khistyaev}},
  \bibinfo {author} {\bibfnamefont {Jaehoon}\ \bibnamefont {Kim}}, \bibinfo
  {author} {\bibfnamefont {Gergely}\ \bibnamefont {Kis}}, \bibinfo {author}
  {\bibfnamefont {Phil}\ \bibnamefont {Klunzinger}}, \bibinfo {author}
  {\bibfnamefont {Zsuzsanna}\ \bibnamefont {Koczor-Benda}}, \bibinfo {author}
  {\bibfnamefont {Joong~Hoon}\ \bibnamefont {Koh}}, \bibinfo {author}
  {\bibfnamefont {Dimitri}\ \bibnamefont {Kosenkov}}, \bibinfo {author}
  {\bibfnamefont {Laura}\ \bibnamefont {Koulias}}, \bibinfo {author}
  {\bibfnamefont {Tim}\ \bibnamefont {Kowalczyk}}, \bibinfo {author}
  {\bibfnamefont {Caroline~M.}\ \bibnamefont {Krauter}}, \bibinfo {author}
  {\bibfnamefont {Karl}\ \bibnamefont {Kue}}, \bibinfo {author} {\bibfnamefont
  {Alexander}\ \bibnamefont {Kunitsa}}, \bibinfo {author} {\bibfnamefont
  {Thomas}\ \bibnamefont {Kus}}, \bibinfo {author} {\bibfnamefont
  {Istv{\ifmmode\acute{a}\else\'{a}\fi}n}\ \bibnamefont
  {Ladj{\ifmmode\acute{a}\else\'{a}\fi}nszki}}, \bibinfo {author}
  {\bibfnamefont {Arie}\ \bibnamefont {Landau}}, \bibinfo {author}
  {\bibfnamefont {Keith~V.}\ \bibnamefont {Lawler}}, \bibinfo {author}
  {\bibfnamefont {Daniel}\ \bibnamefont {Lefrancois}}, \bibinfo {author}
  {\bibfnamefont {Susi}\ \bibnamefont {Lehtola}}, \bibinfo {author}
  {\bibfnamefont {Run~R.}\ \bibnamefont {Li}}, \bibinfo {author} {\bibfnamefont
  {Yi-Pei}\ \bibnamefont {Li}}, \bibinfo {author} {\bibfnamefont {Jiashu}\
  \bibnamefont {Liang}}, \bibinfo {author} {\bibfnamefont {Marcus}\
  \bibnamefont {Liebenthal}}, \bibinfo {author} {\bibfnamefont {Hung-Hsuan}\
  \bibnamefont {Lin}}, \bibinfo {author} {\bibfnamefont {You-Sheng}\
  \bibnamefont {Lin}}, \bibinfo {author} {\bibfnamefont {Fenglai}\ \bibnamefont
  {Liu}}, \bibinfo {author} {\bibfnamefont {Kuan-Yu}\ \bibnamefont {Liu}},
  \bibinfo {author} {\bibfnamefont {Matthias}\ \bibnamefont {Loipersberger}},
  \bibinfo {author} {\bibfnamefont {Arne}\ \bibnamefont {Luenser}}, \bibinfo
  {author} {\bibfnamefont {Aaditya}\ \bibnamefont {Manjanath}}, \bibinfo
  {author} {\bibfnamefont {Prashant}\ \bibnamefont {Manohar}}, \bibinfo
  {author} {\bibfnamefont {Erum}\ \bibnamefont {Mansoor}}, \bibinfo {author}
  {\bibfnamefont {Sam~F.}\ \bibnamefont {Manzer}}, \bibinfo {author}
  {\bibfnamefont {Shan-Ping}\ \bibnamefont {Mao}}, \bibinfo {author}
  {\bibfnamefont {Aleksandr~V.}\ \bibnamefont {Marenich}}, \bibinfo {author}
  {\bibfnamefont {Thomas}\ \bibnamefont {Markovich}}, \bibinfo {author}
  {\bibfnamefont {Stephen}\ \bibnamefont {Mason}}, \bibinfo {author}
  {\bibfnamefont {Simon~A.}\ \bibnamefont {Maurer}}, \bibinfo {author}
  {\bibfnamefont {Peter~F.}\ \bibnamefont {McLaughlin}}, \bibinfo {author}
  {\bibfnamefont {Maximilian F. S.~J.}\ \bibnamefont {Menger}}, \bibinfo
  {author} {\bibfnamefont {Jan-Michael}\ \bibnamefont {Mewes}}, \bibinfo
  {author} {\bibfnamefont {Stefanie~A.}\ \bibnamefont {Mewes}}, \bibinfo
  {author} {\bibfnamefont {Pierpaolo}\ \bibnamefont {Morgante}}, \bibinfo
  {author} {\bibfnamefont {J.~Wayne}\ \bibnamefont {Mullinax}}, \bibinfo
  {author} {\bibfnamefont {Katherine~J.}\ \bibnamefont {Oosterbaan}}, \bibinfo
  {author} {\bibfnamefont {Garrette}\ \bibnamefont {Paran}}, \bibinfo {author}
  {\bibfnamefont {Alexander~C.}\ \bibnamefont {Paul}}, \bibinfo {author}
  {\bibfnamefont {Suranjan~K.}\ \bibnamefont {Paul}}, \bibinfo {author}
  {\bibfnamefont {Fabijan}\ \bibnamefont
  {Pavo{\ifmmode\check{s}\else\v{s}\fi}evi{\ifmmode\acute{c}\else\'{c}\fi}}},
  \bibinfo {author} {\bibfnamefont {Zheng}\ \bibnamefont {Pei}}, \bibinfo
  {author} {\bibfnamefont {Stefan}\ \bibnamefont {Prager}}, \bibinfo {author}
  {\bibfnamefont {Emil~I.}\ \bibnamefont {Proynov}}, \bibinfo {author}
  {\bibfnamefont
  {{\ifmmode\acute{A}\else\'{A}\fi}d{\ifmmode\acute{a}\else\'{a}\fi}m}\
  \bibnamefont {R{\ifmmode\acute{a}\else\'{a}\fi}k}}, \bibinfo {author}
  {\bibfnamefont {Eloy}\ \bibnamefont {Ramos-Cordoba}}, \bibinfo {author}
  {\bibfnamefont {Bhaskar}\ \bibnamefont {Rana}}, \bibinfo {author}
  {\bibfnamefont {Alan~E.}\ \bibnamefont {Rask}}, \bibinfo {author}
  {\bibfnamefont {Adam}\ \bibnamefont {Rettig}}, \bibinfo {author}
  {\bibfnamefont {Ryan~M.}\ \bibnamefont {Richard}}, \bibinfo {author}
  {\bibfnamefont {Fazle}\ \bibnamefont {Rob}}, \bibinfo {author} {\bibfnamefont
  {Elliot}\ \bibnamefont {Rossomme}}, \bibinfo {author} {\bibfnamefont {Tarek}\
  \bibnamefont {Scheele}}, \bibinfo {author} {\bibfnamefont {Maximilian}\
  \bibnamefont {Scheurer}}, \bibinfo {author} {\bibfnamefont {Matthias}\
  \bibnamefont {Schneider}}, \bibinfo {author} {\bibfnamefont {Nickolai}\
  \bibnamefont {Sergueev}}, \bibinfo {author} {\bibfnamefont {Shaama~M.}\
  \bibnamefont {Sharada}}, \bibinfo {author} {\bibfnamefont {Wojciech}\
  \bibnamefont {Skomorowski}}, \bibinfo {author} {\bibfnamefont {David~W.}\
  \bibnamefont {Small}}, \bibinfo {author} {\bibfnamefont {Christopher~J.}\
  \bibnamefont {Stein}}, \bibinfo {author} {\bibfnamefont {Yu-Chuan}\
  \bibnamefont {Su}}, \bibinfo {author} {\bibfnamefont {Eric~J.}\ \bibnamefont
  {Sundstrom}}, \bibinfo {author} {\bibfnamefont {Zhen}\ \bibnamefont {Tao}},
  \bibinfo {author} {\bibfnamefont {Jonathan}\ \bibnamefont {Thirman}},
  \bibinfo {author} {\bibfnamefont {G{\ifmmode\acute{a}\else\'{a}\fi}bor~J.}\
  \bibnamefont {Tornai}}, \bibinfo {author} {\bibfnamefont {Takashi}\
  \bibnamefont {Tsuchimochi}}, \bibinfo {author} {\bibfnamefont {Norm~M.}\
  \bibnamefont {Tubman}}, \bibinfo {author} {\bibfnamefont {Srimukh~Prasad}\
  \bibnamefont {Veccham}}, \bibinfo {author} {\bibfnamefont {Oleg}\
  \bibnamefont {Vydrov}}, \bibinfo {author} {\bibfnamefont {Jan}\ \bibnamefont
  {Wenzel}}, \bibinfo {author} {\bibfnamefont {Jon}\ \bibnamefont {Witte}},
  \bibinfo {author} {\bibfnamefont {Atsushi}\ \bibnamefont {Yamada}}, \bibinfo
  {author} {\bibfnamefont {Kun}\ \bibnamefont {Yao}}, \bibinfo {author}
  {\bibfnamefont {Sina}\ \bibnamefont {Yeganeh}}, \bibinfo {author}
  {\bibfnamefont {Shane~R.}\ \bibnamefont {Yost}}, \bibinfo {author}
  {\bibfnamefont {Alexander}\ \bibnamefont {Zech}}, \bibinfo {author}
  {\bibfnamefont {Igor~Ying}\ \bibnamefont {Zhang}}, \bibinfo {author}
  {\bibfnamefont {Xing}\ \bibnamefont {Zhang}}, \bibinfo {author}
  {\bibfnamefont {Yu}~\bibnamefont {Zhang}}, \bibinfo {author} {\bibfnamefont
  {Dmitry}\ \bibnamefont {Zuev}}, \bibinfo {author} {\bibfnamefont
  {Al{\ifmmode\acute{a}\else\'{a}\fi}n}\ \bibnamefont {Aspuru-Guzik}}, \bibinfo
  {author} {\bibfnamefont {Alexis~T.}\ \bibnamefont {Bell}}, \bibinfo {author}
  {\bibfnamefont {Nicholas~A.}\ \bibnamefont {Besley}}, \bibinfo {author}
  {\bibfnamefont {Ksenia~B.}\ \bibnamefont {Bravaya}}, \bibinfo {author}
  {\bibfnamefont {Bernard~R.}\ \bibnamefont {Brooks}}, \bibinfo {author}
  {\bibfnamefont {David}\ \bibnamefont {Casanova}}, \bibinfo {author}
  {\bibfnamefont {Jeng-Da}\ \bibnamefont {Chai}}, \bibinfo {author}
  {\bibfnamefont {Sonia}\ \bibnamefont {Coriani}}, \bibinfo {author}
  {\bibfnamefont {Christopher~J.}\ \bibnamefont {Cramer}}, \bibinfo {author}
  {\bibfnamefont {Gy{\ifmmode\ddot{o}\else\"{o}\fi}rgy}\ \bibnamefont
  {Cserey}}, \bibinfo {author} {\bibfnamefont {A.~Eugene}\ \bibnamefont
  {DePrince}}, \bibinfo {author} {\bibfnamefont {Robert~A.}\ \bibnamefont
  {DiStasio}}, \bibinfo {author} {\bibfnamefont {Andreas}\ \bibnamefont
  {Dreuw}}, \bibinfo {author} {\bibfnamefont {Barry~D.}\ \bibnamefont
  {Dunietz}}, \bibinfo {author} {\bibfnamefont {Thomas~R.}\ \bibnamefont
  {Furlani}}, \bibinfo {author} {\bibfnamefont {William~A.}\ \bibnamefont
  {Goddard}}, \bibinfo {author} {\bibfnamefont {Sharon}\ \bibnamefont
  {Hammes-Schiffer}}, \bibinfo {author} {\bibfnamefont {Teresa}\ \bibnamefont
  {Head-Gordon}}, \bibinfo {author} {\bibfnamefont {Warren~J.}\ \bibnamefont
  {Hehre}}, \bibinfo {author} {\bibfnamefont {Chao-Ping}\ \bibnamefont {Hsu}},
  \bibinfo {author} {\bibfnamefont {Thomas-C.}\ \bibnamefont {Jagau}}, \bibinfo
  {author} {\bibfnamefont {Yousung}\ \bibnamefont {Jung}}, \bibinfo {author}
  {\bibfnamefont {Andreas}\ \bibnamefont {Klamt}}, \bibinfo {author}
  {\bibfnamefont {Jing}\ \bibnamefont {Kong}}, \bibinfo {author} {\bibfnamefont
  {Daniel~S.}\ \bibnamefont {Lambrecht}}, \bibinfo {author} {\bibfnamefont
  {WanZhen}\ \bibnamefont {Liang}}, \bibinfo {author} {\bibfnamefont
  {Nicholas~J.}\ \bibnamefont {Mayhall}}, \bibinfo {author} {\bibfnamefont
  {C.~William}\ \bibnamefont {McCurdy}}, \bibinfo {author} {\bibfnamefont
  {Jeffrey~B.}\ \bibnamefont {Neaton}}, \bibinfo {author} {\bibfnamefont
  {Christian}\ \bibnamefont {Ochsenfeld}}, \bibinfo {author} {\bibfnamefont
  {John~A.}\ \bibnamefont {Parkhill}}, \bibinfo {author} {\bibfnamefont
  {Roberto}\ \bibnamefont {Peverati}}, \bibinfo {author} {\bibfnamefont
  {Vitaly~A.}\ \bibnamefont {Rassolov}}, \bibinfo {author} {\bibfnamefont
  {Yihan}\ \bibnamefont {Shao}}, \bibinfo {author} {\bibfnamefont
  {Lyudmila~V.}\ \bibnamefont {Slipchenko}}, \bibinfo {author} {\bibfnamefont
  {Tim}\ \bibnamefont {Stauch}}, \bibinfo {author} {\bibfnamefont {Ryan~P.}\
  \bibnamefont {Steele}}, \bibinfo {author} {\bibfnamefont {Joseph~E.}\
  \bibnamefont {Subotnik}}, \bibinfo {author} {\bibfnamefont {Alex J.~W.}\
  \bibnamefont {Thom}}, \bibinfo {author} {\bibfnamefont {Alexandre}\
  \bibnamefont {Tkatchenko}}, \bibinfo {author} {\bibfnamefont {Donald~G.}\
  \bibnamefont {Truhlar}}, \bibinfo {author} {\bibfnamefont {Troy}\
  \bibnamefont {Van~Voorhis}}, \bibinfo {author} {\bibfnamefont {Tomasz~A.}\
  \bibnamefont {Wesolowski}}, \bibinfo {author} {\bibfnamefont {K.~Birgitta}\
  \bibnamefont {Whaley}}, \bibinfo {author} {\bibfnamefont {H.~Lee}\
  \bibnamefont {Woodcock}}, \bibinfo {author} {\bibfnamefont {Paul~M.}\
  \bibnamefont {Zimmerman}}, \bibinfo {author} {\bibfnamefont {Shirin}\
  \bibnamefont {Faraji}}, \bibinfo {author} {\bibfnamefont {Peter M.~W.}\
  \bibnamefont {Gill}}, \bibinfo {author} {\bibfnamefont {Martin}\ \bibnamefont
  {Head-Gordon}}, \bibinfo {author} {\bibfnamefont {John~M.}\ \bibnamefont
  {Herbert}}, \ and\ \bibinfo {author} {\bibfnamefont {Anna~I.}\ \bibnamefont
  {Krylov}},\ }\bibfield  {title} {\enquote {\bibinfo {title} {{Software for
  the frontiers of quantum chemistry: An overview of developments in the Q-Chem
  5 package}},}\ }\href {\doibase 10.1063/5.0055522} {\bibfield  {journal}
  {\bibinfo  {journal} {J. Chem. Phys.}\ }\textbf {\bibinfo {volume} {155}},\
  \bibinfo {pages} {084801} (\bibinfo {year} {2021})}\BibitemShut {NoStop}%
\bibitem [{\citenamefont {Lin}(2016)}]{Lin2016May}%
  \BibitemOpen
  \bibfield  {author} {\bibinfo {author} {\bibfnamefont {Lin}\ \bibnamefont
  {Lin}},\ }\bibfield  {title} {\enquote {\bibinfo {title} {{Adaptively
  Compressed Exchange Operator}},}\ }\href {\doibase 10.1021/acs.jctc.6b00092}
  {\bibfield  {journal} {\bibinfo  {journal} {J. Chem. Theory Comput.}\
  }\textbf {\bibinfo {volume} {12}},\ \bibinfo {pages} {2242--2249} (\bibinfo
  {year} {2016})}\BibitemShut {NoStop}%
\bibitem [{\citenamefont {Lippert}\ \emph {et~al.}(1997)\citenamefont
  {Lippert}, \citenamefont {Hutter},\ and\ \citenamefont
  {Parrinello}}]{lippert1997hybrid}%
  \BibitemOpen
  \bibfield  {author} {\bibinfo {author} {\bibfnamefont {Gerald}\ \bibnamefont
  {Lippert}}, \bibinfo {author} {\bibfnamefont {J{\"u}rg}\ \bibnamefont
  {Hutter}}, \ and\ \bibinfo {author} {\bibfnamefont {Michele}\ \bibnamefont
  {Parrinello}},\ }\bibfield  {title} {\enquote {\bibinfo {title} {A hybrid
  gaussian and plane wave density functional scheme},}\ }\href@noop {}
  {\bibfield  {journal} {\bibinfo  {journal} {Mol. Phys.}\ }\textbf {\bibinfo
  {volume} {92}},\ \bibinfo {pages} {477--488} (\bibinfo {year}
  {1997})}\BibitemShut {NoStop}%
\bibitem [{\citenamefont {VandeVondele}\ \emph {et~al.}(2005)\citenamefont
  {VandeVondele}, \citenamefont {Krack}, \citenamefont {Mohamed}, \citenamefont
  {Parrinello}, \citenamefont {Chassaing},\ and\ \citenamefont
  {Hutter}}]{vandevondele2005quickstep}%
  \BibitemOpen
  \bibfield  {author} {\bibinfo {author} {\bibfnamefont {Joost}\ \bibnamefont
  {VandeVondele}}, \bibinfo {author} {\bibfnamefont {Matthias}\ \bibnamefont
  {Krack}}, \bibinfo {author} {\bibfnamefont {Fawzi}\ \bibnamefont {Mohamed}},
  \bibinfo {author} {\bibfnamefont {Michele}\ \bibnamefont {Parrinello}},
  \bibinfo {author} {\bibfnamefont {Thomas}\ \bibnamefont {Chassaing}}, \ and\
  \bibinfo {author} {\bibfnamefont {J{\"u}rg}\ \bibnamefont {Hutter}},\
  }\bibfield  {title} {\enquote {\bibinfo {title} {Quickstep: Fast and accurate
  density functional calculations using a mixed gaussian and plane waves
  approach},}\ }\href@noop {} {\bibfield  {journal} {\bibinfo  {journal}
  {Comput. Phys. Commun.}\ }\textbf {\bibinfo {volume} {167}},\ \bibinfo
  {pages} {103--128} (\bibinfo {year} {2005})}\BibitemShut {NoStop}%
\bibitem [{\citenamefont {Franchini}\ \emph {et~al.}(2014)\citenamefont
  {Franchini}, \citenamefont {Philipsen}, \citenamefont {van Lenthe},\ and\
  \citenamefont {Visscher}}]{Franchini2014May}%
  \BibitemOpen
  \bibfield  {author} {\bibinfo {author} {\bibfnamefont {Mirko}\ \bibnamefont
  {Franchini}}, \bibinfo {author} {\bibfnamefont {Pierre Herman~Theodoor}\
  \bibnamefont {Philipsen}}, \bibinfo {author} {\bibfnamefont {Erik}\
  \bibnamefont {van Lenthe}}, \ and\ \bibinfo {author} {\bibfnamefont {Lucas}\
  \bibnamefont {Visscher}},\ }\bibfield  {title} {\enquote {\bibinfo {title}
  {{Accurate Coulomb Potentials for Periodic and Molecular Systems through
  Density Fitting}},}\ }\href {\doibase 10.1021/ct500172n} {\bibfield
  {journal} {\bibinfo  {journal} {J. Chem. Theory Comput.}\ }\textbf {\bibinfo
  {volume} {10}},\ \bibinfo {pages} {1994--2004} (\bibinfo {year}
  {2014})}\BibitemShut {NoStop}%
\bibitem [{\citenamefont {Sun}\ \emph {et~al.}(2017)\citenamefont {Sun},
  \citenamefont {Berkelbach}, \citenamefont {McClain},\ and\ \citenamefont
  {Chan}}]{Sun2017Oct}%
  \BibitemOpen
  \bibfield  {author} {\bibinfo {author} {\bibfnamefont {Qiming}\ \bibnamefont
  {Sun}}, \bibinfo {author} {\bibfnamefont {Timothy~C.}\ \bibnamefont
  {Berkelbach}}, \bibinfo {author} {\bibfnamefont {James~D.}\ \bibnamefont
  {McClain}}, \ and\ \bibinfo {author} {\bibfnamefont {Garnet Kin-Lic}\
  \bibnamefont {Chan}},\ }\bibfield  {title} {\enquote {\bibinfo {title}
  {{Gaussian and plane-wave mixed density fitting for periodic systems}},}\
  }\href {\doibase 10.1063/1.4998644} {\bibfield  {journal} {\bibinfo
  {journal} {J. Chem. Phys.}\ }\textbf {\bibinfo {volume} {147}},\ \bibinfo
  {pages} {164119} (\bibinfo {year} {2017})}\BibitemShut {NoStop}%
\bibitem [{\citenamefont {Dovesi}\ \emph {et~al.}(2020)\citenamefont {Dovesi},
  \citenamefont {Pascale}, \citenamefont {Civalleri}, \citenamefont {Doll},
  \citenamefont {Harrison}, \citenamefont {Bush}, \citenamefont {D{'}Arco},
  \citenamefont {No{\ifmmode\ddot{e}\else\"{e}\fi}l}, \citenamefont
  {R{\ifmmode\acute{e}\else\'{e}\fi}rat}, \citenamefont
  {Carbonni{\ifmmode\grave{e}\else\`{e}\fi}re}, \citenamefont
  {Caus{\ifmmode\grave{a}\else\`{a}\fi}}, \citenamefont {Salustro},
  \citenamefont {Lacivita}, \citenamefont {Kirtman}, \citenamefont {Ferrari},
  \citenamefont {Gentile}, \citenamefont {Baima}, \citenamefont {Ferrero},
  \citenamefont {Demichelis},\ and\ \citenamefont
  {De~La~Pierre}}]{Dovesi2020May}%
  \BibitemOpen
  \bibfield  {author} {\bibinfo {author} {\bibfnamefont {Roberto}\ \bibnamefont
  {Dovesi}}, \bibinfo {author} {\bibfnamefont {Fabien}\ \bibnamefont
  {Pascale}}, \bibinfo {author} {\bibfnamefont {Bartolomeo}\ \bibnamefont
  {Civalleri}}, \bibinfo {author} {\bibfnamefont {Klaus}\ \bibnamefont {Doll}},
  \bibinfo {author} {\bibfnamefont {Nicholas~M.}\ \bibnamefont {Harrison}},
  \bibinfo {author} {\bibfnamefont {Ian}\ \bibnamefont {Bush}}, \bibinfo
  {author} {\bibfnamefont {Philippe}\ \bibnamefont {D{'}Arco}}, \bibinfo
  {author} {\bibfnamefont {Yves}\ \bibnamefont
  {No{\ifmmode\ddot{e}\else\"{e}\fi}l}}, \bibinfo {author} {\bibfnamefont
  {Michel}\ \bibnamefont {R{\ifmmode\acute{e}\else\'{e}\fi}rat}}, \bibinfo
  {author} {\bibfnamefont {Philippe}\ \bibnamefont
  {Carbonni{\ifmmode\grave{e}\else\`{e}\fi}re}}, \bibinfo {author}
  {\bibfnamefont {Mauro}\ \bibnamefont {Caus{\ifmmode\grave{a}\else\`{a}\fi}}},
  \bibinfo {author} {\bibfnamefont {Simone}\ \bibnamefont {Salustro}}, \bibinfo
  {author} {\bibfnamefont {Valentina}\ \bibnamefont {Lacivita}}, \bibinfo
  {author} {\bibfnamefont {Bernard}\ \bibnamefont {Kirtman}}, \bibinfo {author}
  {\bibfnamefont {Anna~Maria}\ \bibnamefont {Ferrari}}, \bibinfo {author}
  {\bibfnamefont {Francesco~Silvio}\ \bibnamefont {Gentile}}, \bibinfo {author}
  {\bibfnamefont {Jacopo}\ \bibnamefont {Baima}}, \bibinfo {author}
  {\bibfnamefont {Mauro}\ \bibnamefont {Ferrero}}, \bibinfo {author}
  {\bibfnamefont {Raffaella}\ \bibnamefont {Demichelis}}, \ and\ \bibinfo
  {author} {\bibfnamefont {Marco}\ \bibnamefont {De~La~Pierre}},\ }\bibfield
  {title} {\enquote {\bibinfo {title} {{The CRYSTAL code, 1976{\textendash}2020
  and beyond, a long story}},}\ }\href {\doibase 10.1063/5.0004892} {\bibfield
  {journal} {\bibinfo  {journal} {J. Chem. Phys.}\ }\textbf {\bibinfo {volume}
  {152}},\ \bibinfo {pages} {204111} (\bibinfo {year} {2020})}\BibitemShut
  {NoStop}%
\bibitem [{\citenamefont {K{\ifmmode\ddot{u}\else\"{u}\fi}hne}\ \emph
  {et~al.}(2020)\citenamefont {K{\ifmmode\ddot{u}\else\"{u}\fi}hne},
  \citenamefont {Iannuzzi}, \citenamefont {Del~Ben}, \citenamefont {Rybkin},
  \citenamefont {Seewald}, \citenamefont {Stein}, \citenamefont {Laino},
  \citenamefont {Khaliullin}, \citenamefont
  {Sch{\ifmmode\ddot{u}\else\"{u}\fi}tt}, \citenamefont {Schiffmann},
  \citenamefont {Golze}, \citenamefont {Wilhelm}, \citenamefont {Chulkov},
  \citenamefont {Bani-Hashemian}, \citenamefont {Weber}, \citenamefont
  {Bor{\ifmmode\check{s}\else\v{s}\fi}tnik}, \citenamefont {Taillefumier},
  \citenamefont {Jakobovits}, \citenamefont {Lazzaro}, \citenamefont {Pabst},
  \citenamefont {M{\ifmmode\ddot{u}\else\"{u}\fi}ller}, \citenamefont {Schade},
  \citenamefont {Guidon}, \citenamefont {Andermatt}, \citenamefont {Holmberg},
  \citenamefont {Schenter}, \citenamefont {Hehn}, \citenamefont {Bussy},
  \citenamefont {Belleflamme}, \citenamefont {Tabacchi}, \citenamefont
  {Gl{\ifmmode\ddot{o}\else\"{o}\fi}{\ss}}, \citenamefont {Lass}, \citenamefont
  {Bethune}, \citenamefont {Mundy}, \citenamefont {Plessl}, \citenamefont
  {Watkins}, \citenamefont {VandeVondele}, \citenamefont {Krack},\ and\
  \citenamefont {Hutter}}]{Kuhne2020May}%
  \BibitemOpen
  \bibfield  {author} {\bibinfo {author} {\bibfnamefont {Thomas~D.}\
  \bibnamefont {K{\ifmmode\ddot{u}\else\"{u}\fi}hne}}, \bibinfo {author}
  {\bibfnamefont {Marcella}\ \bibnamefont {Iannuzzi}}, \bibinfo {author}
  {\bibfnamefont {Mauro}\ \bibnamefont {Del~Ben}}, \bibinfo {author}
  {\bibfnamefont {Vladimir~V.}\ \bibnamefont {Rybkin}}, \bibinfo {author}
  {\bibfnamefont {Patrick}\ \bibnamefont {Seewald}}, \bibinfo {author}
  {\bibfnamefont {Frederick}\ \bibnamefont {Stein}}, \bibinfo {author}
  {\bibfnamefont {Teodoro}\ \bibnamefont {Laino}}, \bibinfo {author}
  {\bibfnamefont {Rustam~Z.}\ \bibnamefont {Khaliullin}}, \bibinfo {author}
  {\bibfnamefont {Ole}\ \bibnamefont {Sch{\ifmmode\ddot{u}\else\"{u}\fi}tt}},
  \bibinfo {author} {\bibfnamefont {Florian}\ \bibnamefont {Schiffmann}},
  \bibinfo {author} {\bibfnamefont {Dorothea}\ \bibnamefont {Golze}}, \bibinfo
  {author} {\bibfnamefont {Jan}\ \bibnamefont {Wilhelm}}, \bibinfo {author}
  {\bibfnamefont {Sergey}\ \bibnamefont {Chulkov}}, \bibinfo {author}
  {\bibfnamefont {Mohammad~Hossein}\ \bibnamefont {Bani-Hashemian}}, \bibinfo
  {author} {\bibfnamefont {Val{\ifmmode\acute{e}\else\'{e}\fi}ry}\ \bibnamefont
  {Weber}}, \bibinfo {author} {\bibfnamefont {Urban}\ \bibnamefont
  {Bor{\ifmmode\check{s}\else\v{s}\fi}tnik}}, \bibinfo {author} {\bibfnamefont
  {Mathieu}\ \bibnamefont {Taillefumier}}, \bibinfo {author} {\bibfnamefont
  {Alice~Shoshana}\ \bibnamefont {Jakobovits}}, \bibinfo {author}
  {\bibfnamefont {Alfio}\ \bibnamefont {Lazzaro}}, \bibinfo {author}
  {\bibfnamefont {Hans}\ \bibnamefont {Pabst}}, \bibinfo {author}
  {\bibfnamefont {Tiziano}\ \bibnamefont
  {M{\ifmmode\ddot{u}\else\"{u}\fi}ller}}, \bibinfo {author} {\bibfnamefont
  {Robert}\ \bibnamefont {Schade}}, \bibinfo {author} {\bibfnamefont {Manuel}\
  \bibnamefont {Guidon}}, \bibinfo {author} {\bibfnamefont {Samuel}\
  \bibnamefont {Andermatt}}, \bibinfo {author} {\bibfnamefont {Nico}\
  \bibnamefont {Holmberg}}, \bibinfo {author} {\bibfnamefont {Gregory~K.}\
  \bibnamefont {Schenter}}, \bibinfo {author} {\bibfnamefont {Anna}\
  \bibnamefont {Hehn}}, \bibinfo {author} {\bibfnamefont {Augustin}\
  \bibnamefont {Bussy}}, \bibinfo {author} {\bibfnamefont {Fabian}\
  \bibnamefont {Belleflamme}}, \bibinfo {author} {\bibfnamefont {Gloria}\
  \bibnamefont {Tabacchi}}, \bibinfo {author} {\bibfnamefont {Andreas}\
  \bibnamefont {Gl{\ifmmode\ddot{o}\else\"{o}\fi}{\ss}}}, \bibinfo {author}
  {\bibfnamefont {Michael}\ \bibnamefont {Lass}}, \bibinfo {author}
  {\bibfnamefont {Iain}\ \bibnamefont {Bethune}}, \bibinfo {author}
  {\bibfnamefont {Christopher~J.}\ \bibnamefont {Mundy}}, \bibinfo {author}
  {\bibfnamefont {Christian}\ \bibnamefont {Plessl}}, \bibinfo {author}
  {\bibfnamefont {Matt}\ \bibnamefont {Watkins}}, \bibinfo {author}
  {\bibfnamefont {Joost}\ \bibnamefont {VandeVondele}}, \bibinfo {author}
  {\bibfnamefont {Matthias}\ \bibnamefont {Krack}}, \ and\ \bibinfo {author}
  {\bibfnamefont {J{\ifmmode\ddot{u}\else\"{u}\fi}rg}\ \bibnamefont {Hutter}},\
  }\bibfield  {title} {\enquote {\bibinfo {title} {{CP2K: An electronic
  structure and molecular dynamics software package - Quickstep: Efficient and
  accurate electronic structure calculations}},}\ }\href {\doibase
  10.1063/5.0007045} {\bibfield  {journal} {\bibinfo  {journal} {J. Chem.
  Phys.}\ }\textbf {\bibinfo {volume} {152}},\ \bibinfo {pages} {194103}
  (\bibinfo {year} {2020})}\BibitemShut {NoStop}%
\bibitem [{\citenamefont {Balasubramani}\ \emph {et~al.}(2020)\citenamefont
  {Balasubramani}, \citenamefont {Chen}, \citenamefont {Coriani}, \citenamefont
  {Diedenhofen}, \citenamefont {Frank}, \citenamefont {Franzke}, \citenamefont
  {Furche}, \citenamefont {Grotjahn}, \citenamefont {Harding}, \citenamefont
  {H{\ifmmode\ddot{a}\else\"{a}\fi}ttig}, \citenamefont {Hellweg},
  \citenamefont {Helmich-Paris}, \citenamefont {Holzer}, \citenamefont
  {Huniar}, \citenamefont {Kaupp}, \citenamefont {Marefat~Khah}, \citenamefont
  {Karbalaei~Khani}, \citenamefont {M{\ifmmode\ddot{u}\else\"{u}\fi}ller},
  \citenamefont {Mack}, \citenamefont {Nguyen}, \citenamefont {Parker},
  \citenamefont {Perlt}, \citenamefont {Rappoport}, \citenamefont {Reiter},
  \citenamefont {Roy}, \citenamefont {R{\ifmmode\ddot{u}\else\"{u}\fi}ckert},
  \citenamefont {Schmitz}, \citenamefont {Sierka}, \citenamefont {Tapavicza},
  \citenamefont {Tew}, \citenamefont {van
  W{\ifmmode\ddot{u}\else\"{u}\fi}llen}, \citenamefont {Voora}, \citenamefont
  {Weigend}, \citenamefont {Wody{\ifmmode\acute{n}\else\'{n}\fi}ski},\ and\
  \citenamefont {Yu}}]{Balasubramani2020May}%
  \BibitemOpen
  \bibfield  {author} {\bibinfo {author} {\bibfnamefont {Sree~Ganesh}\
  \bibnamefont {Balasubramani}}, \bibinfo {author} {\bibfnamefont {Guo~P.}\
  \bibnamefont {Chen}}, \bibinfo {author} {\bibfnamefont {Sonia}\ \bibnamefont
  {Coriani}}, \bibinfo {author} {\bibfnamefont {Michael}\ \bibnamefont
  {Diedenhofen}}, \bibinfo {author} {\bibfnamefont {Marius~S.}\ \bibnamefont
  {Frank}}, \bibinfo {author} {\bibfnamefont {Yannick~J.}\ \bibnamefont
  {Franzke}}, \bibinfo {author} {\bibfnamefont {Filipp}\ \bibnamefont
  {Furche}}, \bibinfo {author} {\bibfnamefont {Robin}\ \bibnamefont
  {Grotjahn}}, \bibinfo {author} {\bibfnamefont {Michael~E.}\ \bibnamefont
  {Harding}}, \bibinfo {author} {\bibfnamefont {Christof}\ \bibnamefont
  {H{\ifmmode\ddot{a}\else\"{a}\fi}ttig}}, \bibinfo {author} {\bibfnamefont
  {Arnim}\ \bibnamefont {Hellweg}}, \bibinfo {author} {\bibfnamefont
  {Benjamin}\ \bibnamefont {Helmich-Paris}}, \bibinfo {author} {\bibfnamefont
  {Christof}\ \bibnamefont {Holzer}}, \bibinfo {author} {\bibfnamefont {Uwe}\
  \bibnamefont {Huniar}}, \bibinfo {author} {\bibfnamefont {Martin}\
  \bibnamefont {Kaupp}}, \bibinfo {author} {\bibfnamefont {Alireza}\
  \bibnamefont {Marefat~Khah}}, \bibinfo {author} {\bibfnamefont {Sarah}\
  \bibnamefont {Karbalaei~Khani}}, \bibinfo {author} {\bibfnamefont {Thomas}\
  \bibnamefont {M{\ifmmode\ddot{u}\else\"{u}\fi}ller}}, \bibinfo {author}
  {\bibfnamefont {Fabian}\ \bibnamefont {Mack}}, \bibinfo {author}
  {\bibfnamefont {Brian~D.}\ \bibnamefont {Nguyen}}, \bibinfo {author}
  {\bibfnamefont {Shane~M.}\ \bibnamefont {Parker}}, \bibinfo {author}
  {\bibfnamefont {Eva}\ \bibnamefont {Perlt}}, \bibinfo {author} {\bibfnamefont
  {Dmitrij}\ \bibnamefont {Rappoport}}, \bibinfo {author} {\bibfnamefont
  {Kevin}\ \bibnamefont {Reiter}}, \bibinfo {author} {\bibfnamefont {Saswata}\
  \bibnamefont {Roy}}, \bibinfo {author} {\bibfnamefont {Matthias}\
  \bibnamefont {R{\ifmmode\ddot{u}\else\"{u}\fi}ckert}}, \bibinfo {author}
  {\bibfnamefont {Gunnar}\ \bibnamefont {Schmitz}}, \bibinfo {author}
  {\bibfnamefont {Marek}\ \bibnamefont {Sierka}}, \bibinfo {author}
  {\bibfnamefont {Enrico}\ \bibnamefont {Tapavicza}}, \bibinfo {author}
  {\bibfnamefont {David~P.}\ \bibnamefont {Tew}}, \bibinfo {author}
  {\bibfnamefont {Christoph}\ \bibnamefont {van
  W{\ifmmode\ddot{u}\else\"{u}\fi}llen}}, \bibinfo {author} {\bibfnamefont
  {Vamsee~K.}\ \bibnamefont {Voora}}, \bibinfo {author} {\bibfnamefont
  {Florian}\ \bibnamefont {Weigend}}, \bibinfo {author} {\bibfnamefont {Artur}\
  \bibnamefont {Wody{\ifmmode\acute{n}\else\'{n}\fi}ski}}, \ and\ \bibinfo
  {author} {\bibfnamefont {Jason~M.}\ \bibnamefont {Yu}},\ }\bibfield  {title}
  {\enquote {\bibinfo {title} {{TURBOMOLE: Modular program suite for ab initio
  quantum-chemical and condensed-matter simulations}},}\ }\href {\doibase
  10.1063/5.0004635} {\bibfield  {journal} {\bibinfo  {journal} {J. Chem.
  Phys.}\ }\textbf {\bibinfo {volume} {152}},\ \bibinfo {pages} {184107}
  (\bibinfo {year} {2020})}\BibitemShut {NoStop}%
\bibitem [{\citenamefont {Blum}\ \emph {et~al.}(2022)\citenamefont {Blum},
  \citenamefont {Rossi}, \citenamefont {Kokott},\ and\ \citenamefont
  {Scheffler}}]{Blum2022Aug}%
  \BibitemOpen
  \bibfield  {author} {\bibinfo {author} {\bibfnamefont {Volker}\ \bibnamefont
  {Blum}}, \bibinfo {author} {\bibfnamefont {Mariana}\ \bibnamefont {Rossi}},
  \bibinfo {author} {\bibfnamefont {Sebastian}\ \bibnamefont {Kokott}}, \ and\
  \bibinfo {author} {\bibfnamefont {Matthias}\ \bibnamefont {Scheffler}},\
  }\bibfield  {title} {\enquote {\bibinfo {title} {{The FHI-aims Code:
  All-electron, ab initio materials simulations towards the exascale}},}\
  }\href {\doibase 10.48550/arXiv.2208.12335} {\bibfield  {journal} {\bibinfo
  {journal} {ArXiv}\ } (\bibinfo {year} {2022}),\ 10.48550/arXiv.2208.12335},\
  \Eprint {http://arxiv.org/abs/2208.12335} {2208.12335} \BibitemShut {NoStop}%
\bibitem [{\citenamefont {Sun}\ \emph {et~al.}(2020)\citenamefont {Sun},
  \citenamefont {Zhang}, \citenamefont {Banerjee}, \citenamefont {Bao},
  \citenamefont {Barbry}, \citenamefont {Blunt}, \citenamefont {Bogdanov},
  \citenamefont {Booth}, \citenamefont {Chen}, \citenamefont {Cui} \emph
  {et~al.}}]{sun2020recent}%
  \BibitemOpen
  \bibfield  {author} {\bibinfo {author} {\bibfnamefont {Qiming}\ \bibnamefont
  {Sun}}, \bibinfo {author} {\bibfnamefont {Xing}\ \bibnamefont {Zhang}},
  \bibinfo {author} {\bibfnamefont {Samragni}\ \bibnamefont {Banerjee}},
  \bibinfo {author} {\bibfnamefont {Peng}\ \bibnamefont {Bao}}, \bibinfo
  {author} {\bibfnamefont {Marc}\ \bibnamefont {Barbry}}, \bibinfo {author}
  {\bibfnamefont {Nick~S}\ \bibnamefont {Blunt}}, \bibinfo {author}
  {\bibfnamefont {Nikolay~A}\ \bibnamefont {Bogdanov}}, \bibinfo {author}
  {\bibfnamefont {George~H}\ \bibnamefont {Booth}}, \bibinfo {author}
  {\bibfnamefont {Jia}\ \bibnamefont {Chen}}, \bibinfo {author} {\bibfnamefont
  {Zhi-Hao}\ \bibnamefont {Cui}},  \emph {et~al.},\ }\bibfield  {title}
  {\enquote {\bibinfo {title} {Recent developments in the pyscf program
  package},}\ }\href@noop {} {\bibfield  {journal} {\bibinfo  {journal} {J.
  Chem. Phys.}\ }\textbf {\bibinfo {volume} {153}},\ \bibinfo {pages} {024109}
  (\bibinfo {year} {2020})}\BibitemShut {NoStop}%
\bibitem [{\citenamefont {Giannozzi}\ \emph {et~al.}(2020)\citenamefont
  {Giannozzi}, \citenamefont {Baseggio}, \citenamefont
  {Bonf{\ifmmode\grave{a}\else\`{a}\fi}}, \citenamefont {Brunato},
  \citenamefont {Car}, \citenamefont {Carnimeo}, \citenamefont {Cavazzoni},
  \citenamefont {de~Gironcoli}, \citenamefont {Delugas}, \citenamefont
  {Ferrari~Ruffino}, \citenamefont {Ferretti}, \citenamefont {Marzari},
  \citenamefont {Timrov}, \citenamefont {Urru},\ and\ \citenamefont
  {Baroni}}]{Giannozzi2020Apr}%
  \BibitemOpen
  \bibfield  {author} {\bibinfo {author} {\bibfnamefont {Paolo}\ \bibnamefont
  {Giannozzi}}, \bibinfo {author} {\bibfnamefont {Oscar}\ \bibnamefont
  {Baseggio}}, \bibinfo {author} {\bibfnamefont {Pietro}\ \bibnamefont
  {Bonf{\ifmmode\grave{a}\else\`{a}\fi}}}, \bibinfo {author} {\bibfnamefont
  {Davide}\ \bibnamefont {Brunato}}, \bibinfo {author} {\bibfnamefont
  {Roberto}\ \bibnamefont {Car}}, \bibinfo {author} {\bibfnamefont {Ivan}\
  \bibnamefont {Carnimeo}}, \bibinfo {author} {\bibfnamefont {Carlo}\
  \bibnamefont {Cavazzoni}}, \bibinfo {author} {\bibfnamefont {Stefano}\
  \bibnamefont {de~Gironcoli}}, \bibinfo {author} {\bibfnamefont {Pietro}\
  \bibnamefont {Delugas}}, \bibinfo {author} {\bibfnamefont {Fabrizio}\
  \bibnamefont {Ferrari~Ruffino}}, \bibinfo {author} {\bibfnamefont {Andrea}\
  \bibnamefont {Ferretti}}, \bibinfo {author} {\bibfnamefont {Nicola}\
  \bibnamefont {Marzari}}, \bibinfo {author} {\bibfnamefont {Iurii}\
  \bibnamefont {Timrov}}, \bibinfo {author} {\bibfnamefont {Andrea}\
  \bibnamefont {Urru}}, \ and\ \bibinfo {author} {\bibfnamefont {Stefano}\
  \bibnamefont {Baroni}},\ }\bibfield  {title} {\enquote {\bibinfo {title}
  {{Quantum ESPRESSO toward the exascale}},}\ }\href {\doibase
  10.1063/5.0005082} {\bibfield  {journal} {\bibinfo  {journal} {J. Chem.
  Phys.}\ }\textbf {\bibinfo {volume} {152}},\ \bibinfo {pages} {154105}
  (\bibinfo {year} {2020})}\BibitemShut {NoStop}%
\bibitem [{\citenamefont {Hafner}(2008)}]{Hafner2008Oct}%
  \BibitemOpen
  \bibfield  {author} {\bibinfo {author} {\bibfnamefont
  {J{\ifmmode\ddot{u}\else\"{u}\fi}rgen}\ \bibnamefont {Hafner}},\ }\bibfield
  {title} {\enquote {\bibinfo {title} {{Ab-initio simulations of materials
  using VASP: Density-functional theory and beyond}},}\ }\href {\doibase
  10.1002/jcc.21057} {\bibfield  {journal} {\bibinfo  {journal} {J. Comput.
  Chem.}\ }\textbf {\bibinfo {volume} {29}},\ \bibinfo {pages} {2044--2078}
  (\bibinfo {year} {2008})}\BibitemShut {NoStop}%
\bibitem [{Bib(2022)}]{BibEntry2022Sep}%
  \BibitemOpen
  \href {https://www.flapw.de/MaX-6.0} {\enquote {\bibinfo {title} {{Welcome -
  FLEUR}},}\ } (\bibinfo {year} {2022}),\ \bibinfo {note} {[Online; accessed
  28. Sep. 2022]}\BibitemShut {NoStop}%
\bibitem [{\citenamefont {Blaha}\ \emph {et~al.}(2020)\citenamefont {Blaha},
  \citenamefont {Schwarz}, \citenamefont {Tran}, \citenamefont {Laskowski},
  \citenamefont {Madsen},\ and\ \citenamefont {Marks}}]{Blaha2020Feb}%
  \BibitemOpen
  \bibfield  {author} {\bibinfo {author} {\bibfnamefont {Peter}\ \bibnamefont
  {Blaha}}, \bibinfo {author} {\bibfnamefont {Karlheinz}\ \bibnamefont
  {Schwarz}}, \bibinfo {author} {\bibfnamefont {Fabien}\ \bibnamefont {Tran}},
  \bibinfo {author} {\bibfnamefont {Robert}\ \bibnamefont {Laskowski}},
  \bibinfo {author} {\bibfnamefont {Georg K.~H.}\ \bibnamefont {Madsen}}, \
  and\ \bibinfo {author} {\bibfnamefont {Laurence~D.}\ \bibnamefont {Marks}},\
  }\bibfield  {title} {\enquote {\bibinfo {title} {{WIEN2k: An APW+lo program
  for calculating the properties of solids}},}\ }\href {\doibase
  10.1063/1.5143061} {\bibfield  {journal} {\bibinfo  {journal} {J. Chem.
  Phys.}\ }\textbf {\bibinfo {volume} {152}},\ \bibinfo {pages} {074101}
  (\bibinfo {year} {2020})}\BibitemShut {NoStop}%
\bibitem [{\citenamefont {Fraser}\ \emph {et~al.}(1996)\citenamefont {Fraser},
  \citenamefont {Foulkes}, \citenamefont {Rajagopal}, \citenamefont {Needs},
  \citenamefont {Kenny},\ and\ \citenamefont {Williamson}}]{Fraser1996Jan}%
  \BibitemOpen
  \bibfield  {author} {\bibinfo {author} {\bibfnamefont {Louisa~M.}\
  \bibnamefont {Fraser}}, \bibinfo {author} {\bibfnamefont {W.~M.~C.}\
  \bibnamefont {Foulkes}}, \bibinfo {author} {\bibfnamefont {G.}~\bibnamefont
  {Rajagopal}}, \bibinfo {author} {\bibfnamefont {R.~J.}\ \bibnamefont
  {Needs}}, \bibinfo {author} {\bibfnamefont {S.~D.}\ \bibnamefont {Kenny}}, \
  and\ \bibinfo {author} {\bibfnamefont {A.~J.}\ \bibnamefont {Williamson}},\
  }\bibfield  {title} {\enquote {\bibinfo {title} {{Finite-size effects and
  Coulomb interactions in quantum Monte Carlo calculations for homogeneous
  systems with periodic boundary conditions}},}\ }\href {\doibase
  10.1103/PhysRevB.53.1814} {\bibfield  {journal} {\bibinfo  {journal} {Phys.
  Rev. B}\ }\textbf {\bibinfo {volume} {53}},\ \bibinfo {pages} {1814--1832}
  (\bibinfo {year} {1996})}\BibitemShut {NoStop}%
\bibitem [{\citenamefont {Spencer}\ and\ \citenamefont
  {Alavi}(2008)}]{Spencer2008May}%
  \BibitemOpen
  \bibfield  {author} {\bibinfo {author} {\bibfnamefont {James}\ \bibnamefont
  {Spencer}}\ and\ \bibinfo {author} {\bibfnamefont {Ali}\ \bibnamefont
  {Alavi}},\ }\bibfield  {title} {\enquote {\bibinfo {title} {{Efficient
  calculation of the exact exchange energy in periodic systems using a
  truncated Coulomb potential}},}\ }\href {\doibase 10.1103/PhysRevB.77.193110}
  {\bibfield  {journal} {\bibinfo  {journal} {Phys. Rev. B}\ }\textbf {\bibinfo
  {volume} {77}},\ \bibinfo {pages} {193110} (\bibinfo {year}
  {2008})}\BibitemShut {NoStop}%
\bibitem [{\citenamefont {Kawashima}\ and\ \citenamefont
  {Hirao}(2017)}]{Kawashima2017Mar}%
  \BibitemOpen
  \bibfield  {author} {\bibinfo {author} {\bibfnamefont {Yukio}\ \bibnamefont
  {Kawashima}}\ and\ \bibinfo {author} {\bibfnamefont {Kimihiko}\ \bibnamefont
  {Hirao}},\ }\bibfield  {title} {\enquote {\bibinfo {title} {{Singularity
  Correction for Long-Range-Corrected Density Functional Theory with Plane-Wave
  Basis Sets}},}\ }\href {\doibase 10.1021/acs.jpca.7b00162} {\bibfield
  {journal} {\bibinfo  {journal} {J. Phys. Chem. A}\ }\textbf {\bibinfo
  {volume} {121}},\ \bibinfo {pages} {2035--2045} (\bibinfo {year}
  {2017})}\BibitemShut {NoStop}%
\bibitem [{\citenamefont {Neese}\ and\ \citenamefont
  {Olbrich}(2002)}]{Neese2002Aug}%
  \BibitemOpen
  \bibfield  {author} {\bibinfo {author} {\bibfnamefont {Frank}\ \bibnamefont
  {Neese}}\ and\ \bibinfo {author} {\bibfnamefont {Gottfried}\ \bibnamefont
  {Olbrich}},\ }\bibfield  {title} {\enquote {\bibinfo {title} {{Efficient use
  of the resolution of the identity approximation in time-dependent density
  functional calculations with hybrid density functionals}},}\ }\href {\doibase
  10.1016/S0009-2614(02)01053-9} {\bibfield  {journal} {\bibinfo  {journal}
  {Chem. Phys. Lett.}\ }\textbf {\bibinfo {volume} {362}},\ \bibinfo {pages}
  {170--178} (\bibinfo {year} {2002})}\BibitemShut {NoStop}%
\bibitem [{\citenamefont {Weigend}(2002)}]{Weigend2002Sep}%
  \BibitemOpen
  \bibfield  {author} {\bibinfo {author} {\bibfnamefont {Florian}\ \bibnamefont
  {Weigend}},\ }\bibfield  {title} {\enquote {\bibinfo {title} {{A fully direct
  RI-HF algorithm: Implementation, optimised auxiliary basis sets,
  demonstration of accuracy and efficiency}},}\ }\href {\doibase
  10.1039/B204199P} {\bibfield  {journal} {\bibinfo  {journal} {Phys. Chem.
  Chem. Phys.}\ }\textbf {\bibinfo {volume} {4}},\ \bibinfo {pages}
  {4285--4291} (\bibinfo {year} {2002})}\BibitemShut {NoStop}%
\bibitem [{\citenamefont {Ye}\ and\ \citenamefont
  {Berkelbach}(2022)}]{Ye2022Mar}%
  \BibitemOpen
  \bibfield  {author} {\bibinfo {author} {\bibfnamefont {Hong-Zhou}\
  \bibnamefont {Ye}}\ and\ \bibinfo {author} {\bibfnamefont {Timothy~C.}\
  \bibnamefont {Berkelbach}},\ }\bibfield  {title} {\enquote {\bibinfo {title}
  {{Correlation-Consistent Gaussian Basis Sets for Solids Made Simple}},}\
  }\href {\doibase 10.1021/acs.jctc.1c01245} {\bibfield  {journal} {\bibinfo
  {journal} {J. Chem. Theory Comput.}\ }\textbf {\bibinfo {volume} {18}},\
  \bibinfo {pages} {1595--1606} (\bibinfo {year} {2022})}\BibitemShut {NoStop}%
\bibitem [{\citenamefont {Mostofi}\ \emph {et~al.}(2008)\citenamefont
  {Mostofi}, \citenamefont {Yates}, \citenamefont {Lee}, \citenamefont {Souza},
  \citenamefont {Vanderbilt},\ and\ \citenamefont {Marzari}}]{Mostofi2008May}%
  \BibitemOpen
  \bibfield  {author} {\bibinfo {author} {\bibfnamefont {Arash~A.}\
  \bibnamefont {Mostofi}}, \bibinfo {author} {\bibfnamefont {Jonathan~R.}\
  \bibnamefont {Yates}}, \bibinfo {author} {\bibfnamefont {Young-Su}\
  \bibnamefont {Lee}}, \bibinfo {author} {\bibfnamefont {Ivo}\ \bibnamefont
  {Souza}}, \bibinfo {author} {\bibfnamefont {David}\ \bibnamefont
  {Vanderbilt}}, \ and\ \bibinfo {author} {\bibfnamefont {Nicola}\ \bibnamefont
  {Marzari}},\ }\bibfield  {title} {\enquote {\bibinfo {title} {{wannier90: A
  tool for obtaining maximally-localised Wannier functions}},}\ }\href
  {\doibase 10.1016/j.cpc.2007.11.016} {\bibfield  {journal} {\bibinfo
  {journal} {Comput. Phys. Commun.}\ }\textbf {\bibinfo {volume} {178}},\
  \bibinfo {pages} {685--699} (\bibinfo {year} {2008})}\BibitemShut {NoStop}%
\bibitem [{\citenamefont {Bintrim}\ \emph {et~al.}(2022)\citenamefont
  {Bintrim}, \citenamefont {Berkelbach},\ and\ \citenamefont
  {Ye}}]{Bintrim2022Sep}%
  \BibitemOpen
  \bibfield  {author} {\bibinfo {author} {\bibfnamefont {Sylvia~J.}\
  \bibnamefont {Bintrim}}, \bibinfo {author} {\bibfnamefont {Timothy~C.}\
  \bibnamefont {Berkelbach}}, \ and\ \bibinfo {author} {\bibfnamefont
  {Hong-Zhou}\ \bibnamefont {Ye}},\ }\bibfield  {title} {\enquote {\bibinfo
  {title} {{Integral-Direct Hartree{\textendash}Fock and
  M{\o}ller{\textendash}Plesset Perturbation Theory for Periodic Systems with
  Density Fitting: Application to the Benzene Crystal}},}\ }\href {\doibase
  10.1021/acs.jctc.2c00640} {\bibfield  {journal} {\bibinfo  {journal} {J.
  Chem. Theory Comput.}\ }\textbf {\bibinfo {volume} {18}},\ \bibinfo {pages}
  {5374--5381} (\bibinfo {year} {2022})}\BibitemShut {NoStop}%
\bibitem [{\citenamefont {Becke}(1993)}]{Becke1993Apr}%
  \BibitemOpen
  \bibfield  {author} {\bibinfo {author} {\bibfnamefont {Axel~D.}\ \bibnamefont
  {Becke}},\ }\bibfield  {title} {\enquote {\bibinfo {title}
  {{Density{-}functional thermochemistry. III. The role of exact exchange}},}\
  }\href {\doibase 10.1063/1.464913} {\bibfield  {journal} {\bibinfo  {journal}
  {J. Chem. Phys.}\ }\textbf {\bibinfo {volume} {98}},\ \bibinfo {pages}
  {5648--5652} (\bibinfo {year} {1993})}\BibitemShut {NoStop}%
\bibitem [{\citenamefont {Perdew}\ \emph
  {et~al.}(1996{\natexlab{b}})\citenamefont {Perdew}, \citenamefont
  {Ernzerhof},\ and\ \citenamefont {Burke}}]{Perdew1996Dec}%
  \BibitemOpen
  \bibfield  {author} {\bibinfo {author} {\bibfnamefont {John~P.}\ \bibnamefont
  {Perdew}}, \bibinfo {author} {\bibfnamefont {Matthias}\ \bibnamefont
  {Ernzerhof}}, \ and\ \bibinfo {author} {\bibfnamefont {Kieron}\ \bibnamefont
  {Burke}},\ }\bibfield  {title} {\enquote {\bibinfo {title} {{Rationale for
  mixing exact exchange with density functional approximations}},}\ }\href
  {\doibase 10.1063/1.472933} {\bibfield  {journal} {\bibinfo  {journal} {J.
  Chem. Phys.}\ }\textbf {\bibinfo {volume} {105}},\ \bibinfo {pages}
  {9982--9985} (\bibinfo {year} {1996}{\natexlab{b}})}\BibitemShut {NoStop}%
\bibitem [{\citenamefont {Zhang}\ and\ \citenamefont
  {Yang}(1998)}]{Zhang1998Jan}%
  \BibitemOpen
  \bibfield  {author} {\bibinfo {author} {\bibfnamefont {Yingkai}\ \bibnamefont
  {Zhang}}\ and\ \bibinfo {author} {\bibfnamefont {Weitao}\ \bibnamefont
  {Yang}},\ }\bibfield  {title} {\enquote {\bibinfo {title} {{Comment on
  ``Generalized Gradient Approximation Made Simple''}},}\ }\href {\doibase
  10.1103/PhysRevLett.80.890} {\bibfield  {journal} {\bibinfo  {journal} {Phys.
  Rev. Lett.}\ }\textbf {\bibinfo {volume} {80}},\ \bibinfo {pages} {890}
  (\bibinfo {year} {1998})}\BibitemShut {NoStop}%
\bibitem [{\citenamefont {Keal}\ and\ \citenamefont
  {Tozer}(2005)}]{Keal2005Sep}%
  \BibitemOpen
  \bibfield  {author} {\bibinfo {author} {\bibfnamefont {Thomas~W.}\
  \bibnamefont {Keal}}\ and\ \bibinfo {author} {\bibfnamefont {David~J.}\
  \bibnamefont {Tozer}},\ }\bibfield  {title} {\enquote {\bibinfo {title}
  {{Semiempirical hybrid functional with improved performance in an extensive
  chemical assessment}},}\ }\href {\doibase 10.1063/1.2061227} {\bibfield
  {journal} {\bibinfo  {journal} {J. Chem. Phys.}\ }\textbf {\bibinfo {volume}
  {123}},\ \bibinfo {pages} {121103} (\bibinfo {year} {2005})}\BibitemShut
  {NoStop}%
\bibitem [{\citenamefont {Zhao}\ and\ \citenamefont
  {Truhlar}(2008)}]{Zhao2008May}%
  \BibitemOpen
  \bibfield  {author} {\bibinfo {author} {\bibfnamefont {Yan}\ \bibnamefont
  {Zhao}}\ and\ \bibinfo {author} {\bibfnamefont {Donald~G.}\ \bibnamefont
  {Truhlar}},\ }\bibfield  {title} {\enquote {\bibinfo {title} {{The M06 suite
  of density functionals for main group thermochemistry, thermochemical
  kinetics, noncovalent interactions, excited states, and transition elements:
  two new functionals and systematic testing of four M06-class functionals and
  12 other functionals}},}\ }\href {\doibase 10.1007/s00214-007-0310-x}
  {\bibfield  {journal} {\bibinfo  {journal} {Theor. Chem. Acc.}\ }\textbf
  {\bibinfo {volume} {120}},\ \bibinfo {pages} {215--241} (\bibinfo {year}
  {2008})}\BibitemShut {NoStop}%
\bibitem [{\citenamefont {Yu}\ \emph {et~al.}(2016{\natexlab{b}})\citenamefont
  {Yu}, \citenamefont {He}, \citenamefont {Li},\ and\ \citenamefont
  {Truhlar}}]{Yu2016Jul}%
  \BibitemOpen
  \bibfield  {author} {\bibinfo {author} {\bibfnamefont {Haoyu~S.}\
  \bibnamefont {Yu}}, \bibinfo {author} {\bibfnamefont {Xiao}\ \bibnamefont
  {He}}, \bibinfo {author} {\bibfnamefont {Shaohong~L.}\ \bibnamefont {Li}}, \
  and\ \bibinfo {author} {\bibfnamefont {Donald~G.}\ \bibnamefont {Truhlar}},\
  }\bibfield  {title} {\enquote {\bibinfo {title} {{MN15: A
  Kohn{\textendash}Sham global-hybrid exchange{\textendash}correlation density
  functional with broad accuracy for multi-reference and single-reference
  systems and noncovalent interactions}},}\ }\href {\doibase
  10.1039/C6SC00705H} {\bibfield  {journal} {\bibinfo  {journal} {Chem. Sci.}\
  }\textbf {\bibinfo {volume} {7}},\ \bibinfo {pages} {5032--5051} (\bibinfo
  {year} {2016}{\natexlab{b}})}\BibitemShut {NoStop}%
\bibitem [{\citenamefont {Hui}\ and\ \citenamefont {Chai}(2016)}]{Hui2016Jan}%
  \BibitemOpen
  \bibfield  {author} {\bibinfo {author} {\bibfnamefont {Kerwin}\ \bibnamefont
  {Hui}}\ and\ \bibinfo {author} {\bibfnamefont {Jeng-Da}\ \bibnamefont
  {Chai}},\ }\bibfield  {title} {\enquote {\bibinfo {title} {{SCAN-based hybrid
  and double-hybrid density functionals from models without fitted
  parameters}},}\ }\href {\doibase 10.1063/1.4940734} {\bibfield  {journal}
  {\bibinfo  {journal} {J. Chem. Phys.}\ }\textbf {\bibinfo {volume} {144}},\
  \bibinfo {pages} {044114} (\bibinfo {year} {2016})}\BibitemShut {NoStop}%
\bibitem [{\citenamefont {Heyd}\ \emph {et~al.}(2003)\citenamefont {Heyd},
  \citenamefont {Scuseria},\ and\ \citenamefont {Ernzerhof}}]{Heyd2003May}%
  \BibitemOpen
  \bibfield  {author} {\bibinfo {author} {\bibfnamefont {Jochen}\ \bibnamefont
  {Heyd}}, \bibinfo {author} {\bibfnamefont {Gustavo~E.}\ \bibnamefont
  {Scuseria}}, \ and\ \bibinfo {author} {\bibfnamefont {Matthias}\ \bibnamefont
  {Ernzerhof}},\ }\bibfield  {title} {\enquote {\bibinfo {title} {{Hybrid
  functionals based on a screened Coulomb potential}},}\ }\href {\doibase
  10.1063/1.1564060} {\bibfield  {journal} {\bibinfo  {journal} {J. Chem.
  Phys.}\ }\textbf {\bibinfo {volume} {118}},\ \bibinfo {pages} {8207--8215}
  (\bibinfo {year} {2003})}\BibitemShut {NoStop}%
\bibitem [{\citenamefont {Krukau}\ \emph {et~al.}(2006)\citenamefont {Krukau},
  \citenamefont {Vydrov}, \citenamefont {Izmaylov},\ and\ \citenamefont
  {Scuseria}}]{Krukau2006Dec}%
  \BibitemOpen
  \bibfield  {author} {\bibinfo {author} {\bibfnamefont {Aliaksandr~V.}\
  \bibnamefont {Krukau}}, \bibinfo {author} {\bibfnamefont {Oleg~A.}\
  \bibnamefont {Vydrov}}, \bibinfo {author} {\bibfnamefont {Artur~F.}\
  \bibnamefont {Izmaylov}}, \ and\ \bibinfo {author} {\bibfnamefont
  {Gustavo~E.}\ \bibnamefont {Scuseria}},\ }\bibfield  {title} {\enquote
  {\bibinfo {title} {{Influence of the exchange screening parameter on the
  performance of screened hybrid functionals}},}\ }\href {\doibase
  10.1063/1.2404663} {\bibfield  {journal} {\bibinfo  {journal} {J. Chem.
  Phys.}\ }\textbf {\bibinfo {volume} {125}},\ \bibinfo {pages} {224106}
  (\bibinfo {year} {2006})}\BibitemShut {NoStop}%
\bibitem [{\citenamefont {Heyd}\ \emph {et~al.}(2006)\citenamefont {Heyd},
  \citenamefont {Scuseria},\ and\ \citenamefont {Ernzerhof}}]{Heyd2006Jun}%
  \BibitemOpen
  \bibfield  {author} {\bibinfo {author} {\bibfnamefont {Jochen}\ \bibnamefont
  {Heyd}}, \bibinfo {author} {\bibfnamefont {Gustavo~E.}\ \bibnamefont
  {Scuseria}}, \ and\ \bibinfo {author} {\bibfnamefont {Matthias}\ \bibnamefont
  {Ernzerhof}},\ }\bibfield  {title} {\enquote {\bibinfo {title} {{Erratum:
  {\textquotedblleft}Hybrid functionals based on a screened Coulomb
  potential{\textquotedblright} [J. Chem. Phys. 118, 8207 (2003)]}},}\ }\href
  {\doibase 10.1063/1.2204597} {\bibfield  {journal} {\bibinfo  {journal} {J.
  Chem. Phys.}\ }\textbf {\bibinfo {volume} {124}},\ \bibinfo {pages} {219906}
  (\bibinfo {year} {2006})}\BibitemShut {NoStop}%
\bibitem [{\citenamefont {Henderson}\ \emph
  {et~al.}(2008{\natexlab{a}})\citenamefont {Henderson}, \citenamefont
  {Janesko},\ and\ \citenamefont {Scuseria}}]{Henderson2008May}%
  \BibitemOpen
  \bibfield  {author} {\bibinfo {author} {\bibfnamefont {Thomas~M.}\
  \bibnamefont {Henderson}}, \bibinfo {author} {\bibfnamefont {Benjamin~G.}\
  \bibnamefont {Janesko}}, \ and\ \bibinfo {author} {\bibfnamefont
  {Gustavo~E.}\ \bibnamefont {Scuseria}},\ }\bibfield  {title} {\enquote
  {\bibinfo {title} {{Generalized gradient approximation model exchange holes
  for range-separated hybrids}},}\ }\href {\doibase 10.1063/1.2921797}
  {\bibfield  {journal} {\bibinfo  {journal} {J. Chem. Phys.}\ }\textbf
  {\bibinfo {volume} {128}},\ \bibinfo {pages} {194105} (\bibinfo {year}
  {2008}{\natexlab{a}})}\BibitemShut {NoStop}%
\bibitem [{\citenamefont {Yanai}\ \emph {et~al.}(2004)\citenamefont {Yanai},
  \citenamefont {Tew},\ and\ \citenamefont {Handy}}]{Yanai2004Jul}%
  \BibitemOpen
  \bibfield  {author} {\bibinfo {author} {\bibfnamefont {Takeshi}\ \bibnamefont
  {Yanai}}, \bibinfo {author} {\bibfnamefont {David~P.}\ \bibnamefont {Tew}}, \
  and\ \bibinfo {author} {\bibfnamefont {Nicholas~C.}\ \bibnamefont {Handy}},\
  }\bibfield  {title} {\enquote {\bibinfo {title} {{A new hybrid
  exchange{\textendash}correlation functional using the Coulomb-attenuating
  method (CAM-B3LYP)}},}\ }\href {\doibase 10.1016/j.cplett.2004.06.011}
  {\bibfield  {journal} {\bibinfo  {journal} {Chem. Phys. Lett.}\ }\textbf
  {\bibinfo {volume} {393}},\ \bibinfo {pages} {51--57} (\bibinfo {year}
  {2004})}\BibitemShut {NoStop}%
\bibitem [{\citenamefont {Jin}\ and\ \citenamefont
  {Bartlett}(2016)}]{Jin2016Jul}%
  \BibitemOpen
  \bibfield  {author} {\bibinfo {author} {\bibfnamefont {Yifan}\ \bibnamefont
  {Jin}}\ and\ \bibinfo {author} {\bibfnamefont {Rodney~J.}\ \bibnamefont
  {Bartlett}},\ }\bibfield  {title} {\enquote {\bibinfo {title} {{The QTP
  family of consistent functionals and potentials in Kohn-Sham density
  functional theory}},}\ }\href {\doibase 10.1063/1.4955497} {\bibfield
  {journal} {\bibinfo  {journal} {J. Chem. Phys.}\ }\textbf {\bibinfo {volume}
  {145}},\ \bibinfo {pages} {034107} (\bibinfo {year} {2016})}\BibitemShut
  {NoStop}%
\bibitem [{\citenamefont {Grimme}(2006)}]{Grimme2006Nov}%
  \BibitemOpen
  \bibfield  {author} {\bibinfo {author} {\bibfnamefont {Stefan}\ \bibnamefont
  {Grimme}},\ }\bibfield  {title} {\enquote {\bibinfo {title} {{Semiempirical
  GGA-type density functional constructed with a long-range dispersion
  correction}},}\ }\href {\doibase 10.1002/jcc.20495} {\bibfield  {journal}
  {\bibinfo  {journal} {J. Comput. Chem.}\ }\textbf {\bibinfo {volume} {27}},\
  \bibinfo {pages} {1787--1799} (\bibinfo {year} {2006})}\BibitemShut {NoStop}%
\bibitem [{\citenamefont {Grimme}\ \emph {et~al.}(2010)\citenamefont {Grimme},
  \citenamefont {Antony}, \citenamefont {Ehrlich},\ and\ \citenamefont
  {Krieg}}]{Grimme2010Apr}%
  \BibitemOpen
  \bibfield  {author} {\bibinfo {author} {\bibfnamefont {Stefan}\ \bibnamefont
  {Grimme}}, \bibinfo {author} {\bibfnamefont {Jens}\ \bibnamefont {Antony}},
  \bibinfo {author} {\bibfnamefont {Stephan}\ \bibnamefont {Ehrlich}}, \ and\
  \bibinfo {author} {\bibfnamefont {Helge}\ \bibnamefont {Krieg}},\ }\bibfield
  {title} {\enquote {\bibinfo {title} {{A consistent and accurate ab initio
  parametrization of density functional dispersion correction (DFT-D) for the
  94 elements H-Pu}},}\ }\href {\doibase 10.1063/1.3382344} {\bibfield
  {journal} {\bibinfo  {journal} {J. Chem. Phys.}\ }\textbf {\bibinfo {volume}
  {132}},\ \bibinfo {pages} {154104} (\bibinfo {year} {2010})}\BibitemShut
  {NoStop}%
\bibitem [{\citenamefont {Grimme}\ \emph {et~al.}(2011)\citenamefont {Grimme},
  \citenamefont {Ehrlich},\ and\ \citenamefont {Goerigk}}]{Grimme2011May}%
  \BibitemOpen
  \bibfield  {author} {\bibinfo {author} {\bibfnamefont {Stefan}\ \bibnamefont
  {Grimme}}, \bibinfo {author} {\bibfnamefont {Stephan}\ \bibnamefont
  {Ehrlich}}, \ and\ \bibinfo {author} {\bibfnamefont {Lars}\ \bibnamefont
  {Goerigk}},\ }\bibfield  {title} {\enquote {\bibinfo {title} {{Effect of the
  damping function in dispersion corrected density functional theory}},}\
  }\href {\doibase 10.1002/jcc.21759} {\bibfield  {journal} {\bibinfo
  {journal} {J. Comput. Chem.}\ }\textbf {\bibinfo {volume} {32}},\ \bibinfo
  {pages} {1456--1465} (\bibinfo {year} {2011})}\BibitemShut {NoStop}%
\bibitem [{\citenamefont {Borlido}\ \emph {et~al.}(2019)\citenamefont
  {Borlido}, \citenamefont {Aull}, \citenamefont {Huran}, \citenamefont {Tran},
  \citenamefont {Marques},\ and\ \citenamefont {Botti}}]{Borlido2019Sep}%
  \BibitemOpen
  \bibfield  {author} {\bibinfo {author} {\bibfnamefont {Pedro}\ \bibnamefont
  {Borlido}}, \bibinfo {author} {\bibfnamefont {Thorsten}\ \bibnamefont
  {Aull}}, \bibinfo {author} {\bibfnamefont {Ahmad~W.}\ \bibnamefont {Huran}},
  \bibinfo {author} {\bibfnamefont {Fabien}\ \bibnamefont {Tran}}, \bibinfo
  {author} {\bibfnamefont {Miguel A.~L.}\ \bibnamefont {Marques}}, \ and\
  \bibinfo {author} {\bibfnamefont {Silvana}\ \bibnamefont {Botti}},\
  }\bibfield  {title} {\enquote {\bibinfo {title} {{Large-Scale Benchmark of
  Exchange{\textendash}Correlation Functionals for the Determination of
  Electronic Band Gaps of Solids}},}\ }\href {\doibase
  10.1021/acs.jctc.9b00322} {\bibfield  {journal} {\bibinfo  {journal} {J.
  Chem. Theory Comput.}\ }\textbf {\bibinfo {volume} {15}},\ \bibinfo {pages}
  {5069--5079} (\bibinfo {year} {2019})}\BibitemShut {NoStop}%
\bibitem [{\citenamefont {Goedecker}\ \emph {et~al.}(1996)\citenamefont
  {Goedecker}, \citenamefont {Teter},\ and\ \citenamefont
  {Hutter}}]{Goedecker1996Jul}%
  \BibitemOpen
  \bibfield  {author} {\bibinfo {author} {\bibfnamefont {S.}~\bibnamefont
  {Goedecker}}, \bibinfo {author} {\bibfnamefont {M.}~\bibnamefont {Teter}}, \
  and\ \bibinfo {author} {\bibfnamefont {J.}~\bibnamefont {Hutter}},\
  }\bibfield  {title} {\enquote {\bibinfo {title} {{Separable dual-space
  Gaussian pseudopotentials}},}\ }\href {\doibase 10.1103/PhysRevB.54.1703}
  {\bibfield  {journal} {\bibinfo  {journal} {Phys. Rev. B}\ }\textbf {\bibinfo
  {volume} {54}},\ \bibinfo {pages} {1703--1710} (\bibinfo {year}
  {1996})}\BibitemShut {NoStop}%
\bibitem [{\citenamefont {Hartwigsen}\ \emph {et~al.}(1998)\citenamefont
  {Hartwigsen}, \citenamefont {Goedecker},\ and\ \citenamefont
  {Hutter}}]{Hartwigsen1998Aug}%
  \BibitemOpen
  \bibfield  {author} {\bibinfo {author} {\bibfnamefont {C.}~\bibnamefont
  {Hartwigsen}}, \bibinfo {author} {\bibfnamefont {S.}~\bibnamefont
  {Goedecker}}, \ and\ \bibinfo {author} {\bibfnamefont {J.}~\bibnamefont
  {Hutter}},\ }\bibfield  {title} {\enquote {\bibinfo {title} {{Relativistic
  separable dual-space Gaussian pseudopotentials from H to Rn}},}\ }\href
  {\doibase 10.1103/PhysRevB.58.3641} {\bibfield  {journal} {\bibinfo
  {journal} {Phys. Rev. B}\ }\textbf {\bibinfo {volume} {58}},\ \bibinfo
  {pages} {3641--3662} (\bibinfo {year} {1998})}\BibitemShut {NoStop}%
\bibitem [{\citenamefont {VandeVondele}\ and\ \citenamefont
  {Hutter}(2007)}]{VandeVondele2007Sep}%
  \BibitemOpen
  \bibfield  {author} {\bibinfo {author} {\bibfnamefont {Joost}\ \bibnamefont
  {VandeVondele}}\ and\ \bibinfo {author} {\bibfnamefont
  {J{\ifmmode\ddot{u}\else\"{u}\fi}rg}\ \bibnamefont {Hutter}},\ }\bibfield
  {title} {\enquote {\bibinfo {title} {{Gaussian basis sets for accurate
  calculations on molecular systems in gas and condensed phases}},}\ }\href
  {\doibase 10.1063/1.2770708} {\bibfield  {journal} {\bibinfo  {journal} {J.
  Chem. Phys.}\ }\textbf {\bibinfo {volume} {127}},\ \bibinfo {pages} {114105}
  (\bibinfo {year} {2007})}\BibitemShut {NoStop}%
\bibitem [{\citenamefont {Civalleri}\ \emph {et~al.}(2012)\citenamefont
  {Civalleri}, \citenamefont {Presti}, \citenamefont {Dovesi},\ and\
  \citenamefont {Savin}}]{Civalleri2012Oct}%
  \BibitemOpen
  \bibfield  {author} {\bibinfo {author} {\bibfnamefont {Bartolomeo}\
  \bibnamefont {Civalleri}}, \bibinfo {author} {\bibfnamefont {Davide}\
  \bibnamefont {Presti}}, \bibinfo {author} {\bibfnamefont {Roberto}\
  \bibnamefont {Dovesi}}, \ and\ \bibinfo {author} {\bibfnamefont {Andreas}\
  \bibnamefont {Savin}},\ }\bibfield  {title} {\enquote {\bibinfo {title} {{On
  choosing the best density functional approximation}},}\ }in\ \href {\doibase
  10.1039/9781849734790-00168} {\emph {\bibinfo {booktitle} {{Chemical
  Modelling: Applications and Theory Volume 9}}}},\ Vol.~\bibinfo {volume} {9}\
  (\bibinfo  {publisher} {The Royal Society of Chemistry},\ \bibinfo {year}
  {2012})\ pp.\ \bibinfo {pages} {168--185}\BibitemShut {NoStop}%
\bibitem [{\citenamefont {Koller}\ \emph {et~al.}(2013)\citenamefont {Koller},
  \citenamefont {Blaha},\ and\ \citenamefont {Tran}}]{Koller2013Oct}%
  \BibitemOpen
  \bibfield  {author} {\bibinfo {author} {\bibfnamefont {David}\ \bibnamefont
  {Koller}}, \bibinfo {author} {\bibfnamefont {Peter}\ \bibnamefont {Blaha}}, \
  and\ \bibinfo {author} {\bibfnamefont {Fabien}\ \bibnamefont {Tran}},\
  }\bibfield  {title} {\enquote {\bibinfo {title} {{Hybrid functionals for
  solids with an optimized Hartree{\textendash}Fock mixing parameter}},}\
  }\href {\doibase 10.1088/0953-8984/25/43/435503} {\bibfield  {journal}
  {\bibinfo  {journal} {J. Phys.: Condens. Matter}\ }\textbf {\bibinfo {volume}
  {25}},\ \bibinfo {pages} {435503} (\bibinfo {year} {2013})}\BibitemShut
  {NoStop}%
\bibitem [{\citenamefont {Skone}\ \emph {et~al.}(2014)\citenamefont {Skone},
  \citenamefont {Govoni},\ and\ \citenamefont {Galli}}]{Skone2014May}%
  \BibitemOpen
  \bibfield  {author} {\bibinfo {author} {\bibfnamefont {Jonathan~H.}\
  \bibnamefont {Skone}}, \bibinfo {author} {\bibfnamefont {Marco}\ \bibnamefont
  {Govoni}}, \ and\ \bibinfo {author} {\bibfnamefont {Giulia}\ \bibnamefont
  {Galli}},\ }\href {\doibase 10.1103/PhysRevB.89.195112} {\enquote {\bibinfo
  {title} {{Self-consistent hybrid functional for condensed systems}},}\ }
  (\bibinfo {year} {2014})\BibitemShut {NoStop}%
\bibitem [{\citenamefont {Henderson}\ \emph {et~al.}(2007)\citenamefont
  {Henderson}, \citenamefont {Izmaylov}, \citenamefont {Scuseria},\ and\
  \citenamefont {Savin}}]{Henderson2007Dec}%
  \BibitemOpen
  \bibfield  {author} {\bibinfo {author} {\bibfnamefont {Thomas~M.}\
  \bibnamefont {Henderson}}, \bibinfo {author} {\bibfnamefont {Artur~F.}\
  \bibnamefont {Izmaylov}}, \bibinfo {author} {\bibfnamefont {Gustavo~E.}\
  \bibnamefont {Scuseria}}, \ and\ \bibinfo {author} {\bibfnamefont {Andreas}\
  \bibnamefont {Savin}},\ }\bibfield  {title} {\enquote {\bibinfo {title} {{The
  importance of middle-range Hartree-Fock-type exchange for hybrid density
  functionals}},}\ }\href {\doibase 10.1063/1.2822021} {\bibfield  {journal}
  {\bibinfo  {journal} {J. Chem. Phys.}\ }\textbf {\bibinfo {volume} {127}},\
  \bibinfo {pages} {221103} (\bibinfo {year} {2007})}\BibitemShut {NoStop}%
\bibitem [{\citenamefont {Henderson}\ \emph
  {et~al.}(2008{\natexlab{b}})\citenamefont {Henderson}, \citenamefont
  {Izmaylov}, \citenamefont {Scuseria},\ and\ \citenamefont
  {Savin}}]{Henderson2008Aug}%
  \BibitemOpen
  \bibfield  {author} {\bibinfo {author} {\bibfnamefont {Thomas~M.}\
  \bibnamefont {Henderson}}, \bibinfo {author} {\bibfnamefont {Artur~F.}\
  \bibnamefont {Izmaylov}}, \bibinfo {author} {\bibfnamefont {Gustavo~E.}\
  \bibnamefont {Scuseria}}, \ and\ \bibinfo {author} {\bibfnamefont {Andreas}\
  \bibnamefont {Savin}},\ }\bibfield  {title} {\enquote {\bibinfo {title}
  {{Assessment of a Middle-Range Hybrid Functional}},}\ }\href {\doibase
  10.1021/ct800149y} {\bibfield  {journal} {\bibinfo  {journal} {J. Chem.
  Theory Comput.}\ }\textbf {\bibinfo {volume} {4}},\ \bibinfo {pages}
  {1254--1262} (\bibinfo {year} {2008}{\natexlab{b}})}\BibitemShut {NoStop}%
\bibitem [{\citenamefont {Janesko}\ \emph {et~al.}(2009)\citenamefont
  {Janesko}, \citenamefont {Henderson},\ and\ \citenamefont
  {Scuseria}}]{Janesko2009Jan}%
  \BibitemOpen
  \bibfield  {author} {\bibinfo {author} {\bibfnamefont {Benjamin~G.}\
  \bibnamefont {Janesko}}, \bibinfo {author} {\bibfnamefont {Thomas~M.}\
  \bibnamefont {Henderson}}, \ and\ \bibinfo {author} {\bibfnamefont
  {Gustavo~E.}\ \bibnamefont {Scuseria}},\ }\bibfield  {title} {\enquote
  {\bibinfo {title} {{Screened hybrid density functionals for solid-state
  chemistry and physics}},}\ }\href {\doibase 10.1039/B812838C} {\bibfield
  {journal} {\bibinfo  {journal} {Phys. Chem. Chem. Phys.}\ }\textbf {\bibinfo
  {volume} {11}},\ \bibinfo {pages} {443--454} (\bibinfo {year}
  {2009})}\BibitemShut {NoStop}%
\bibitem [{\citenamefont {Brawand}\ \emph {et~al.}(2016)\citenamefont
  {Brawand}, \citenamefont
  {V{\ifmmode\ddot{o}\else\"{o}\fi}r{\ifmmode\ddot{o}\else\"{o}\fi}s},
  \citenamefont {Govoni},\ and\ \citenamefont {Galli}}]{Brawand2016Oct}%
  \BibitemOpen
  \bibfield  {author} {\bibinfo {author} {\bibfnamefont {Nicholas~P.}\
  \bibnamefont {Brawand}}, \bibinfo {author} {\bibfnamefont
  {M{\ifmmode\acute{a}\else\'{a}\fi}rton}\ \bibnamefont
  {V{\ifmmode\ddot{o}\else\"{o}\fi}r{\ifmmode\ddot{o}\else\"{o}\fi}s}},
  \bibinfo {author} {\bibfnamefont {Marco}\ \bibnamefont {Govoni}}, \ and\
  \bibinfo {author} {\bibfnamefont {Giulia}\ \bibnamefont {Galli}},\ }\bibfield
   {title} {\enquote {\bibinfo {title} {{Generalization of Dielectric-Dependent
  Hybrid Functionals to Finite Systems}},}\ }\href {\doibase
  10.1103/PhysRevX.6.041002} {\bibfield  {journal} {\bibinfo  {journal} {Phys.
  Rev. X}\ }\textbf {\bibinfo {volume} {6}},\ \bibinfo {pages} {041002}
  (\bibinfo {year} {2016})}\BibitemShut {NoStop}%
\bibitem [{\citenamefont {Atalla}\ \emph {et~al.}(2013)\citenamefont {Atalla},
  \citenamefont {Yoon}, \citenamefont {Caruso}, \citenamefont {Rinke},\ and\
  \citenamefont {Scheffler}}]{Atalla2013Oct}%
  \BibitemOpen
  \bibfield  {author} {\bibinfo {author} {\bibfnamefont {Viktor}\ \bibnamefont
  {Atalla}}, \bibinfo {author} {\bibfnamefont {Mina}\ \bibnamefont {Yoon}},
  \bibinfo {author} {\bibfnamefont {Fabio}\ \bibnamefont {Caruso}}, \bibinfo
  {author} {\bibfnamefont {Patrick}\ \bibnamefont {Rinke}}, \ and\ \bibinfo
  {author} {\bibfnamefont {Matthias}\ \bibnamefont {Scheffler}},\ }\bibfield
  {title} {\enquote {\bibinfo {title} {{Hybrid density functional theory meets
  quasiparticle calculations: A consistent electronic structure approach}},}\
  }\href {\doibase 10.1103/PhysRevB.88.165122} {\bibfield  {journal} {\bibinfo
  {journal} {Phys. Rev. B}\ }\textbf {\bibinfo {volume} {88}},\ \bibinfo
  {pages} {165122} (\bibinfo {year} {2013})}\BibitemShut {NoStop}%
\bibitem [{\citenamefont {Pinheiro}\ \emph {et~al.}(2015)\citenamefont
  {Pinheiro}, \citenamefont {Caldas}, \citenamefont {Rinke}, \citenamefont
  {Blum},\ and\ \citenamefont {Scheffler}}]{Pinheiro2015Nov}%
  \BibitemOpen
  \bibfield  {author} {\bibinfo {author} {\bibfnamefont {Max}\ \bibnamefont
  {Pinheiro}}, \bibinfo {author} {\bibfnamefont {Marilia~J.}\ \bibnamefont
  {Caldas}}, \bibinfo {author} {\bibfnamefont {Patrick}\ \bibnamefont {Rinke}},
  \bibinfo {author} {\bibfnamefont {Volker}\ \bibnamefont {Blum}}, \ and\
  \bibinfo {author} {\bibfnamefont {Matthias}\ \bibnamefont {Scheffler}},\
  }\bibfield  {title} {\enquote {\bibinfo {title} {{Length dependence of
  ionization potentials of transacetylenes: Internally consistent DFT/$GW$
  approach}},}\ }\href {\doibase 10.1103/PhysRevB.92.195134} {\bibfield
  {journal} {\bibinfo  {journal} {Phys. Rev. B}\ }\textbf {\bibinfo {volume}
  {92}},\ \bibinfo {pages} {195134} (\bibinfo {year} {2015})}\BibitemShut
  {NoStop}%
\bibitem [{\citenamefont {Marom}(2017)}]{Marom2017Jan}%
  \BibitemOpen
  \bibfield  {author} {\bibinfo {author} {\bibfnamefont {Noa}\ \bibnamefont
  {Marom}},\ }\bibfield  {title} {\enquote {\bibinfo {title} {{Accurate
  description of the electronic structure of organic semiconductors by GW
  methods}},}\ }\href {\doibase 10.1088/1361-648x/29/10/103003} {\bibfield
  {journal} {\bibinfo  {journal} {J. Phys.: Condens. Matter}\ }\textbf
  {\bibinfo {volume} {29}},\ \bibinfo {pages} {103003} (\bibinfo {year}
  {2017})}\BibitemShut {NoStop}%
\bibitem [{\citenamefont {Hohenstein}\ \emph {et~al.}(2012)\citenamefont
  {Hohenstein}, \citenamefont {Parrish},\ and\ \citenamefont
  {Mart{\'{i}}nez}}]{Hohenstein2012}%
  \BibitemOpen
  \bibfield  {author} {\bibinfo {author} {\bibfnamefont {Edward~G.}\
  \bibnamefont {Hohenstein}}, \bibinfo {author} {\bibfnamefont {Robert~M.}\
  \bibnamefont {Parrish}}, \ and\ \bibinfo {author} {\bibfnamefont {Todd~J.}\
  \bibnamefont {Mart{\'{i}}nez}},\ }\bibfield  {title} {\enquote {\bibinfo
  {title} {{Tensor hypercontraction density fitting. I. Quartic scaling second-
  and third-order M{\o}ller-Plesset perturbation theory}},}\ }\href {\doibase
  10.1063/1.4732310} {\bibfield  {journal} {\bibinfo  {journal} {J. Chem.
  Phys.}\ }\textbf {\bibinfo {volume} {137}},\ \bibinfo {pages} {1085}
  (\bibinfo {year} {2012})}\BibitemShut {NoStop}%
\bibitem [{\citenamefont {Lee}\ \emph {et~al.}(2019)\citenamefont {Lee},
  \citenamefont {Lin},\ and\ \citenamefont
  {Head-Gordon}}]{lee2019systematically}%
  \BibitemOpen
  \bibfield  {author} {\bibinfo {author} {\bibfnamefont {Joonho}\ \bibnamefont
  {Lee}}, \bibinfo {author} {\bibfnamefont {Lin}\ \bibnamefont {Lin}}, \ and\
  \bibinfo {author} {\bibfnamefont {Martin}\ \bibnamefont {Head-Gordon}},\
  }\bibfield  {title} {\enquote {\bibinfo {title} {Systematically improvable
  tensor hypercontraction: Interpolative separable density-fitting for
  molecules applied to exact exchange, second-and third-order m{\o}ller-plesset
  perturbation theory},}\ }\href {\doibase 10.1021/acs.jctc.9b00820} {\bibfield
   {journal} {\bibinfo  {journal} {Journal of Chemical Theory and Computation}\
  }\textbf {\bibinfo {volume} {16}},\ \bibinfo {pages} {243--263} (\bibinfo
  {year} {2019})}\BibitemShut {NoStop}%
\bibitem [{\citenamefont {Wu}\ \emph {et~al.}(2022)\citenamefont {Wu},
  \citenamefont {Qin}, \citenamefont {Hu},\ and\ \citenamefont
  {Yang}}]{Wu2022Jan}%
  \BibitemOpen
  \bibfield  {author} {\bibinfo {author} {\bibfnamefont {Kai}\ \bibnamefont
  {Wu}}, \bibinfo {author} {\bibfnamefont {Xinming}\ \bibnamefont {Qin}},
  \bibinfo {author} {\bibfnamefont {Wei}\ \bibnamefont {Hu}}, \ and\ \bibinfo
  {author} {\bibfnamefont {Jinlong}\ \bibnamefont {Yang}},\ }\bibfield  {title}
  {\enquote {\bibinfo {title} {{Low-Rank Approximations Accelerated Plane-Wave
  Hybrid Functional Calculations with k-Point Sampling}},}\ }\href {\doibase
  10.1021/acs.jctc.1c00874} {\bibfield  {journal} {\bibinfo  {journal} {J.
  Chem. Theory Comput.}\ }\textbf {\bibinfo {volume} {18}},\ \bibinfo {pages}
  {206--218} (\bibinfo {year} {2022})}\BibitemShut {NoStop}%
\bibitem [{\citenamefont {Euwema}\ \emph {et~al.}(1973)\citenamefont {Euwema},
  \citenamefont {Wilhite},\ and\ \citenamefont {Surratt}}]{Euwema1973Jan}%
  \BibitemOpen
  \bibfield  {author} {\bibinfo {author} {\bibfnamefont {R.~N.}\ \bibnamefont
  {Euwema}}, \bibinfo {author} {\bibfnamefont {D.~L.}\ \bibnamefont {Wilhite}},
  \ and\ \bibinfo {author} {\bibfnamefont {G.~T.}\ \bibnamefont {Surratt}},\
  }\bibfield  {title} {\enquote {\bibinfo {title} {{General Crystalline
  Hartree-Fock Formalism: Diamond Results}},}\ }\href {\doibase
  10.1103/PhysRevB.7.818} {\bibfield  {journal} {\bibinfo  {journal} {Phys.
  Rev. B}\ }\textbf {\bibinfo {volume} {7}},\ \bibinfo {pages} {818--831}
  (\bibinfo {year} {1973})}\BibitemShut {NoStop}%
\bibitem [{\citenamefont {Baraille}\ \emph {et~al.}(1998)\citenamefont
  {Baraille}, \citenamefont {Pouchan}, \citenamefont {Caus{\'a}},\ and\
  \citenamefont {Marinelli}}]{Baraille1998Dec}%
  \BibitemOpen
  \bibfield  {author} {\bibinfo {author} {\bibfnamefont {I.}~\bibnamefont
  {Baraille}}, \bibinfo {author} {\bibfnamefont {C.}~\bibnamefont {Pouchan}},
  \bibinfo {author} {\bibfnamefont {M.}~\bibnamefont {Caus{\'a}}}, \ and\
  \bibinfo {author} {\bibfnamefont {F.}~\bibnamefont {Marinelli}},\ }\bibfield
  {title} {\enquote {\bibinfo {title} {{Comparison between Hartree-Fock and
  Kohn-Sham electronic and structural properties for hexagonal-close-packed
  magnesium}},}\ }\href {\doibase 10.1088/0953-8984/10/48/017} {\bibfield
  {journal} {\bibinfo  {journal} {J. Phys.: Condens. Matter}\ }\textbf
  {\bibinfo {volume} {10}},\ \bibinfo {pages} {10969--10977} (\bibinfo {year}
  {1998})}\BibitemShut {NoStop}%
\bibitem [{\citenamefont {Gillan}\ \emph {et~al.}(2008)\citenamefont {Gillan},
  \citenamefont {Alf{\'e}}, \citenamefont {de~Gironcoli},\ and\ \citenamefont
  {Manby}}]{Gillan2008Oct}%
  \BibitemOpen
  \bibfield  {author} {\bibinfo {author} {\bibfnamefont {M.~J.}\ \bibnamefont
  {Gillan}}, \bibinfo {author} {\bibfnamefont {D.}~\bibnamefont {Alf{\'e}}},
  \bibinfo {author} {\bibfnamefont {S.}~\bibnamefont {de~Gironcoli}}, \ and\
  \bibinfo {author} {\bibfnamefont {F.~R.}\ \bibnamefont {Manby}},\ }\bibfield
  {title} {\enquote {\bibinfo {title} {{High-precision calculation of
  Hartree-Fock energy of crystals}},}\ }\href {\doibase 10.1002/jcc.21033}
  {\bibfield  {journal} {\bibinfo  {journal} {J. Comput. Chem.}\ }\textbf
  {\bibinfo {volume} {29}},\ \bibinfo {pages} {2098--2106} (\bibinfo {year}
  {2008})}\BibitemShut {NoStop}%
\bibitem [{\citenamefont {Paier}\ \emph {et~al.}(2009)\citenamefont {Paier},
  \citenamefont {Diaconu}, \citenamefont {Scuseria}, \citenamefont {Guidon},
  \citenamefont {VandeVondele},\ and\ \citenamefont {Hutter}}]{Paier2009Nov}%
  \BibitemOpen
  \bibfield  {author} {\bibinfo {author} {\bibfnamefont {Joachim}\ \bibnamefont
  {Paier}}, \bibinfo {author} {\bibfnamefont {Cristian~V.}\ \bibnamefont
  {Diaconu}}, \bibinfo {author} {\bibfnamefont {Gustavo~E.}\ \bibnamefont
  {Scuseria}}, \bibinfo {author} {\bibfnamefont {Manuel}\ \bibnamefont
  {Guidon}}, \bibinfo {author} {\bibfnamefont {Joost}\ \bibnamefont
  {VandeVondele}}, \ and\ \bibinfo {author} {\bibfnamefont {J{\"u}rg}\
  \bibnamefont {Hutter}},\ }\bibfield  {title} {\enquote {\bibinfo {title}
  {{Accurate Hartree-Fock energy of extended systems using large Gaussian basis
  sets}},}\ }\href {\doibase 10.1103/PhysRevB.80.174114} {\bibfield  {journal}
  {\bibinfo  {journal} {Phys. Rev. B}\ }\textbf {\bibinfo {volume} {80}},\
  \bibinfo {pages} {174114} (\bibinfo {year} {2009})}\BibitemShut {NoStop}%
\bibitem [{\citenamefont {Civalleri}\ \emph {et~al.}(2010)\citenamefont
  {Civalleri}, \citenamefont {Orlando}, \citenamefont {Zicovich-Wilson},
  \citenamefont {Roetti}, \citenamefont {Saunders}, \citenamefont {Pisani},\
  and\ \citenamefont {Dovesi}}]{Civalleri2010Mar}%
  \BibitemOpen
  \bibfield  {author} {\bibinfo {author} {\bibfnamefont {Bartolomeo}\
  \bibnamefont {Civalleri}}, \bibinfo {author} {\bibfnamefont {Roberto}\
  \bibnamefont {Orlando}}, \bibinfo {author} {\bibfnamefont {Claudio~M.}\
  \bibnamefont {Zicovich-Wilson}}, \bibinfo {author} {\bibfnamefont {Carla}\
  \bibnamefont {Roetti}}, \bibinfo {author} {\bibfnamefont {Victor~R.}\
  \bibnamefont {Saunders}}, \bibinfo {author} {\bibfnamefont {Cesare}\
  \bibnamefont {Pisani}}, \ and\ \bibinfo {author} {\bibfnamefont {Roberto}\
  \bibnamefont {Dovesi}},\ }\bibfield  {title} {\enquote {\bibinfo {title}
  {{Comment on ``Accurate Hartree-Fock energy of extended systems using large
  Gaussian basis sets''}},}\ }\href {\doibase 10.1103/PhysRevB.81.106101}
  {\bibfield  {journal} {\bibinfo  {journal} {Phys. Rev. B}\ }\textbf {\bibinfo
  {volume} {81}},\ \bibinfo {pages} {106101} (\bibinfo {year}
  {2010})}\BibitemShut {NoStop}%
\bibitem [{\citenamefont {Tran}\ and\ \citenamefont
  {Blaha}(2017)}]{Tran2017May}%
  \BibitemOpen
  \bibfield  {author} {\bibinfo {author} {\bibfnamefont {Fabien}\ \bibnamefont
  {Tran}}\ and\ \bibinfo {author} {\bibfnamefont {Peter}\ \bibnamefont
  {Blaha}},\ }\bibfield  {title} {\enquote {\bibinfo {title} {{Importance of
  the Kinetic Energy Density for Band Gap Calculations in Solids with Density
  Functional Theory}},}\ }\href {\doibase 10.1021/acs.jpca.7b02882} {\bibfield
  {journal} {\bibinfo  {journal} {J. Phys. Chem. A}\ }\textbf {\bibinfo
  {volume} {121}},\ \bibinfo {pages} {3318--3325} (\bibinfo {year}
  {2017})}\BibitemShut {NoStop}%
\bibitem [{\citenamefont {Tran}\ \emph {et~al.}(2018)\citenamefont {Tran},
  \citenamefont {Ehsan},\ and\ \citenamefont {Blaha}}]{Tran2018Feb}%
  \BibitemOpen
  \bibfield  {author} {\bibinfo {author} {\bibfnamefont {Fabien}\ \bibnamefont
  {Tran}}, \bibinfo {author} {\bibfnamefont {Sohaib}\ \bibnamefont {Ehsan}}, \
  and\ \bibinfo {author} {\bibfnamefont {Peter}\ \bibnamefont {Blaha}},\
  }\bibfield  {title} {\enquote {\bibinfo {title} {{Assessment of the GLLB-SC
  potential for solid-state properties and attempts for improvement}},}\ }\href
  {\doibase 10.1103/PhysRevMaterials.2.023802} {\bibfield  {journal} {\bibinfo
  {journal} {Phys. Rev. Mater.}\ }\textbf {\bibinfo {volume} {2}},\ \bibinfo
  {pages} {023802} (\bibinfo {year} {2018})}\BibitemShut {NoStop}%
\bibitem [{\citenamefont {Mallikarjun~Sharada}\ \emph
  {et~al.}(2017)\citenamefont {Mallikarjun~Sharada}, \citenamefont {Bligaard},
  \citenamefont {Luntz}, \citenamefont {Kroes},\ and\ \citenamefont
  {N{\o}rskov}}]{MallikarjunSharada2017Sep}%
  \BibitemOpen
  \bibfield  {author} {\bibinfo {author} {\bibfnamefont {Shaama}\ \bibnamefont
  {Mallikarjun~Sharada}}, \bibinfo {author} {\bibfnamefont {Thomas}\
  \bibnamefont {Bligaard}}, \bibinfo {author} {\bibfnamefont {Alan~C.}\
  \bibnamefont {Luntz}}, \bibinfo {author} {\bibfnamefont {Geert-Jan}\
  \bibnamefont {Kroes}}, \ and\ \bibinfo {author} {\bibfnamefont {Jens~K.}\
  \bibnamefont {N{\o}rskov}},\ }\bibfield  {title} {\enquote {\bibinfo {title}
  {{SBH10: A Benchmark Database of Barrier Heights on Transition Metal
  Surfaces}},}\ }\href {\doibase 10.1021/acs.jpcc.7b05677} {\bibfield
  {journal} {\bibinfo  {journal} {J. Phys. Chem. C}\ }\textbf {\bibinfo
  {volume} {121}},\ \bibinfo {pages} {19807--19815} (\bibinfo {year}
  {2017})}\BibitemShut {NoStop}%
\bibitem [{\citenamefont {Maier}\ \emph {et~al.}(2019)\citenamefont {Maier},
  \citenamefont {Arbuznikov},\ and\ \citenamefont {Kaupp}}]{Maier2019Jan}%
  \BibitemOpen
  \bibfield  {author} {\bibinfo {author} {\bibfnamefont {Toni~M.}\ \bibnamefont
  {Maier}}, \bibinfo {author} {\bibfnamefont {Alexei~V.}\ \bibnamefont
  {Arbuznikov}}, \ and\ \bibinfo {author} {\bibfnamefont {Martin}\ \bibnamefont
  {Kaupp}},\ }\bibfield  {title} {\enquote {\bibinfo {title} {{Local hybrid
  functionals: Theory, implementation, and performance of an emerging new tool
  in quantum chemistry and beyond}},}\ }\href {\doibase 10.1002/wcms.1378}
  {\bibfield  {journal} {\bibinfo  {journal} {WIREs Comput. Mol. Sci.}\
  }\textbf {\bibinfo {volume} {9}},\ \bibinfo {pages} {e1378} (\bibinfo {year}
  {2019})}\BibitemShut {NoStop}%
\bibitem [{\citenamefont {Dahvyd}\ \emph {et~al.}(2021)\citenamefont {Dahvyd},
  \citenamefont {Guy}, \citenamefont {Haber~Jonah}, \citenamefont
  {Filip~Marina}, \citenamefont {Gant~Stephen}, \citenamefont
  {Neaton~Jeffrey},\ and\ \citenamefont {Leeor}}]{Dahvyd2021Aug}%
  \BibitemOpen
  \bibfield  {author} {\bibinfo {author} {\bibfnamefont {Wing}\ \bibnamefont
  {Dahvyd}}, \bibinfo {author} {\bibfnamefont {Ohad}\ \bibnamefont {Guy}},
  \bibinfo {author} {\bibfnamefont {B.}~\bibnamefont {Haber~Jonah}}, \bibinfo
  {author} {\bibfnamefont {R.}~\bibnamefont {Filip~Marina}}, \bibinfo {author}
  {\bibfnamefont {E.}~\bibnamefont {Gant~Stephen}}, \bibinfo {author}
  {\bibfnamefont {B.}~\bibnamefont {Neaton~Jeffrey}}, \ and\ \bibinfo {author}
  {\bibfnamefont {Kronik}\ \bibnamefont {Leeor}},\ }\bibfield  {title}
  {\enquote {\bibinfo {title} {{Band gaps of crystalline solids from
  Wannier-localization{\textendash}based optimal tuning of a screened
  range-separated hybrid functional}},}\ }\href {\doibase
  10.1073/pnas.2104556118} {\bibfield  {journal} {\bibinfo  {journal} {Proc.
  Natl. Acad. Sci. U.S.A.}\ }\textbf {\bibinfo {volume} {118}},\ \bibinfo
  {pages} {e2104556118} (\bibinfo {year} {2021})}\BibitemShut {NoStop}%
\bibitem [{\citenamefont {Sharma}\ \emph {et~al.}()\citenamefont {Sharma},
  \citenamefont {White},\ and\ \citenamefont {Beylkin}}]{Sharma2022Jul}%
  \BibitemOpen
  \bibfield  {author} {\bibinfo {author} {\bibfnamefont {Sandeep}\ \bibnamefont
  {Sharma}}, \bibinfo {author} {\bibfnamefont {Alec~F.}\ \bibnamefont {White}},
  \ and\ \bibinfo {author} {\bibfnamefont {Gregory}\ \bibnamefont {Beylkin}},\
  }\bibfield  {title} {\enquote {\bibinfo {title} {{Fast exchange with Gaussian
  basis set using robust pseudospectral method}},}\ }\href@noop {} {\ }\bibinfo
  {note} {ArXiv.2207.04636}\BibitemShut {NoStop}%
\end{thebibliography}%

\end{document}